\documentclass[pdflatex,sn-nature]{sn-jnl}

\usepackage{graphicx}
\usepackage{tabularx}
\usepackage{multirow}
\usepackage{amsmath,amssymb,amsfonts}
\usepackage{amsthm}
\usepackage{mathrsfs}
\usepackage{xcolor}
\usepackage{textcomp}
\usepackage{manyfoot}
\usepackage{booktabs}
\usepackage{algorithm}
\usepackage{algorithmicx}
\usepackage{algpseudocode}
\usepackage{listings}
\usepackage{subcaption}
\usepackage{longtable}
\usepackage{makecell}
\usepackage{rotating}
\usepackage{tikz}
\usetikzlibrary{mindmap, shapes.geometric, positioning, arrows.meta, shapes.callouts}
\usepackage{bm}
\usepackage{cleveref}

\crefname{figure}{Fig.}{Figs.}     \Crefname{figure}{Fig.}{Figs.}
\crefname{table}{Table}{Tables}    \Crefname{table}{Table}{Tables}
\crefname{sfigure}{fig.}{figs.}    \Crefname{sfigure}{Fig.}{Figs.}
\crefname{stable}{table}{tables}   \Crefname{stable}{Table}{Tables}

\newcommand{\EMM}{\operatorname{EMM}}
\newcommand{\SIref}[1]{section~\ref{#1}} %
\newcommand{\secref}[1]{section~\ref{#1}} %
\newcommand{\figscale}{0.51} %

\newcommand{\anim}{\text{anim}}
\newcommand{\stance}{\text{stance}}

\newcommand{\pre}{\text{pre}}
\newcommand{\pos}{\text{pos}}
\newcommand{\fol}{\text{fol}}
\providecommand{\dg}[1]{} %
\providecommand{\todo}[2][]{} %
\providecommand{\Endparasplit}{} %
\makeatletter
\newcommand\gs[1]{\@ifnextchar\bgroup{\gsdrop}{}} %
\newcommand\gsdrop[1]{}
\makeatother
\newcommand\td[1]{}

\newcommand{\Wrt}{\text{Wrt}}
\newcommand{\Dbt}{\text{Dbt}}
\newcommand{\Own}{\text{Own}}
\newcommand{\Opp}{\text{Opp}}

\newcommand{\win}{\text{win}}
\newcommand{\lose}{\text{lose}}

\newcommand{\arm}[2]{(#1,#2)}
\newcommand{\armtxt}{\text{arm}}

\newcommand{\delt}[2]{\Delta^{#1}(#2)}
\newcommand{\diffarms}[3]{\Delta^{#1}(#2) - \Delta^{#1}(#3)}

\newcommand{\DiD}{\operatorname{DiD}}
\newcommand{\did}[2]{\DiD^{#1}(#2)}

\graphicspath{ {figures/} }

\renewenvironment{table}[1][]%
{\begin{tableorg}[htbp]%
\begin{center}
\begin{threeparttable}
\tablebodyfont%
\renewcommand\footnotetext[2][]{{\removelastskip\vskip3pt%
\let\tablebodyfont\tablefootnotefont%
\hskip0pt\if!##1!\else{\smash{$^{##1}$}}\fi##2\par}}%
}{\end{threeparttable}\end{center}\end{tableorg}}
\renewenvironment{sidewaystable}[1][]%
{\begin{sidewaystableorg}[p]%
\begin{center}
\begin{threeparttable}
\tablebodyfont%
\renewcommand\footnotetext[2][]{{\removelastskip\vskip3pt%
\let\tablebodyfont\tablefootnotefont%
\hskip0pt\if!##1!\else{\smash{$^{##1}$}}\fi##2\par}}%
}{\end{threeparttable}\end{center}\end{sidewaystableorg}}

\begin{document}

\title[The Ideological Turing Test]{The Ideological Turing Test for moderation of issue-based affective animosity}

\author*[1]{\fnm{David} \sur{Gamba}}\email{gamba@umich.edu}
\author[1,2,3]{\fnm{Daniel M.} \sur{Romero}}\email{drom@umich.edu}
\author[1]{\fnm{Grant} \sur{Schoenebeck}}\email{schoeneb@umich.edu}

\affil*[1]{\orgdiv{School of Information}, \orgname{University of Michigan}, \orgaddress{\city{Ann Arbor}, \state{MI}, \country{USA}}}
\affil[2]{\orgdiv{Center for the Study of Complex Systems}, \orgname{University of Michigan}, \orgaddress{\city{Ann Arbor}, \state{MI}, \country{USA}}}
\affil[3]{\orgdiv{Department of Electrical Engineering and Computer Science}, \orgname{University of Michigan}, \orgaddress{\city{Ann Arbor}, \state{MI}, \country{USA}}}

\abstract{Rising animosity between political opponents presents a major societal challenge. We introduce the ``Ideological Turing Test,'' a gamified intervention in which participants argue opposing viewpoints on divisive political issues. In an experiment ($N=203$) varying interaction modality (debate vs. writing) and assigned perspective (own vs. opposite side), arguing the opposite side reduced issue-based affective animosity and shifted stance, but durability differed by modality. Writing produced a large immediate reduction in animosity that faded by follow-up, whereas debate produced a reduction that persisted at follow-up about two weeks later. Stance shifts persisted in both modalities. Contrary to concerns that adversarial formats entrench views, debate sustained the gains longer. Our framework offers a practical blueprint and design principles for durably reducing issue-based animosity.

}


\maketitle

\section{Introduction}\label{sec:intro}
Animosity toward people with different views, political opinions, or party affiliations is a growing concern in contemporary democracies.
When this animosity becomes prevalent across a population, it manifests as affective polarization: a society-level pattern where people systematically dislike and distrust those on the opposing political side \citep{iyengarOriginsConsequencesAffective2019, bakkerPuttingAffectAffective2024}. This collective phenomenon is pervasive across the United States \cite{hollidayAffectivePolarizationUniformly2024}, reshapes social behavior, erodes trust in democratic institutions \citep{druckmanWhatWeMeasure2019}, and exacerbates responses to crises \citep{druckmanAffectivePolarizationLocal2020}. These patterns have spurred research on interventions aimed at reducing animosity (and affective polarization), ranging from misperception correction \citep{voelkelMegastudyIdentifyingEffective2023} to perspective-taking exercises \citep{saveskiPerspectiveTakingReduceAffective2022}; yet, sustained effects remain elusive, as highlighted in recent evaluations of depolarization attempts \citep{hollidayWhyDepolarizationHard2025}. 

This party-based conceptualization captures only one form of political animosity. In this paper, we focus on \emph{issue-based animosity}: negative feelings toward people who hold opposing views on specific political issues, regardless of their party identification. This distinction matters because political conflict increasingly crystallizes around specific issues---such as animal rights, public health mandates during the COVID-19 pandemic, immigration enforcement, or Indigenous rights---that mobilize animosity across traditional partisan lines \citep{bliucOnlineIntergroupPolarization2021, stevensUsingEmotionsFrame2020, iannelliCitizensPolarizationPandemic2022, esauDivisionNotReconciliation2025}. Critically, we note that this conceptualization of animosity is contingent upon the \emph{stances} individuals adopt on the issues themselves \citep{iyengarConceptualizingAffectivePolarization2025, baldassarriEmergencePerilsPolarization2021}. However, while related, these dimensions can diverge in behavior; people may hold strong views on an issue without animosity toward opponents, or feel animosity while remaining moderate in their own position. This matters for intervention design because different mechanisms may affect these components differently, and durable change may require addressing both the cognitive component, reflected in stances, and the affective component, reflected in animosity \citep{hartmanInterventionsReducePartisan2022}.\Endparasplit

\begin{figure*}[tbp]
    \centering
    \includegraphics[width=.95\textwidth]{supporting/debate_disguise.pdf}
    \caption{\textbf{Design and incentives of the Ideological Turing Test.} \textbf{Left}: Intervention arms, 2x2 design including participant counts assigned to each arm. Here, the focus participant has four different possibilities once they engage in the in-person sessions. There is an issue at hand for which we have the participant's true opinion based on a pre-intervention survey. In this case, the question is: ``Should pineapple be on pizza?'' (only for pictorial purposes, as topics in our experiment are more serious and potentially divisive). Imagine that the participant agrees with the statement; they will be a true PRO. The participant can engage in either a debating modality against a peer (top row) or a writing-only modality (bottom row). However, the participant is randomly assigned to engage either in defending their own position (Own) or the opposite position (Opp), in which case they would be considered as ``a pretender''. We also note that debating and writing happen via an anonymized chat interface. \textbf{Right}: Judging and incentives. After the intervention occurs, a panel of judges will assess either the debate log or the written statements of paired participants. The judges assess not only who presents the best argument but also who is a pretender. An intervention participant only wins when both conditions are fulfilled. We use this future judgment as an incentivizing method for participants.}
    \label{fig:intervention}
\end{figure*}

Previous research on reducing animosity has broadly followed two currents.

The first current uses \textit{perspective-taking}: asking individuals to imagine the experiences of someone on the opposing side \citep{batsonPerspectiveTakingImagining2016, broockmanDurablyReducingTransphobia2016, saveskiPerspectiveTakingReduceAffective2022}. Following the dispositional empathy literature, we treat perspective-taking as the cognitive component of empathy, distinct from its affective component, empathic concern (feelings of compassion for others) \citep{davisMeasuringIndividualDifferences1983}. Although perspective-taking is a cognitive mechanism, interventions often deploy it with the goal of generating empathic concern as a pathway to reducing prejudice, and this approach has produced lasting reductions in prejudice in the field \citep{broockmanDurablyReducingTransphobia2016}. However, recent political science research suggests that targeting empathic concern can be parochial: it may strengthen ingroup bonds rather than bridge divides, and in some cases even increase affective polarization \citep{gillissenEmpathicConcernPerspectiveTaking2024, simasEmpathicConcernsPolarization2020}. Moreover, even when perspective-taking successfully reduces animosity, its effects tend to decay rapidly \citep{hollidayWhyDepolarizationHard2025}. We argue that this is because the cognitive engagement involved is shallow and does not require systematic elaboration over the opposite viewpoint.

The second current uses cognitive engagement through \textit{debate and deliberation}: requiring systematic processing of counter-attitudinal information through argument generation and exchange \citep{schwardmannSelfPersuasionEvidenceField2021, greenwaldOpenmindednessCounterattitudinalRole1969, moslehCognitiveReflectionCorrelates2021}. These interventions produce cognitive change that persists beyond the exchange itself---shifting factual beliefs, issue positions, and confidence in those positions---through self-persuasion. Yet they face a structural backfire risk: when debate becomes adversarial without guardrails, emotional escalation can replace reasoned argument, entrenching rather than reducing polarization \citep{loweTypesContactField2021, santoroPromisePitfallsCrosspartisan2022}. And because these studies typically measure cognitive rather than affective outcomes, we do not know whether durable cognitive change translates into reduced animosity toward opponents.

This creates a theoretical tension: the first current engages cognition shallowly and risks psychological backfire, while the second current achieves durable cognitive change but may not affect animosity and risks structural backfire. Neither current, on its own, has been reliably shown to reduce animosity in a durable way. We hypothesize that combining these two currents---explicitly incentivizing perspective-taking within a debate format---can overcome these limitations by simultaneously deepening cognitive engagement, which renders change long-lasting (debate), and directing that engagement at understanding opponents, which targets animosity (perspective-taking). This combination may also mitigate both backfire risks. Because participants argue an assigned position rather than defend their own identity, and interact anonymously under explicit rules, the exchange is less likely to escalate emotionally. And because the task rewards representing the opposing view convincingly rather than feeling compassion for its holders, it does not depend on empathic concern, which can be parochial.

We test this hypothesis by introducing \textit{structural perspective-taking}: embedding perspective-taking within a semi-structured debate format, thereby foregrounding perspective-taking as the central cognitive task, while requiring participants to construct and defend opposing views under incentives for authenticity and argument quality. We implement it as an \emph{Ideological Turing Test} \citep{mastroianniBothDemocratsRepublicans2024, brandIdeologicalTuringTest2025}: just as the original Turing Test assesses whether an observer can distinguish a human from an AI \citep{turingComputingMachineryIntelligence1950}, here judges evaluate whether participants arguing a position genuinely hold it or are ``pretenders''. This design deepens cognitive elaboration and explicitly activates perspective-taking as a practiced skill. First, it triggers self-persuasion: generating arguments for a counter-attitudinal position shifts one's own beliefs \citep{schwardmannSelfPersuasionEvidenceField2021}. Second, the interactive defense (adapting to a live opponent, maintaining authenticity) forces deeper processing than solitary writing. Elaboration likelihood theory predicts that deeper processing produces stronger, more lasting attitudes \citep{pettyElaborationLikelihoodModel1986}. If opponents' positions become more cognitively accessible, animosity toward them may decrease---even if agreement does not change. Thus, structural perspective-taking offers a pathway to durable reductions in animosity that neither perspective-taking alone nor debate alone has reliably achieved.

We use a $2\times2$ design varying modality (writing vs. debate) and perspective (own vs. opposite), with participants engaging on contentious issues such as reproductive rights and public health mandates. \Cref{fig:intervention} illustrates the four conditions. Each condition serves a distinct theoretical role:

\begin{description}
    \item[(Writing, Own)]: Participants express their own views without perspective-taking or interaction. This provides a baseline against which to assess any intervention effect. Control condition.
    \item[(Writing, Opposite)]: Perspective-taking alone. Participants construct written arguments for the opposing side. Prior research on such writing tasks shows they reduce animosity and extremity immediately \citep{tullerSeeingOtherSide2015, warnerReducingPoliticalPolarization2020}, but, like most depolarization interventions, such effects tend to decay within weeks \citep{hollidayWhyDepolarizationHard2025}.
    \item[(Debate, Own)]: Debate alone. Participants argue their own side in a live interaction. Self-persuasion studies, which assign debate sides at random, find that arguing a position moves beliefs toward it, at least until the next day \citep{schwardmannSelfPersuasionEvidenceField2021}; for one's own side this predicts reinforcement rather than moderation, and the effect on animosity remains untested.
    \item[(Debate, Opposite)]: Structural perspective-taking. Participants defend the opposing side interactively under authenticity incentives. We hypothesize that the combination of PT, interactive defense, and authenticity pressure yields deeper elaboration and thereby durable reductions in animosity.
\end{description}

This design separates the contribution of perspective (own vs. opposite) and modality (writing vs. debate), allowing us to test whether adding perspective-taking to debate produces outcomes beyond debate alone, and whether interactive defense adds value beyond static perspective-taking. To distinguish transient from sustained effects, we measure animosity and stance before the intervention, immediately after, and at a follow-up a median of two weeks later (typically 1--4 weeks), around the point by which most depolarization effects have decayed \citep{hollidayWhyDepolarizationHard2025}. We also ask whether participants would repeat the activity without compensation, since a cognitively demanding intervention can only scale if people are willing to take part.

Participants are incentivized to be convincing on two dimensions, argument quality and perceived authenticity, through three gamification elements: performance-based bonuses \citep{schwardmannSelfPersuasionEvidenceField2021, rajadesinganGuesSyncOnlineCasual2023}, evaluation by human judges, and a score that combines the two dimensions \citep{hamariDoesGamificationWork2014}. Authenticity pressure forces deeper internalization of the opposing perspective, while argument quality rewards systematic elaboration; the game format may also foster intrinsic motivation \citep{sailerHowGamificationMotivates2017}. Together, these elements turn perspective-taking from a momentary imaginative act into a practiced skill sustained under evaluative pressure. We therefore expect participants who score higher on these dimensions to show greater movement toward the opposite side in both animosity and stance \citep{greenwaldOpenmindednessCounterattitudinalRole1969, schwardmannSelfPersuasionEvidenceField2021}.

We test four hypotheses using our $2\times2$ design:

\begin{description}
    \item[H1 (Structural perspective-taking)] The \textit{Debate, Opposite} condition will produce larger reductions in \emph{issue-based animosity} than the other three conditions, particularly at follow-up, because it uniquely combines explicit perspective-taking with the interactive defense that deepens cognitive elaboration.
    \item[H2 (Perspective-taking necessity)] Both opposite-perspective conditions (\textit{Writing/Opposite} and \textit{Debate/Opposite}) will outperform their own-perspective counterparts on both outcomes, confirming that adopting the opposing view is necessary for change.
    \item[H3 (Depth-of-processing)] Participants’ performance (argument quality and perceived authenticity) will positively predict attitudinal change, reflecting the deeper engagement induced by the incentive structure.
    \item[H4 (Engagement)] Debate will maintain willingness to re-engage comparable to writing despite higher cognitive demands, suggesting that the intervention’s cognitive load does not undermine its scalability potential.
\end{description}

Our results, evaluated both immediately after the intervention (post) and again at a follow-up typically 1--4 weeks later (median about 2 weeks), reveal a time-dependent pattern that yields mixed support for \textbf{H1}. In the short term, engaging from the opposite perspective (Opp) reduced animosity and shifted stance in both modalities (writing and debate). However, durability diverged: for \emph{issue-based animosity}, only \emph{Debate/Opp} (debating from the opposite perspective; i.e., structural perspective-taking within debate) retained a detectable reduction at follow-up, whereas the effect of \emph{Write/Opp} attenuated. In contrast, stance shifts in both opposite-perspective arms persisted to follow-up, smaller than at post but still significant. This pattern suggests that stated positions may adjust more readily than animosity. Such divergence underscores the distinction between animosity and stance when designing and evaluating interventions.

Overall, we do not detect one intervention arm as consistently superior across outcomes; however, for stance, Writing/Opposite outperformed Writing/Own. Importantly, argument quality and perceived authenticity were not predictive of attitudinal change, suggesting that preparation and reflection processes, rather than performance, drive the effects of interventions. Finally, when asked whether they would re-engage in the same activity without compensation, participants across all arms expressed similar willingness (roughly 20--36\%), with no detectable difference between debate and writing despite debate's higher cognitive demands, although the estimates are too imprecise to rule out a modest reduction for debate.  

Our study advances polarization intervention research through four contributions. First, we introduce \emph{structural perspective-taking} through debates, showing how embedding perspective-taking into adversarial yet gamified exchanges provides a framework in which perspective-taking reduces animosity through cognitive elaboration rather than empathic concern. Second, we identify a divergence between immediate and follow-up outcomes: unexpectedly, writing from the opposite perspective produced the largest immediate reduction in animosity, but only debating sustained the reduction at follow-up, as hypothesized, bridging previous studies \citep{amsalemDoesTalkingOther2022, schwardmannSelfPersuasionEvidenceField2021, brierleyModeratingEffectDebates2020, levenduskyWhenEffortsDepolarize2018, saveskiPerspectiveTakingReduceAffective2022, santoroPromisePitfallsCrosspartisan2022} while building on prior work demonstrating durable effects through deliberation \citep{fishkinDeliberationAntidoteExtreme2021} and corrective information \citep{voelkelMegastudyIdentifyingEffective2023}. At the same time, \emph{stance change} appeared more stable across arms, underscoring a distinction between interventions targeting stance versus animosity. Third, debate partners' changes were uncorrelated, suggesting that outcomes did not hinge on a particular partner. This makes the intervention easier to scale, for example with large language models scaffolding or simulating the exchange \citep{costelloDurablyReducingConspiracy2024}. Finally, we provide implementable design principles: perspective-taking through semi-structured debate, gamified scoring that balances authenticity and rhetorical quality, and anonymous online implementation. These demonstrate feasibility in high-conflict settings where face-to-face dialogue proves impractical \citep{santoroPromisePitfallsCrosspartisan2022, amsalemDoesTalkingOther2022, nyhanWhenCorrectionsFail2010}.

\section{Results}\label{sec:results}
\subsection{Interventions and incentives}
Participants were recruited from a university's behavioral economics experiment pools (registered volunteer students) and undergraduate courses. Of the 224 participants who took part (attending in-person sessions), 21 were excluded for failing attention checks or other criteria, yielding a final analytic sample of $N = 203$ (details in \SIref{app:data}). Participants completed pre-intervention surveys that identified their positions on divisive issues, such as statewide policy proposals for abortion regulation (surveys and topics described in \SIref{app:surveys}). We employed an experimental design with two activity types (debating vs. writing) crossed with two perspective conditions (engaging from own vs. opposing position). The modality was not randomized but implemented across two sequential studies: 13 debate sessions followed by 4 writing-only sessions. Counts for participants in the final analysis sample are in \cref{fig:intervention}. Random assignment occurred at two levels: 1) perspective (own vs. opposing) was individually randomized within sessions, and 2) the issue of discussion (This is not fully randomized due to matching constraints, thus we account for it in models). We acknowledge that modality was not randomly assigned but implemented in sequential waves (debate sessions in Fall 2023 and Spring 2024, writing sessions in Fall 2024). This introduces potential confounds, including the November 2024 U.S. presidential election, which took place shortly before the writing sessions. We address this through: (1) balance tests showing covariate equivalence across waves, and (2) robustness checks using within-wave comparisons where possible. Demographic and political distributional equivalence between the debate ($N = 153$) and writing ($N = 50$) cohorts was confirmed through balance tests on pre-treatment survey responses (see \SIref{app:data} for full details).

In the debate conditions, participants were asked to: (1) prepare written arguments to be used during the debate for 25 minutes, then (2) engage in real-time, chat-based exchanges via an anonymized custom deployment of the RocketChat messaging platform \citep{RocketChatRocketChat2025}(platform details, including screenshots of tool in \SIref{app:platform}). 
The instructions emphasized conversational engagement (``Maintain a conversational tone'') rather than formal debate---a discussion mode successful in other studies \citep{hoffmanCivilDialogueInterventions2025,broockmanDurablyReducingTransphobia2016, kallaVoterOutreachCampaigns2022} and validated through pilot tests, which showed that informal formats increased message volume and participants felt less confused about how to engage in the experiment. There are several reasons for facilitating debates via an anonymous chat interface. First, we wanted to maintain participants' anonymity to avoid potential reputational harm. It also disallowed the judges from using simple cues to identify authenticity (for example, associating demographic factors with political positions). Second, online spaces with similar interfaces also hold a significant number of political conversations, and we wanted to set our experiment within this important frame \citep{mungerFrenemiesHowSocial2019, rathjeOutgroupAnimosityDrives2021, tuckerSocialMediaPolitical2018}. Third, online interaction offers greater opportunities to scale up the intervention in the future.

Writing sessions replicated the preparation of arguments but included no debate. Participants were instructed to craft persuasive statements defending their assigned positions without any interaction with other participants. We argue that this isolates perspective-taking effects from the additional effects of interactions with others during the debate. Preparation instructions (e.g., ``Prepare main points in defense of your position'') and writing prompts remained consistent across conditions (full protocols in \SIref{app:surveys}). Although for the writing modality, language in the materials was modified so that no reference was made to debating being part of the activity.  

\paragraph*{Incentives}
We used two different sources to recruit participants (lab participant pools and courses). Participants received compensation depending on the pool they registered: monetary payment (averaging \$15/hour) for 36\% of the participants and extra credit points for a course for 64\%. We maintained identical proportional payout structures for both groups: half of total compensation depended on task completion, which required full attendance and completion of pre- and post-surveys, while the other half rewarded performance across two components. Performance incentives prioritized intervention engagement: 75\% of this portion was allocated to argumentation success, determined by peer evaluations of both argument quality (``best argument'') and perspective authenticity (``not pretending''). We emphasized in both the design and our communication to participants that 'winning' meant achieving both goals simultaneously. We argue that this disincentivizes strategies in which participants focus only on rhetoric (by presenting just facts without engaging with the assigned perspective) or only on appearance (without paying attention to arguments). In addition, participants judged other participants' writing and transcripts anonymously after the activity. The remaining 25\% of the performance incentives were based on peers' evaluations of arguments. This means correctly assessing who was ``flipping'' (not authentic) and who had the best argument. Detailed incentive specification in \secref{sec:methods:compensation}.

\subsection{Measuring issue-based animosity and stance change}
\label{sec:results:measures}

We track two participant-level outcomes across three time points: before the intervention (pre, baseline), immediately after (post), and at a two-week to one-month follow-up. 

The first outcome we track is \textbf{issue-based stance} on the focal issue. Before participants engage in their assigned activity, we present them with a policy scenario (e.g., the state introducing legislation concerning abortion rights) and ask them to rate their agreement or disagreement on a 5-point Likert scale. We then reassess this position immediately after the intervention and again at follow-up. We recode so that positive change reflects movement toward the assigned opposing stance or toward moderation when applicable.

The second outcome is \textbf{issue-based animosity}, measured via a feeling thermometer (0--100) toward someone who disagrees with the participant on this same policy issue. Higher values indicate warmer feelings toward the opponent, and thus lower issue-based animosity \citep[e.g.,][]{iyengarAffectNotIdeology2012, iyengarFearLoathingParty2015, iyengarOriginsConsequencesAffective2019, nelsonFeelingThermometer2008}. By assessing how participants feel toward their issue opponents at each time point, we track whether animosity decreases following the intervention.

Full question wording and coding details appear in \SIref{app:surveys}. Although participants engaged in paired activities, outcomes are recorded individually, with these dyadic dependencies accounted for in our statistical models.

\noindent\textbf{Primary measures of change.}
Our core quantities are changes from baseline: $\delt{\stance}{t|\armtxt}$ and $\delt{\anim}{t|\armtxt}$, where $t$ denotes either the change at post-intervention or follow-up. For both change measures, positive values indicate `improvement'. For stance, a positive $\Delta^{\stance}$ indicates movement toward the opposing position or moderation. For issue-based animosity, a positive $\Delta^{\anim}$ indicates warmer feelings toward the issue opponent. 

These measures are computed via \emph{estimated marginal mean} change within each experimental arm, derived from mixed-effects models that adjust for demographical covariates and account for repeated measures and session structure \citep[cf.][]{costelloDurablyReducingConspiracy2024}. 

\noindent\textbf{Key comparisons.}
We evaluate our hypotheses through the following contrasts:

\begin{description}
  \item[H1] \emph{(structural perspective-taking).} We report each $\delt{\stance/\anim}{t|\armtxt}$ for each of the four experimental arms \{(Write, Own), (Write, Opp), (Debate, Own), (Debate, Opp)\} at both post and follow-up, testing whether the change from baseline (pre) differs from zero, and compare arms pairwise to assess whether (Debate, Opp) produces larger reductions in animosity than the other three arms, particularly at follow-up.

  \item[H2] \emph{(perspective-taking necessity).} We contrast Own versus Opp perspective within each modality (Writing, Debate) by taking differences of $\Delta$'s. This tests specifically whether arguing the opposing view produces greater change in each outcome than articulating one's own view.

  \item[H3] \emph{(depth of processing).} Within each arm, we compare participants whom judges identified as having the best argument or authenticity (coded 1 when at least two of three judges select the participant) against those not so identified, pooling across arms, to test whether judged performance predicts larger change.

  \item[H4] \emph{(engagement).} We analyze the binary outcome of whether participants would ``probably'' or ``definitely'' participate again without compensation, estimating covariate-adjusted probabilities by arm using logistic mixed models, and test whether debate is non-inferior to writing using a pre-specified margin.
\end{description}

\medskip
\noindent\textit{Statistical approach.}
All models adjust for debate topic, demographics (gender and ethnicity), political leaning, and opinion strength. Random intercepts account for repeated measures at the participant level and for session or debate-block structure, including dyadic pair intercepts where applicable. This specification separates pairing effects from individual-level change. We also tested directly whether debate partners changed alike. After covariate adjustment, partners' changes were essentially uncorrelated (partner correlation $\rho$ between $-0.13$ and $0.12$ across outcomes and time points, all $p>0.3$; \SIref{app:analysis:partner}). We compute contrasts as estimated marginal means on the response scale, with unadjusted 95\% confidence intervals; $p$-values for families of comparisons are Holm-adjusted. Further model details are in \secref{sec:methods:analysis}.

\subsection{H1: Effects on issue-based animosity and stance}

\begin{figure*}[htbp]
    \centering
    \includegraphics[width=0.9\textwidth]{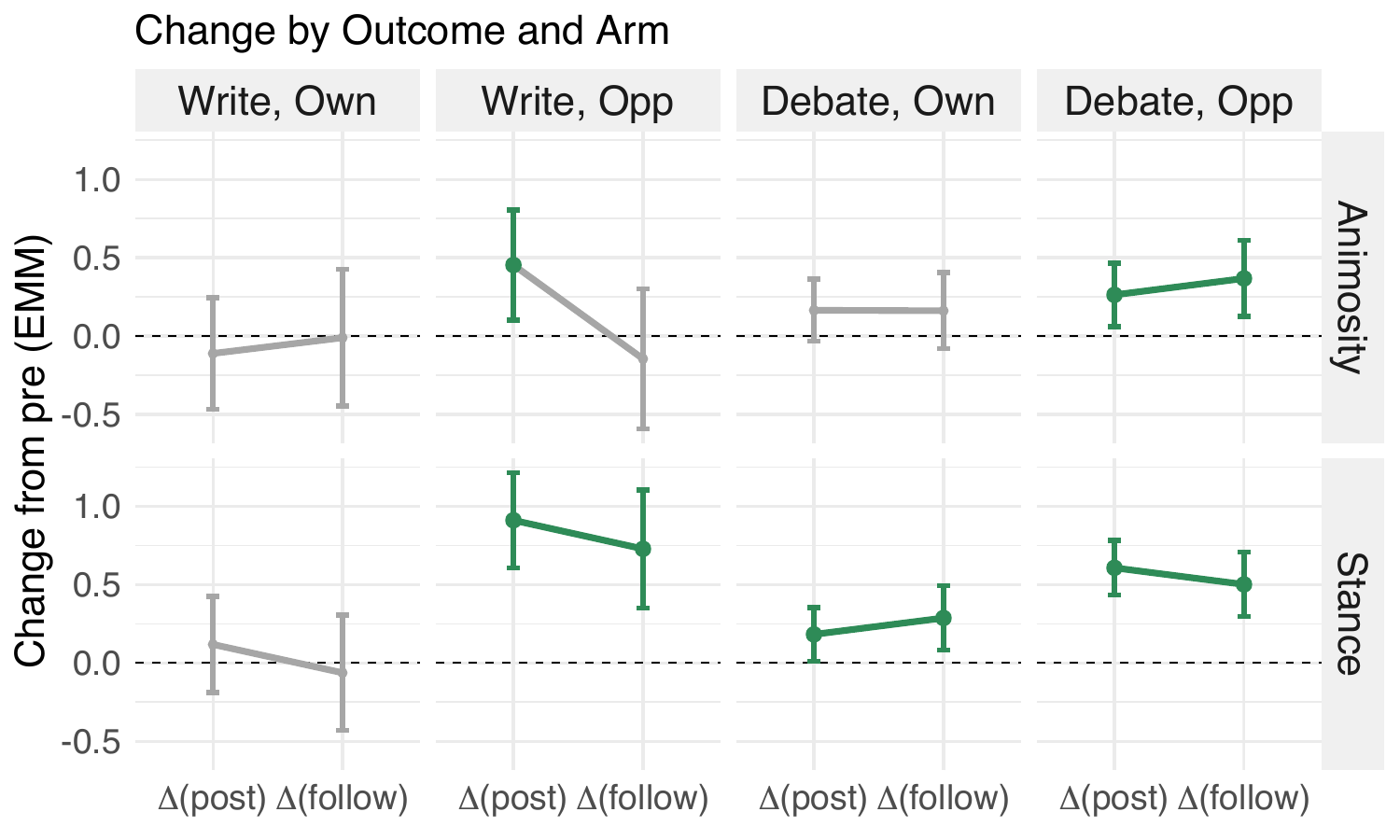}
    \caption{\textbf{Change in animosity and stance.} Adjusted within-arm change $\Delta_{t}$ from \emph{pre} to \emph{pos} ($t{=}\pos$) and \emph{follow} ($t{=}\fol$) for issue-based animosity (top; higher means warmer feelings toward issue opponents) and stance (bottom; higher means movement toward the opposite position or moderation when applicable). Points are estimated marginal means (EMMs) with 95\% CIs; the dashed line marks no change. \emph{Key remark:} the largest immediate gains appear for (Write, Opp) at $t{=}\pos$, while the sustained animosity reduction at $t{=}\fol$ is primarily (Debate, Opp); stance shifts persist for both perspective-taking arms.}
    \label{fig:h1_change_time}
\end{figure*}

We estimate changes in issue-based animosity and stance using within-arm EMM contrasts, as described in the previous section. For simplicity, we write $\Delta$ as shorthand for the within-arm EMM estimate defined in \secref{sec:results:measures}; where $Y, t$ and the arm $a$ are made specific by the surrounding context. For example: ``$\Delta$ animosity in the modality writing for the opposite perspective'' refers to $\delt{Y}{t=\pos|\armtxt=\arm{\Wrt}{\Opp}}$. 

Results (\cref{fig:h1_change_time}) reveal patterns where perspective-taking conditions (Opposite arms) outperformed same-perspective controls, with writing modalities showing particular strength. We also note that wider estimates for writing activity type conditions are likely associated with the smaller number of participants in those conditions.

\paragraph*{Issue-based Animosity ($Y{=}\anim$)}
Immediately after the intervention, both perspective-taking arms reduced animosity: Writing for the opposite had the largest point estimate ($\Delta = 0.45$, 95\% CI [0.10, 0.80]) followed by debating the opposite perspective ($\Delta = 0.26$ [0.06, 0.47]). Own-perspective arms were smaller and not distinguishable from zero. At follow-up, debating for the opposite perspective showed promise in durability, retained a detectable reduction ($\Delta = 0.37$ [0.12, 0.61]), whereas (Write, Opp) attenuated and was no longer statistically distinguishable from zero ($\Delta = -0.15$ [--0.59, 0.30]). Thus, for issue-based animosity, the evidence suggests that writing from the opposite perspective is most effective immediately, while debate-based perspective-taking is more durable.

\paragraph*{Stance ($Y{=}\stance$)}
Positive values indicate movement toward moderation (e.g., a participant who strongly disagreed at pre-intervention now agrees at post-intervention). For change measured at post-intervention, writing for the opposite perspective showed the largest shift ($\Delta = 0.91$ [0.61, 1.21]), followed by debating for the opposite ($\Delta = 0.61$ [0.43, 0.78]), and a smaller but significant change for debating the own perspective ($\Delta = 0.18$ [0.01, 0.35]). At follow-up, perspective-taking arms remained significant: (Write, Opp): $\Delta = 0.73$ [0.35, 1.11]; (Debate, Opp): $\Delta = 0.50$ [0.30, 0.71]. This indicates that, unlike animosity, stance change persists across the different arm conditions.

We also estimated pairwise arm comparisons (differences in $\Delta_{t}$ across arms), which are reported in \SIref{app:analysis:h1}. In general, we don't have enough data (samples) to differentiate particular arms for animosity change; however, for stance, we have evidence that (Write, Opp) has a significantly higher estimate than (Write, Own). Thus, differences are more pronounced for stance change than for animosity change.

\begin{figure}[htbp]
    \centering
    \includegraphics[scale=\figscale]{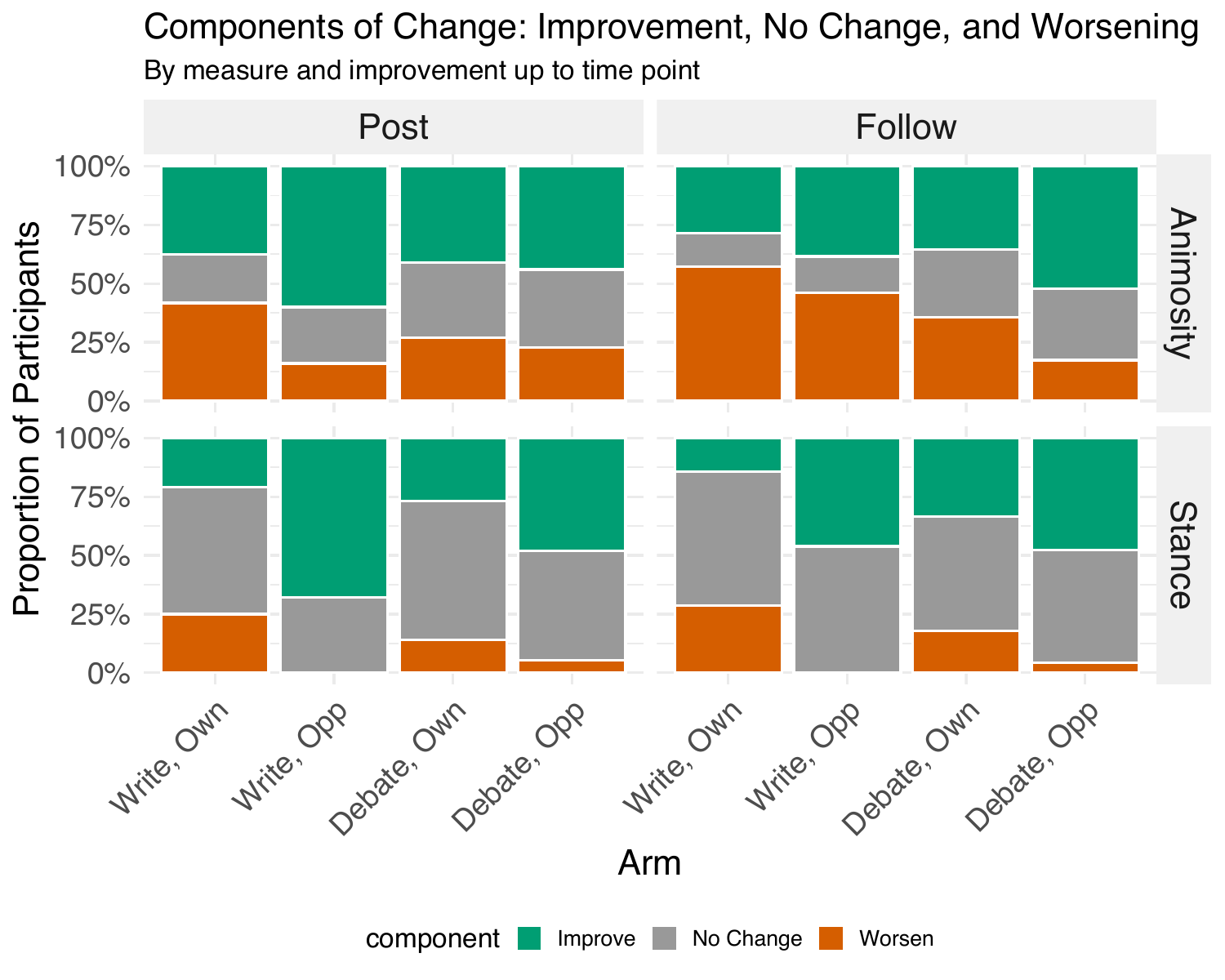}
    \caption{\textbf{Shares of participants who improve, do not change, or worsen.} Change is measured relative to pre, for each arm and outcome. Top panel: Issue-based animosity, where ``Improve'' corresponds to warmer feelings toward the issue's opponent. Bottom panel: Stance, where ``Improve'' corresponds to movement toward the assigned opposite stance or toward moderation, when applicable. Bars show stacked proportions at post and at follow for each arm.}
    \label{fig:h1_improvements_change_components_full}
\end{figure}

After the main analysis, we noticed that because both outcomes are recorded on ordinal scales, a substantial share of participants remain in the same category before and after treatment. To complement mean-based contrasts, we classify each person at each time point as ``Improve'', ``No Change'', or ``Worsen'' relative to pre-intervention. The stacked bars in \cref{fig:h1_improvements_change_components_full} summarize these shares.

Overall, the categorical view aligns with the EMM results: immediate gains for self-paced opposite perspective writing, and more durable reductions in animosity when opposite perspective engagement is embedded in debate. At post-intervention, (Write, Opp) shows the largest improvement, about 60\% for animosity and 68\% for stance; (Debate, Opp) is next, in the mid-40s on both outcomes; and No Change is common in own-perspective arms, near one-half for stance. By follow-up, animosity diverges, with Write/Opp’s Improve share falling to 38\%, while Debate/Opp remains near one-half. For stance, both opposing perspective arms retain large improvement shares at follow-up: (Write, Opp) at 46\% and (Debate, Opp) at 48\%; (Debate, Own) is moderate, and (Write, Own) is the lowest at about 14\%. Pairwise differences in improvement rates are not statistically distinguishable after adjustment, likely reflecting limited precision in the writing cohorts.

\subsection{H2: Perspective effects across modalities}
\label{sec:results:h2}

In our $2\times2$, we can decompose perspective-taking (PT) effects within modality (debate versus writing). For this hypothesis, we are interested in whether perspective-taking consistently has an advantage across different modalities, as measured by consistently higher animosity and stance change estimates compared to arms without perspective-taking.

We test whether the advantage of engaging from the opposite perspective depends on modality. We use $\delt{Y}{t}$ to denote the within-arm EMM change from pre to time $t\in\{\pos,\fol\}$. Here, we also define the perspective advantage $\delta$, which represents the contrast in changes in outcome between perspectives. That is, for each modality $m\in\{\Wrt,\Dbt\}$ and time $t$, we define

\begin{equation}
    \delta^Y(t|m)=
\delt{Y}{t|\arm{m}{\Opp}}-\delt{Y}{t|\arm{m}{\Own}}.
\label{eq:perspective_advantage}
\end{equation}

Here, $\delta$ is the incremental pre-to-$t$ improvement attributable to assigning the opposite perspective rather than one's own perspective within the same modality $m$ on the EMM (response) scale. So $\delta>0$ means the opposite perspective produced a larger gain, $\delta=0$ means equal change, and $\delta<0$ means an own-perspective advantage. Estimates utilize the same adjusted specification as H1 (topic, demographics, political viewpoint, opinion strength; random effects are included as in H1). Design and attrition checks (balance, IPW for session-by-modality scheduling, and follow-up attrition adjustments) yield qualitatively similar patterns; see \SIref{app:analysis:h2}.

\begin{figure}[htbp]
    \centering
    \includegraphics[scale=\figscale]{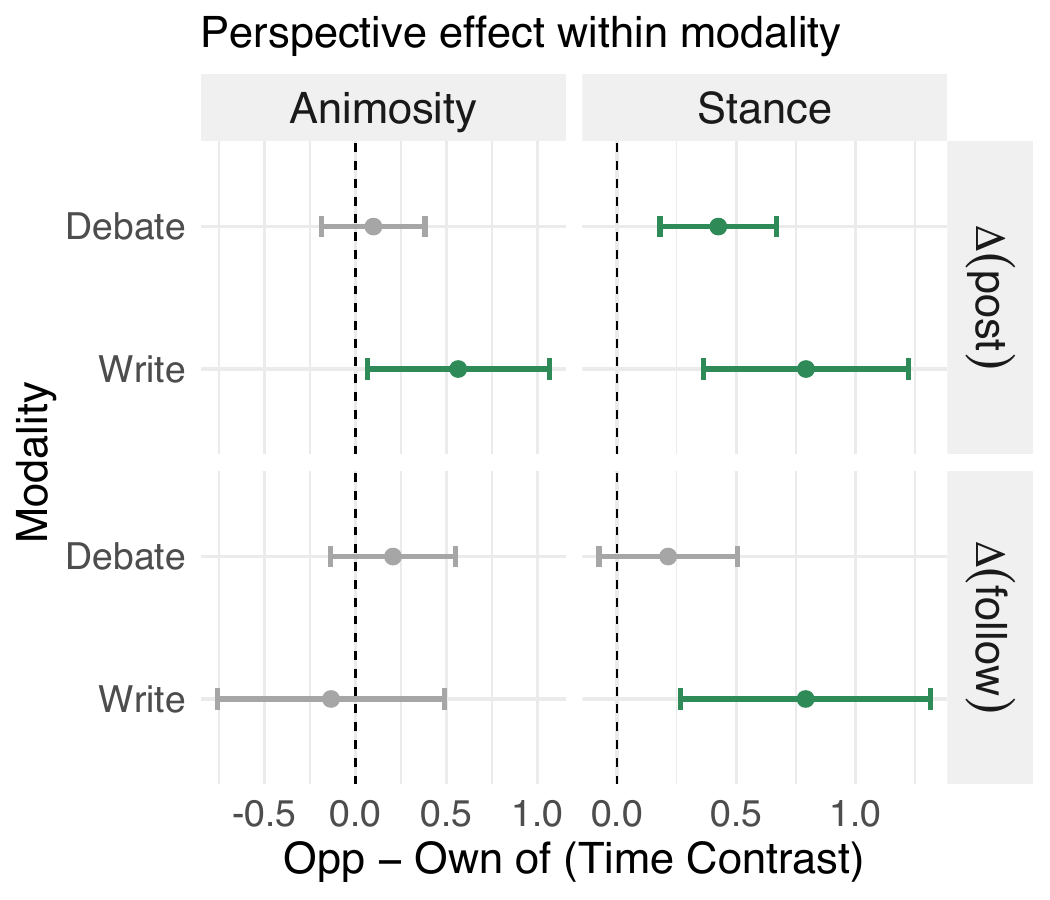}
    \caption{\textbf{Effect of arguing the opposite side within each modality.} Perspective contrasts within modality: $\delta^Y(t|m)$ which is the incremental pre-to-$t$ improvement attributable to assigning the opposite perspective rather than one's own perspective within the same modality $m$ on the EMM (response) scale. So $\delta>0$ means the opposite perspective produced a larger gain, $\delta=0$ means equal change, and $\delta<0$ means an own-perspective advantage. Points are EMMs with 95\% CIs; the dashed line indicates no change. \emph{Key remarks:} contrasts are generally positive, with weaker detectability for animosity and clearer, more persistent advantages for stance.}
    \label{fig:h2_did}
\end{figure}

We plot these contrasts in \cref{fig:h2_did}. Our results suggest mixed evidence for H2, where we observe a more consistent advantage for engaging from the opposite side when referring to stance change. For animosity, we detect the advantage of engagement from the opposite side only at post-intervention.

\paragraph*{Issue-based Animosity ($Y{=}\anim$)}
Perspective advantage estimates are generally positive, but the difference is often not statistically detectable. For the modality of writing at post-intervention, $\delta = 0.56$ [0.06, 1.07] indicates a clear advantage for opposite-perspective engagement; however, the follow-up contrast is small and non-significant ($\delta = -0.13$ [--0.76, 0.49]). In the debating modality, both post and follow-up estimates are positive but non-significant ($\delta = 0.10$ [--0.19, 0.38]; $\delta = 0.21$ [--0.14, 0.55]).

\paragraph*{Stance ($Y{=}\stance$)}
Stance change follows a distinctive pattern in which engaging from the opposite side (which we argue operationalizes perspective-taking) is more advantageous. Opposite-perspective engagement outperforms own-perspective at post in both modalities (Writing: $\delta = 0.79$ [0.36, 1.22]; Debate: $\delta = 0.42$ [0.18, 0.67]) and remains detectable at follow-up in writing ($\delta = 0.79$ [0.27, 1.32]). Debate's follow-up advantage is directionally positive but not significant ($\delta = 0.21$ [--0.08, 0.50]).

\begin{figure}[htbp]
    \centering
    \includegraphics[scale=\figscale]{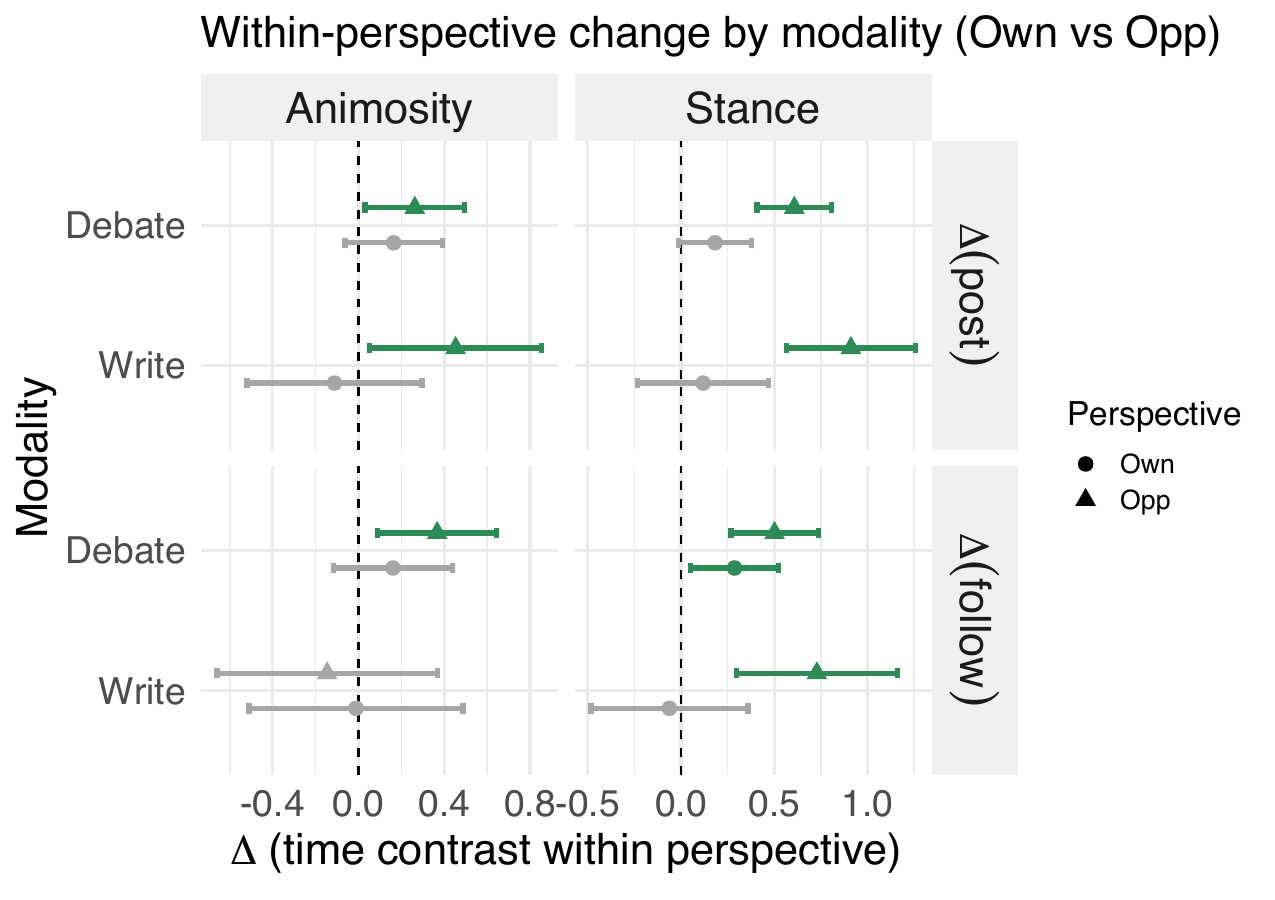}
    \caption{\textbf{Change by modality for own and opposite perspectives.} Within-perspective estimated changes ($\Delta_{\pos}$, $\Delta_{\fol}$) by modality for \emph{Own} vs.\ \emph{Opp}. Points are EMMs with 95\% CIs; the dashed line indicates no change. \emph{Key remark:} the perspective advantage for \emph{Opp} is driven by \emph{Own} lying near zero---particularly in writing---while \emph{Opp} shows clear movement, especially on stance.}
    \label{fig:h2_marginal}
\end{figure}

To unpack the driver of those contrasts, we also show the marginal (non-differenced) within-perspective changes in \cref{fig:h2_marginal}. The decomposition shows that own tends to lie near zero (especially in writing), so the positive perspective advantage is largely carried by opposite gains. For instance, in writing at post-intervention, own-perspective change is near zero for both animosity ($\Delta = -0.11$ [--0.52, 0.30]) and stance ($\Delta = 0.12$ [--0.23, 0.47]), while opposite-perspective shows substantial movement (animosity: $\Delta = 0.45$ [0.05, 0.86]; stance: $\Delta = 0.91$ [0.56, 1.26]; intervals here are Bonferroni-adjusted over the two time points, as in the figure). This suggests that engaging from the opposite perspective confers a consistent within-modality advantage, most notably for stance change, and less reliably for animosity. The decomposition indicates that the pattern arises because own-perspective engagement seldom deviates from baseline---especially in writing---whereas opposite-perspective engagement does.

\subsection{H3: Does winning moderate change?}
\label{sec:results:h3}

As mentioned, participants were incentivized to appear both as having a solid argument and as authentic (to avoid being picked as the pretender). This performance was evaluated by a panel of judges. We assess whether judged performance during the activity (winning on Best Argument, Authenticity, or Both) modulates the change from pre- to post- or follow-up. To study moderation by performance, we compare winners to nonwinners within the same arm. For a given win type, the subgroup contrast is:

\begin{equation}
\delta^Y(t|\armtxt)
=\delt{Y}{t|\armtxt,\win}-\delt{Y}{t|\armtxt,\lose},
\end{equation}

where $\win$ indicates participants for whom at least two of three judges selected for that mechanism, and $\lose$ indicates all others. We then report pooled effects across arms using prespecified weights $v_a$:
\[
\bar{\delta}_{t}^{(\text{win type})}
=\sum_{a} v_a \,\delta^Y(t|a).
\]
This last average $\bar{\delta}_{t}^{(\text{win type})}$ is the average incremental improvement on the EMM response scale for participants who won a given mechanism relative to those who did not. This calculation is done per arm and then pooled across all arms. Positive values for this contrast indicate that winners see greater changes (in animosity or stance). 

We analyze three win types: Authentic, Best Argument, and Both (which is both previous types simultaneously). Fixed effects for controls (topic, demographics, political viewpoint, opinion strength) and random intercepts (participant, debate block) match H1 and H2; arm-by-time terms are included. Arm-specific moderation of winning is not supported, as models become more unstable to compute and unreliable; therefore, we pool across arms.

\begin{figure}[htbp]
    \centering
    \includegraphics[scale=\figscale]{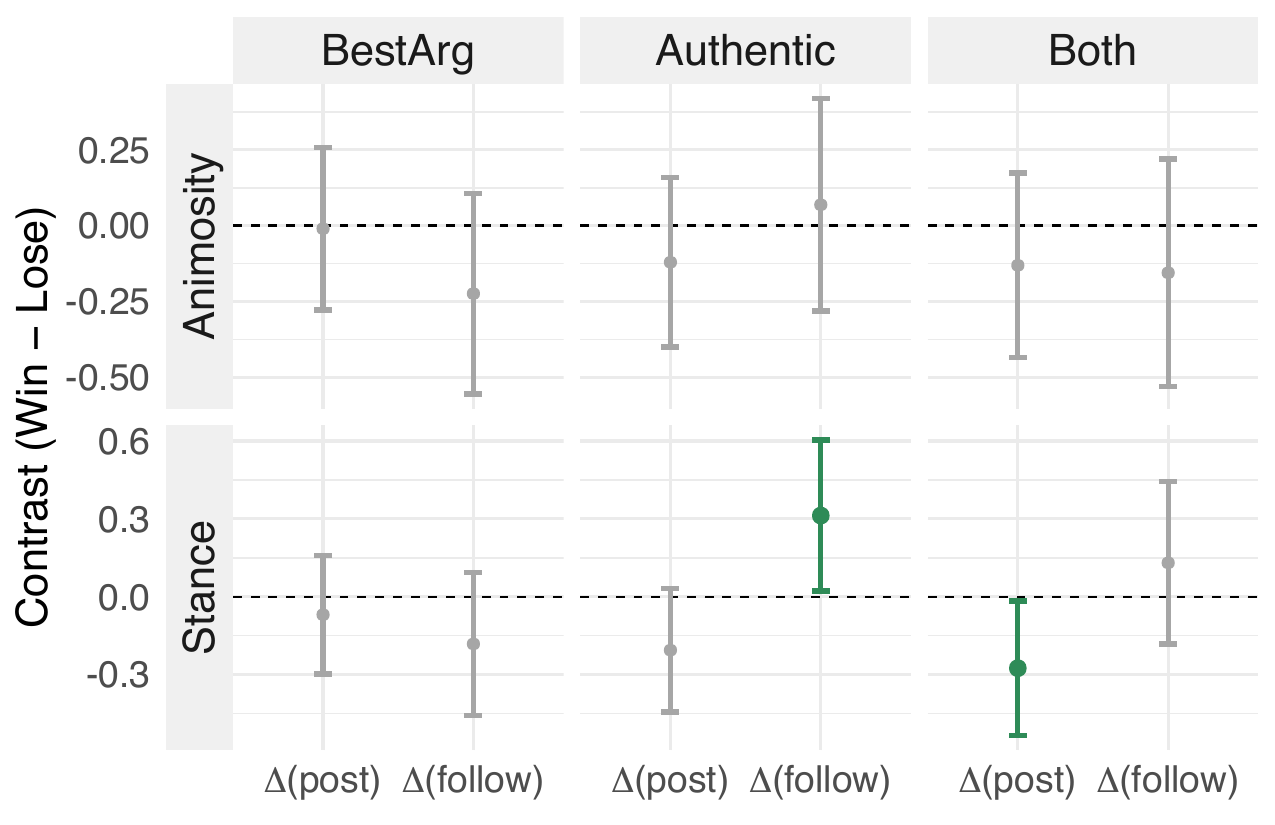}
    \caption{\textbf{Change among winners relative to losers.} Pooled subgroup contrasts ($\bar{\delta}_{t}^{(\text{win type})}$): Win - Lose at post and follow as Lose is the reference level in the analysis. Points represent EMM contrasts with 95\% CIs; the dashed line indicates no difference. Contrasts are generally small and imprecise, indicating no reliable evidence that winning systematically amplifies change (in animosity or stance) beyond the base effects of the activities.}
    \label{fig:h3_mech_plot}
\end{figure}

Overall, across win types, we find no reliable evidence of moderation by winning (\cref{fig:h3_mech_plot}). For animosity change, pooled contrasts are overall quite small and imprecise (post-intervention: BestArg $\bar{\delta} = -0.01$ [--0.28, 0.26], Authentic $\bar{\delta} = -0.12$ [--0.40, 0.16], Both $\bar{\delta} = -0.13$ [--0.44, 0.17]; follow-up: BestArg $\bar{\delta} = -0.22$ [--0.56, 0.11], Authentic $\bar{\delta} = 0.07$ [--0.28, 0.42], Both $\bar{\delta} = -0.16$ [--0.53, 0.22]). This indicates no evidence for any differential change in attitudes contingent on performance during the intervention. For stance change, post-intervention contrasts are modestly negative, with winners changing less than nonwinners (BestArg $\bar{\delta} = -0.07$ [--0.30, 0.16], Authentic $\bar{\delta} = -0.21$ [--0.44, 0.03], Both $\bar{\delta} = -0.28$ [--0.53, --0.02], Holm-adjusted $p = 0.11$), while the Authentic contrast at follow-up is directionally positive but not statistically detectable after adjustment ($\bar{\delta} = 0.31$ [0.02, 0.60], Holm-adjusted $p = 0.11$). 

To contextualize these differences, we also examined marginal estimates that are not differences. These indicate that in some scenarios, the `Lose' group shows slightly larger improvements in animosity (warmer feelings toward opponents) than the `Win' groups. In contrast, for stance, the losing group tends to change less, which is consistent with the contrasts but not conclusive. Nevertheless, judged winning does not systematically amplify attitude or position change beyond the base effects of the activities.

\subsection{H4: Willingness to participate again}
\label{subsec:engagement}

We assessed whether participants would voluntarily repeat the activity without compensation. Participants rated their willingness on a five-point scale from ``definitely no'' to ``definitely yes''; we focus on those who answered ``probably yes'' or ``definitely yes.'' Our goal is to estimate the probability of willing re-engagement and test whether this differs meaningfully across experimental conditions.

We fit a logistic mixed model predicting willingness from modality (Debate vs. Write), perspective (Own vs. Opp), and their interaction, adjusting for topic, demographics, political viewpoint, and opinion strength, with random intercepts for debate pairs. From this model, we extracted covariate-adjusted probabilities for each arm via estimated marginal means. Let $p_a$ denote the probability of re-engagement in arm $a$. We then computed two key pooled risk differences: Debate minus Write (averaged across perspectives) and Opp minus Own (averaged across modalities). These contrasts quantify whether one condition systematically increases or decreases willingness relative to the other.

\begin{figure}[htbp]
    \centering
    \includegraphics[scale=\figscale]{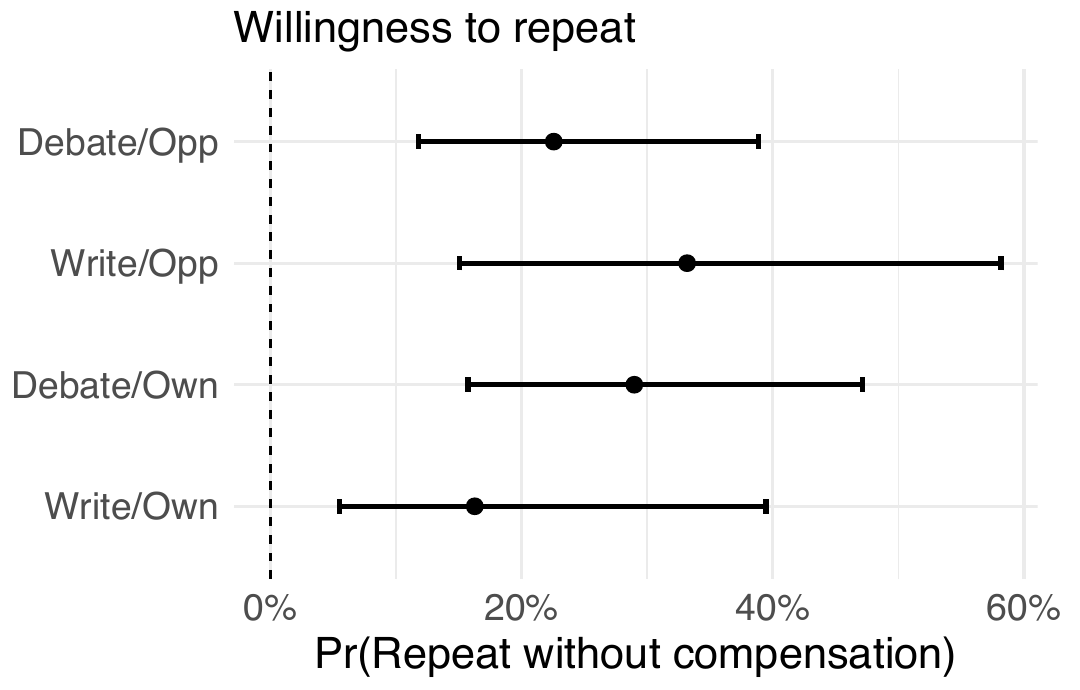}
    \caption{\textbf{Willingness to repeat the activity without compensation.} Adjusted probability of repeating without compensation by modality and perspective. Points are estimated marginal means from the logistic mixed model with 95\% confidence intervals; dashed line marks 50\% probability of more likely than not likely to repeat without compensation.}
    \label{fig:h4_willingness_by_modality_perspective}
\end{figure}

\Cref{fig:h4_willingness_by_modality_perspective} displays the estimated probability of re-engagement in each arm. Willingness rates are modest across all conditions, ranging from approximately 16\% to 33\%, with substantial uncertainty reflected in wide confidence intervals. Write/Own shows the lowest point estimate ($\hat{p} = 0.16$ [95\% CI: 0.05, 0.39]), while Write/Opp shows the highest ($\hat{p} = 0.33$ [0.15, 0.58]). The debate arms fall between these extremes: Debate/Own at $\hat{p} = 0.29$ [0.16, 0.47] and Debate/Opp at $\hat{p} = 0.23$ [0.12, 0.39]. The overlapping intervals indicate no strong evidence that any single arm differs substantially from the others.

\begin{figure}[htbp]
  \centering
  \includegraphics[scale=\figscale]{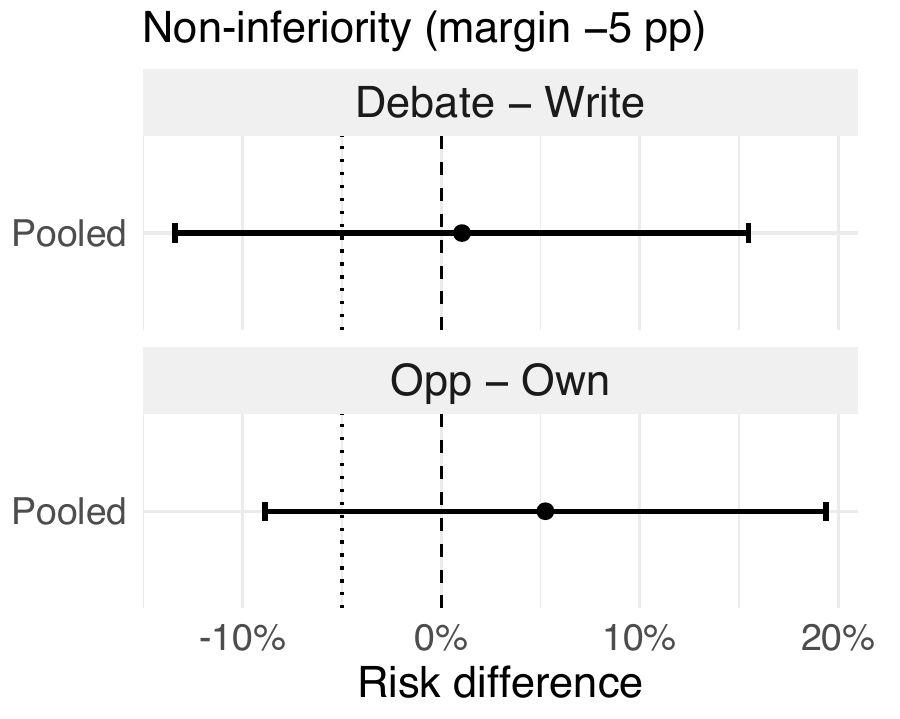}
  \caption{\textbf{Differences in willingness to re-engage by modality and perspective.} Pooled risk differences for re-engagement willingness. Top panel: Debate minus Write (averaged over perspective). Bottom panel: Opp minus Own (averaged over modality). Dots indicate point estimates; whiskers show 95\% confidence intervals; dotted vertical line marks the pre-specified non-inferiority margin of $-0.05$; dashed line marks zero difference.}
  \label{fig:h4_pooled_rd}
\end{figure}

Our question is whether debating differs from writing, and whether arguing the opposing view differs from one's own position, when we average across the complementary dimension. \Cref{fig:h4_pooled_rd} presents these pooled risk differences. The Debate vs. Write contrast yields an estimated difference of 0.01 [-0.13, 0.15], indicating essentially no difference in willingness between modalities. Similarly, the Opp vs. Own contrast produces an estimated difference of 0.05 [-0.09, 0.19], suggesting no meaningful effect of perspective assignment on re-engagement. Both intervals are wide and include zero, consistent with no detectable effect of either experimental factor on willingness to participate again.

We also examined contrasts stratified by the complementary factor. These stratified differences were similarly small and imprecise: within Own perspective, Debate exceeded Write by 0.13; within Opp perspective, Debate trailed Write by 0.11. Correspondingly, within Write modality, Opp exceeded Own by 0.17, while within Debate modality, Opp trailed Own by 0.06. None of these differences approached statistical significance, and pooling across strata does not change the overall picture of null effects.

We pre-specified a non-inferiority margin of $-0.05$ for each pooled contrast, meaning we would conclude that a condition is not substantially worse if its lower 95\% confidence bound exceeded this threshold. Because both intervals extend below $-0.05$ (to $-0.13$ for Debate$-$Write and $-0.09$ for Opp$-$Own), we cannot establish non-inferiority. The data remain compatible with modest harms (up to roughly 9--13 percentage points lower willingness) as well as modest benefits or no effect.

Two sensitivity analyses confirmed these findings. First, to address potential bias from missing responses on the willingness item, we re-estimated the pooled contrasts using stabilized inverse-probability weighting based on observed covariates. This yielded nearly identical results: Debate$-$Write difference of 0.019 [$-0.126$, 0.164]; Opp$-$Own difference of 0.051 [$-0.091$, 0.193]. Second, we fit an ordinal cumulative logit model treating the original five-level response as a collapsed three-category outcome (No/Indifferent/Yes) and recovered the probability of ``Yes'' via category-specific marginal means. This approach produced slightly narrower intervals but the same substantive conclusions. Contrasts from these sensitivity analyses appear in \SIref{app:analysis:h4}.

\section{Discussion}\label{sec:discussion}
Our results show that engaging from an opposing perspective drives change in both animosity and stance, but the modality of engagement determines how these changes unfold over time. Writing from the opposite side produced immediate shifts in both stance and animosity, yet only the stance change persisted at follow-up. Debating from the opposite side showed a different pattern: smaller immediate effects but reductions in animosity still detectable at follow-up about two weeks later. These patterns map directly onto the two currents we identified earlier: perspective-taking alone (writing) yields immediate but decaying reductions in animosity, while embedding perspective-taking into debate, what we term \textit{structural perspective-taking}, sustains reductions in animosity. This helps reconcile two often-parallel literatures. Perspective-taking research shows that constructing counter-attitudinal arguments outperforms purely empathic prompts \citep{greenwaldOpenmindednessCounterattitudinalRole1969, saveskiPerspectiveTakingReduceAffective2022, santoroPromisePitfallsCrosspartisan2022}, but typically examines self-paced, solitary exercises like writing. Research on debate formats reveals more variable effects \citep{schwardmannSelfPersuasionEvidenceField2021, loweTypesContactField2021}, depending on whether the design incentivizes genuine perspective-taking or competitive point-scoring. Our results suggest both approaches work but follow different temporal trajectories: writing excels at immediate stance recalibration, while debate appears necessary for durable reductions in animosity.

Table~\ref{tab:arm-constructs} summarizes the design features present in each experimental condition. These features correspond to the theoretical constructs discussed earlier: perspective-taking (a cognitive mechanism), adversarial engagement (the interactive format), and exposure to opposing or own-side ideas. (Write, Opp) combines cognitive perspective-taking with a non-adversarial format, which may facilitate immediate stance change without social pressure. (Debate, Opp) adds adversarial engagement to perspective-taking, a combination that appears to sustain affective benefits when incentives align with argument quality and authentic representation rather than merely winning.

\begin{table*}[ht]
    \centering
    \caption{\textbf{Design features by experimental condition.}}
    \label{tab:arm-constructs}
    \small\setlength{\tabcolsep}{4pt}
    \begin{tabular}{lcccc}
        \toprule
        \textbf{Feature / Condition} & Debate, Opp & Debate, Own & Write, Opp & Write, Own \\
        \midrule
        Perspective taking & X & & X & \\
        Adversarial engagement & X & X & & \\
        Exposure to other-side ideas  & X & X & X & \\
        Exposure to own-side ideas & X & X &  & X \\
        \bottomrule
    \end{tabular}
    \smallskip\par\noindent
    {\small Note. (Write, Opp) combines cognitive perspective-taking with a non-adversarial setting, consistent with its stronger immediate change. (Debate, Opp) adds responsive exchange to the same cognitive demand, a combination that appears to support durable reductions in animosity when incentives emphasize perspective-taking and argument quality rather than point-scoring.}
\end{table*}

The advantage of opposite-side engagement aligns with research on counter-attitudinal interventions that utilize perspective-taking \citep{saveskiPerspectiveTakingReduceAffective2022,warnerReducingPoliticalPolarization2020,greenwaldOpenmindednessCounterattitudinalRole1969, pettyElaborationLikelihoodModel1986, broockmanDurablyReducingTransphobia2016}. In our study, both writing and debating from the opposing side required participants to anticipate objections and construct coherent counter-attitudinal arguments, inducing accuracy goals. This core cognitive demand operates whether the task is solitary or interactive. Crucially, this cognitive work appears sufficient to shift stated stances, but animosity requires the additional element of interactive rehearsal to endure.

We found that evaluations of issue opponents were more resistant to change than stated positions. Stances shifted more consistently and durably, with the largest effects where people engaged from the opposite perspective. \emph{Issue-based animosity} (feelings toward those who disagree on a specific policy) appears tied to processes requiring more than cognitive reframing, in contrast to stated positions which shifted more readily. This asymmetry connects to research locating affective polarization in identity and status dynamics rather than purely policy disagreements \citep{iyengarOriginsConsequencesAffective2019, druckmanWhatWeMeasure2019}. That work shows that people's stated positions on issues do not necessarily predict how they respond in social scenarios, such as whether they would welcome their child marrying someone from the opposing side \citep{iyengarOriginsConsequencesAffective2019}; our results suggest this asymmetry extends to interventions designed to reduce polarization. It also underscores why interventions focused primarily on correcting cognitive misperceptions may fail to improve democratic attitudes \citep{diasCorrectingMisperceptionsPartisan2024}, as they do not directly engage the affective drivers of animosity. Arguments can shift stated positions relatively easily, but improvements in affective evaluations do not necessarily follow. We note that we measure issue-based animosity as feelings toward issue opponents rather than as the traditional party-based affective polarization gap (the difference between in-party like and out-party dislike) \citep{iyengarOriginsConsequencesAffective2019}. This choice reflects our theoretical focus on issue-level dynamics, but it means our findings may not directly generalize to partisan animosity.

The sustained reduction in animosity for (Debate, Opp) compared to (Write, Opp) reveals how modality influences durability. Self-paced writing allows cognitive work without the pressure of managing impressions or real-time responses, explaining larger immediate shifts \citep{tullerSeeingOtherSide2015, warnerReducingPoliticalPolarization2020}. Real-time debate adds concurrent demands---monitoring a partner's arguments and managing self-presentation---which may tax cognitive resources and dampen immediate change \citep{binnquistZoomSolutionPromoting2022}. However, the need to defend an adopted position against live scrutiny deepens engagement with the counter-attitudinal perspective \citep{schwardmannSelfPersuasionEvidenceField2021, munnekeEffectsSynchronousAsynchronous2007}. This interactive rehearsal consolidates more durable shifts in evaluations of issue opponents, aligning with accounts where repeatedly accessing and defending updated attitudes strengthens their persistence \citep{greenwaldOpenmindednessCounterattitudinalRole1969}, particularly when incentives emphasize genuine perspective-taking \citep{schwardmannSelfPersuasionEvidenceField2021, santoroPromisePitfallsCrosspartisan2022}. Thus, it provides a pathway to more persistent change that holds at the point where many interventions have already decayed \citep{hollidayWhyDepolarizationHard2025}, while complementing prior work demonstrating durable effects through deliberation \citep{fishkinDeliberationAntidoteExtreme2021}, multi-session interventions \citep{voelkelMegastudyIdentifyingEffective2023}, and extended contact \citep{broockmanDurablyReducingTransphobia2016}. For practice, this means self-paced, opposite-perspective tasks may suffice for short-term stance recalibration, while durable improvement in evaluations of issue opponents requires perspective-taking embedded in structured, interactive engagement like debate.

Does adversarial engagement via debate help or harm? Our results suggest the answer depends on design. While adversarial formats can entrench attitudes \citep{loweTypesContactField2021, amsalemDoesTalkingOther2022}, we observed no net negative effect when rules, anonymity, and incentives centered on perspective-taking and argument quality. Even debating from one's own position did not damage outcomes. This aligns with research showing structured bipartisan networks can improve social learning and reduce bias when partisan identities are not primed \citep{guilbeaultSocialLearningPartisan2018}. Willingness to re-engage was also not reduced. Two design features likely contributed: the absence of a public audience removed pressures for performative hardening \citep{binnquistZoomSolutionPromoting2022}, and participants debating from their own side gained incidental exposure to opposing arguments (perspective-getting) \citep{kallaVoterOutreachCampaigns2022}. This contrasts with writing from one's own perspective, where this channel was unavailable.

Judged performance during debates told us little about who changed. Being rated as most convincing or authentic did not systematically amplify stance or animosity movement. This suggests preparation and engagement matter more than performative success, aligning with self-persuasion accounts where constructing counter-attitudinal reasons drives updating regardless of competitive success \citep{schwardmannSelfPersuasionEvidenceField2021}. For designers, this implies rewarding features that guarantee depth---such as preparation time, requirements to articulate opposing views, and structured feedback---rather than competitive dominance.

These findings have practical implications for scalable interventions. First, we detected no decline in willingness to re-engage under our cognitively demanding debate design when anonymity, clear rules, and aligned incentives were in place, although our estimates cannot rule out a modest reduction. This is relevant for online deployments where political discussion is text-based and episodic. Second, emerging human-AI systems can reduce coordination costs by providing structured prompts for counter-attitudinal articulation and critique without the social pressures of live debate \citep{costelloDurablyReducingConspiracy2024, baiLLMgeneratedMessagesCan2025}. The null effect of judged performance cautions against optimizing for persuasive victories; systems should instead optimize for depth by making the retrieval, elaboration, and critical testing of arguments unavoidable.

Several limitations warrant consideration. First, our student sample and issue-focused design mean generalizability to broader populations requires further testing \citep{henrichWeirdestPeopleWorld2010}. Second, modality was implemented in sequential waves (debate in Fall 2023 and Spring 2024, writing in Fall 2024), with the 2024 U.S. presidential election shortly before the writing sessions. However, balance tests confirmed covariate equivalence across waves, and all models include wave fixed effects to absorb time-varying heterogeneity. Third, the writing cohort---particularly Write/Opp at follow-up (n=13)---is smaller than the debate cohort, meaning null findings in this condition should be interpreted with appropriate caution. Likewise, debate partners' changes were uncorrelated, but with 35 complete dyads at follow-up this test can exclude only large partner effects (\SIref{app:analysis:partner}). Fourth, because we did not experimentally vary the incentive structure, its independent contribution remains to be isolated. Fifth, participants had uncurated access to information during preparation, including occasional use of language models; a realistic feature that, if anything, worked against detecting effects. Yet our interventions still produced change. Sixth, our follow-up extends to about four weeks; effects may still erode over longer horizons as participants return to their information environments and polarized social networks \citep{levenduskyWhenEffortsDepolarize2018, broockmanDurablyReducingTransphobia2016, rathjeOutgroupAnimosityDrives2021}. Finally, our measures focus on self-reported attitudes rather than behavioral outcomes like news selection or cross-cutting conversation \citep{falkenbergPatternsPartisanToxicity2024, druckmanWhatWeMeasure2019}, an important direction for future work.

Overall, our results suggest perspective-taking is necessary but not sufficient; the modality of engagement significantly influences the trajectory of change. Self-paced writing from the opposing perspective reliably produces immediate shifts in stated positions. Embedding the same cognitive task in structured debate (structural perspective-taking) is crucial for sustaining reductions in animosity toward opponents. By demonstrating a pathway to more durable affective change, our work contributes to the critical search for solutions to the ``perils of polarization'' threatening democratic functioning \citep{baldassarriEmergencePerilsPolarization2021}. Interventions that reward depth of engagement over performative success, leveraging structured interaction without priming partisan conflict, offer design principles that could inform scalable tools---though significant barriers remain, including participant recruitment, coordination costs, and potential need for automated evaluation systems \citep{costelloDurablyReducingConspiracy2024}.

\section{Methods}\label{sec:methods}

\subsection{Design}
\label{sec:methods:design}
Our experimental design engages participants in what we term the Ideological Turing Test. This test involves a participant temporarily adopting a stance that diverges from their actual beliefs in a debate or writing setting to persuade an external judge that they genuinely uphold the opposing viewpoint. This, in particular, is what is key to the reward scheme and the intervention itself. For instance, in a debate revolving around abortion, a person who is pro-life would attempt to convincingly present themselves as pro-choice. To motivate the debaters, we establish an incentive structure that rewards those who adeptly embody the opposing viewpoint and construct compelling arguments, as evaluated by the judge. 

Our study employed a modified $2\times2$ design crossing perspective-taking (own vs. opposite position) with engagement modality (debate vs. writing). We note that practical constraints necessitated a sequential implementation: Debate sessions ($N = 13$ batches) were conducted in Fall 2023 and Spring 2024, followed by writing sessions ($N = 4$ batches) in Fall 2024. The perspective assignment remained randomized within sessions (50\% own/opposite), creating a partially crossed design where activity types were confounded with temporal phase. We addressed this in two ways. First, stratified recruitment by maintaining identical pools/screening across phases, and second, covariate adjustment in modeling by including batch timing in estimation models.

\subsection{Ethics Information}
In accordance with the ethical requirements for research involving human participants, we confirm that our study complies with all relevant ethical regulations and guidelines. The study protocol has been reviewed and approved by the Institutional Review Board at the University of Michigan (HUM00232834). Furthermore, we obtained informed consent from all human participants involved in the research. Details regarding participant compensation will also be provided as part of the consent process.

\subsection{Participant Recruitment}
\label{sec:methods:recruitment}
Our recruitment strategy comprised adult participants. Considering the demographic characteristics of our pools: students attending courses at a large public university in the U.S. and students registered for the experiments pool (largely the same population).

A total of 203 participants were recruited in two waves:

\begin{itemize}
    \item \textbf{Debate cohort}: 153 participants (Fall 2023 and Spring 2024)
    \item \textbf{Writing cohort}: 50 participants (Fall 2024)
\end{itemize}

No significant differences emerged between cohorts in gender, age, political leaning, or baseline animosity (all $p > .05$, see details in \SIref{app:data}).

Eligibility was determined via a pre-intervention online survey. This survey collected demographics and, critically, identified participants with non-neutral, strongly held views on specific contentious topics who were also willing to discuss them. This ensured our sample comprised individuals open to dialogue. The full survey structure is detailed in \SIref{app:surveys}. The pre-intervention survey includes a set of demographic questions, queries about political affiliation, and inquiries into political media consumption habits. 

\subsection{Debate Topics}
\label{sec:methods:topics}
Six contentious topics (e.g., reproductive rights, public health mandates) were identified through pilot focus groups and adapted to student-relevant framings (full statements in \SIref{app:surveys:issues}). Participants were eligible for a topic if they held a non-neutral position (on a 5-point Likert scale) and expressed willingness to engage in discussion (rated $\geq$ 3 on a 5-point scale). Assignment and matching protocols are detailed in \SIref{app:protocols:logistics}.

\subsection{Experimental Procedure}
\label{sec:methods:procedure}
Upon arrival, participants received anonymized credentials and a standardized introduction explaining the activity, compensation for argument quality and authenticity, and communication rules. They were not informed of their assigned perspective (own or opposite).

\paragraph*{Debate Intervention}
A matching algorithm paired participants into dyads, randomly assigning 50\% to defend their \textit{opposite} pre-survey position. Pairs shared the same baseline stance on a topic, creating debates with identical views but divergent assigned roles. After a 25-minute preparation phase, pairs engaged in a 25-minute free-form chat debate via a messaging interface. Judges later evaluated anonymized chat logs.

\paragraph*{Writing Intervention}
The writing condition mirrored debate in recruitment, matching, and preparation. Critically, session framing avoided references to debate or adversarial interaction. Following preparation, participants composed a standalone essay defending their assigned stance, which was later evaluated by judges. This design isolated perspective-taking from social debate dynamics.

\paragraph*{Judging}
After the main intervention, participants judged anonymized outputs (debates or essays) from their session immediately after completing their own task. Each output was evaluated by at least three peers on perceived authenticity and persuasiveness. Performance bonuses for the activity depended on these peer judgments. Full procedural details are in \SIref{app:protocols:judging}.

Further details on protocols can be found under \SIref{app:protocols:debate}, \SIref{app:protocols:writing}, and \SIref{app:protocols:judging}, as well as logistics details during the sessions in \SIref{app:protocols:logistics}.

\subsection{Experiment Arms}  
\label{sec:methods:arms}  
The study's modified $2\times2$ design generated four experimental arms:  
\begin{itemize}  
    \item \textbf{Debate/Own}: Defend pre-survey position through real-time chat 
    \item \textbf{Debate/Opposite}: Argue opposing stance via debate
    \item \textbf{Writing/Own}: Write essay supporting original position.
    \item \textbf{Writing/Opposite}: Write essay adopting counter-attitudinal stance
\end{itemize}  

\paragraph*{Design Rationale}  
We focused on comparing interventions rather than including a passive control arm (e.g., a survey-only approach), as prior work demonstrates that mere reflection yields minimal reduction in animosity \citep{warnerReducingPoliticalPolarization2020}. Instead, within-subject comparisons, where participants' pre- and post-attitude shifts served as the primary counterfactual, were used. For example, a participant defending their own position (Debate/Own) provided a baseline against which their counterpart debating the opposite stance (Debate/Opposite) could be contrasted, thereby isolating the incremental effect of perspective-taking.  

\paragraph*{Temporal Confounds \& Mitigation}  
The modality of the activity (debate/writing) was confounded with the semester timing (2023--24 academic year versus Fall 2024). To address this, we maintained identical recruitment pools (same courses, screening criteria), balanced topic assignments across semesters (e.g., abortion debated in Spring and written in Fall), and included batch timing as a covariate in all models. Sensitivity analyses showed no detectable time-specific differences in the data distributions (\SIref{app:analysis:balance} contains details on distributions and balance analysis for Modality).  

\subsection{Surveys and Data Collection}  
\label{sec:methods:surveys}  
Participants completed three surveys administered through Qualtrics:  
\begin{itemize}  
    \item \textbf{Pre-intervention}: Baseline attitudes and demographics, completed during recruitment. 1-2 weeks before intervention (activity in the lab)
    \item \textbf{Post-intervention}: Immediate attitude reassessment and intervention feedback, completed after debates/writing.  
    \item \textbf{Follow-up}: Delayed attitude measurement, typically 1--4 weeks post-intervention (median about 2 weeks).  
\end{itemize}  

During sessions, judges evaluated debates/writings via dedicated Qualtrics surveys, assessing (1) perceived authenticity of assigned positions and (2) argument persuasiveness. Each judge evaluated 1-2 randomly assigned interactions from the same session, completing judging evaluations after their post-intervention survey. Additional details about the survey are in \SIref{app:surveys}

\subsection{Participant Compensation}
\label{sec:methods:compensation}
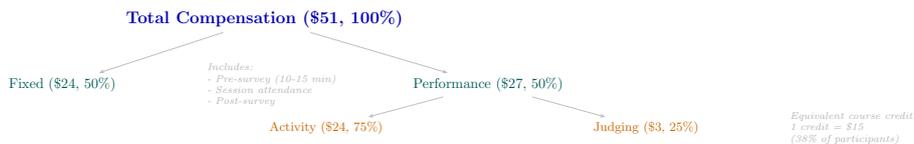
\begin{figure*}[htbp]
\centering
\resizebox{\linewidth}{!}{%
\begin{tikzpicture}[
    node/.style={font=\sffamily},
    l1/.style={node, font=\large\bfseries, blue!70!black},
    l2/.style={node, font=\normalsize, teal!70!black},
    l3/.style={node, font=\small, orange!80!black},
    arrow/.style={-{Stealth[length=3pt]}, gray!50, thin}
]

\node[l1] (total) {Total Compensation (\$51, 100\%)};
    
\node[l2, below left=1cm and 0cm of total] (fixed) 
    {Fixed (\$24, 50\%)};
\node[l2, below right=1cm and 0cm of total] (performance) 
    {Performance (\$27, 50\%)};

\node[l3, below left=0.7cm of performance] (activity) 
    {Activity (\$24, 75\%)};
\node[l3, below right=0.7cm of performance] (judging) 
    {Judging (\$3, 25\%)};

\draw[arrow] (total) -- (fixed);
\draw[arrow] (total) -- (performance);
\draw[arrow] (performance) -- (activity);
\draw[arrow] (performance) -- (judging);

\node[anchor=west, font=\footnotesize\itshape, gray!60, text width=5cm]
    at ([xshift=2cm]fixed.east) {
        Includes:\\
        - Pre-survey (10-15 min)\\
        - Session attendance\\
        - Post-survey
    };

\node[anchor=west, font=\footnotesize\itshape, gray!60, text width=5cm]
    at ([xshift=2cm]judging.east) {
        Equivalent course credit:\\
        1 credit = \$15\\
        (38\% of participants)
    };

\end{tikzpicture}%
}
\caption{\textbf{Participant compensation structure.} Clean text-based compensation structure showing hierarchical relationships and proportional breakdown. Color and font weight differentiate compensation components, with annotations explaining key elements.}
\label{fig:compensation_clean}
\end{figure*}

Participants were recruited from two pools at a large U.S. public university: 36\% received monetary compensation via mailed checks, while 64\% earned course credit scaled to equivalent hourly rates. Both pools followed identical proportional structures to ensure behavioral equivalence, with total compensation comprising a fixed base and performance-based bonuses (c.f. \cref{fig:compensation_clean}).

The fixed compensation of \$24 (approximately 2 hours at \$12/hr) rewarded completion of three components: the pre-intervention survey (10-15 minutes), full session attendance (debate/writing and judging tasks), and the post-intervention survey. This structure intentionally integrated the pre-survey into the base payment to avoid differential attrition between recruitment phases.  

Performance bonuses (up to \$27) were awarded through two channels. First, debaters and writers could earn \$24 (75\% of potential bonus) if a majority of judges rated their arguments both persuasive (top 30\% of submissions) and authentic (not identified as role-playing). Second, judges received a \$3 bonus (25\%) for each evaluation where their pretender detection aligned with the majority consensus. Maximum earnings reached \$51 (\$24 base + \$27 bonus), with extra credit participants receiving proportional course points.  

Compensation was distributed 3-4 weeks after the intervention via mailed checks or gradebook updates, contingent upon completion of the post-survey. This delay ensured data integrity while mirroring real-world incentive timelines. The design prioritized three behavioral drivers: (1) authentic perspective-taking over rhetorical dominance, (2) careful evaluation during judging tasks, and (3) equitable engagement across compensation modalities. Cross-pool equivalence documentation appears in \SIref{app:analysis:balance}.

\subsection{Measures of Issue-based Animosity and Stance}
\label{sec:methods:measures}

Consistent with our theoretical focus on issue-level dynamics, we measure what we term \textbf{issue-based animosity}---feelings toward people who disagree on a specific policy---rather than party-based affective polarization. We also measure the participant's \textbf{stance} on the issue. These primary outcomes were measured at three waves: pre-intervention (\textit{pre}), post-intervention (\textit{post}), and a follow-up typically 1--4 weeks later (\textit{follow}). We analyzed change using estimated marginal mean (EMM) contrasts from mixed-effects models rather than raw change scores. This approach minimizes error propagation from subtracting two noisy measurements and properly accounts for repeated measures correlation \citep{vickersAnalysingControlledTrials2001,fitzmauriceAppliedLongitudinalAnalysis2012,lenthEmmeansEstimatedMarginal2025}. For any outcome $Y$ and arm $a$, we define change at time $t \in \{\pos, \fol\}$ as:
\[
\Delta_{t}(a) \;=\; \EMM\!\big(Y \mid a,t\big)\;-\;\EMM\!\big(Y \mid a,\pre\big),
\]
on the original response scale of $Y$.

\paragraph*{Issue-based Animosity ($Y=\anim$)}
At each wave $s\in\{\pre,\pos,\fol\}$, participants rated warmth (0=``Very cold/unfavorable'' to 100=``Very warm/favorable'') toward people who agree and disagree with the policy position using the Feeling Thermometer \citep{wilcoxItHotIndividual1989, nelsonFeelingThermometer2008}. To conservatively guard against post-wave neutralization, we defined animosity as the within-wave minimum of the two ratings:
\[
\anim_i(s) \;=\; \min\!\big\{T_{i,\text{agree}}(s),\; T_{i,\text{disagree}}(s)\big\}, \qquad \anim_i(s)\in[0,100].
\]
Higher values indicate warmer feelings toward the least-liked group, i.e., lower issue-based animosity. Models are fit to $\anim_i(s)$.

\paragraph*{Stance ($Y=\stance$)}
Agreement with the policy statement was measured on a 5-point Likert scale (1=``Strongly Disagree'' to 5=``Strongly Agree''), mapped to $\tilde S_i(s)\in[-2,2]$ (0=neutral). To ensure interpretability, we reoriented using baseline sign:
\begin{gather*}
s_{i,\pre}=\mathrm{sign}\!\big(\tilde S_i(\pre)\big)\in\{-1,+1\},\\
\stance_i(s)= -\,s_{i,\pre}\,\tilde S_i(s)\in[-2,2].
\end{gather*}
Thus, higher $\stance_i(s)$ always indicates movement toward the opposite stance (or toward moderation). Models are fit to $\stance_i(s)$.

\paragraph*{Performance Indicators}
We created binary indicators for judged performance: Best Argument and Authenticity (each =1 if at least 2 of 3 judges selected the participant). We also considered ``Both'' when both indicators equaled 1. These indicators were used to test whether judged performance moderates attitudinal change (H3).

\paragraph*{Exclusions and Covariates}
We excluded participants (3.3\%) who failed consistency checks (e.g., reporting colder feelings toward their own side) and one session ($N=12$) due to technical failures. All models included covariates: topic fixed effects, demographic variables (gender, ethnicity, political viewpoint, opinion strength), and random intercepts for participant and session block.

\subsection{Analysis}
\label{sec:methods:analysis}

All analyses used mixed-effects models fit with \texttt{lme4} \citep{batesFittingLinearMixedeffects2015}, with estimated marginal means and contrasts computed via \texttt{emmeans} \citep{lenthEmmeansEstimatedMarginal2025}. We report unadjusted 95\% confidence intervals and adjust $p$-values for families of comparisons using Holm's method \citep{holmSimpleSequentiallyRejective1979} (Tukey for pairwise arm differences).

\textbf{Hypothesis 1 (Within-arm Change).} For each outcome $Y\in\{\anim,\stance\}$, we fit a linear mixed model with arm × time factorial structure, covariates, and random intercepts for participant and session block. The model for observation $i$ in arm $a$ at wave $s$ is:
\[
Y_{i,a}(s) = \mu + \alpha_a + \tau_s + (\alpha\tau)_{a,s} + \mathbf{x}_i^\top\beta + u_i + b_{\text{block}(i)} + \varepsilon_{ias},
\]
where $\mathbf{x}_i$ includes the covariates noted above. From this, we compute $\Delta_t(a)$ as defined in \secref{sec:methods:measures}.

\textbf{Hypothesis 2 (Perspective Advantage).} To estimate the advantage of arguing the opposite perspective within each modality, we fit a model with perspective × modality × time factorial structure:
\[
Y_i(s) = \mu + \alpha_p + \beta_m + \tau_s + \text{interactions} + \mathbf{x}_i^\top\gamma + u_i + b_{\text{block}(i)} + \varepsilon_{is},
\]
where $p\in\{\Own,\Opp\}$ and $m\in\{\Wrt,\Dbt\}$. The perspective advantage for modality $m$ at time $t$ is $\delta_{m,t} = \Delta_t(\Opp,m) - \Delta_t(\Own,m)$, a difference-in-differences contrast.

\textbf{Hypothesis 3 (Performance Moderation).} We extended the H1 model to include win-type indicators and their interactions with time, estimating the average difference in $\Delta_t$ between winners (Best Argument, Authenticity, or Both) and losers, pooled across arms.

\textbf{Hypothesis 4 (Willingness to Repeat).} Willingness (binary: ``Probably/Definitely yes'') was analyzed via logistic mixed model with modality × perspective fixed effects, covariates, and random intercept for debate pair. We report pooled risk differences for modality and perspective effects, assessing non-inferiority against $\delta=0.05$ following standard guidelines \citep{piaggioReportingNoninferiorityEquivalence2012}. As a robustness check, we also fit a cumulative logit mixed model \citep{christensenOrdinalRegressionModels2023}.

We conducted sensitivity analyses including: (1) inverse-probability weighting for modality scheduling and follow-up attrition, (2) change-score models as an alternative to EMM contrasts, (3) ordinal models for Likert-scale outcomes, (4) alternative coding of performance indicators, and (5) a test of whether debate partners' changes were correlated. Complete model specifications, formulas, and detailed robustness checks are in \SIref{app:analysis}.

\backmatter

\bmhead{Funding}
National Science Foundation grant CCF-2208662 (G.S., D.G.); University of Michigan Democracy \& Civic Empowerment Initiative, MDemocracy grant (D.G., D.M.R.).

\bmhead{Ethics approval and consent to participate}
The study was approved by the University of Michigan Institutional Review Board (HUM00232834). All participants gave informed consent.

\bmhead{Competing interests}
The authors declare no competing interests.

\bmhead{Data and code availability}
The anonymized data will be available on Zenodo (\url{https://doi.org/10.5281/zenodo.24506839}) \cite{gambaIdeologicalTuringTestData2026}, and the analysis code will be archived on Zenodo (\url{https://doi.org/10.5281/zenodo.24507086}) \cite{gambaIdeologicalTuringTestCode2026}.

\bmhead{Author contributions}
D.G., D.M.R. and G.S. conceived and designed the study. D.G. conducted the experiment, curated and analyzed the data, and wrote the original draft. D.M.R. and G.S. supervised the project and reviewed and edited the manuscript.

\bibliography{IdeoTuring}

\clearpage
\renewcommand{\thefigure}{S\arabic{figure}}
\renewcommand{\thetable}{S\arabic{table}}
\renewcommand{\theequation}{S\arabic{equation}}
\renewcommand{\thesection}{S\arabic{section}}
\setcounter{figure}{0}
\setcounter{table}{0}
\setcounter{equation}{0}
\setcounter{section}{0}
\crefalias{figure}{sfigure}
\crefalias{table}{stable}
\let\cref\Cref %

\begin{center}
{\Large\bfseries Supplementary Materials}
\end{center}
\bigskip

\section{Data inclusion, demographics, and attrition}
\label{app:data}

\subsection{Inclusion and analysis sample}
A total of 315 individuals completed the pre-survey. Of them, 91 did not register for a session or were unable to participate, leaving 224 participants. After excluding one session with technical issues, as well as a small number of debate pairs with assignment problems, and applying the pre-specified consistency check (requiring that affect toward the same side of the argument is greater than toward the opposite side at pre-survey), the analysis sample comprises 203 participants. The flow of participants through these filters is summarized in \cref{tab:cohort_flow}.

\begin{table}[!h]
\centering
\caption{\label{tab:cohort_flow}\textbf{Cohort Flow.} Sequential filters: debated → !exclude (technical/assignment) → consistency. Unit: participant.}
\centering
\begin{tabular}[t]{lrl}
\toprule
Stage & N & Percent of total\\
\midrule
All unique participants & 315 & 100.0\%\\
Has debated row & 224 & 71.1\%\\
Passed technical/assignment & 210 & 66.7\%\\
Passed consistency (pre affect check) & 203 & 64.4\%\\
Included in analysis sample & 203 & 64.4\%\\
\bottomrule
\end{tabular}
\end{table}

Overall, 203 participants (64.4\%) were included, 91 (28.9\%) were excluded at the debated step, 14 (4.4\%) were removed due to technical or assignment issues, and 7 (2.2\%) failed the consistency check (\cref{tab:exclusion_reasons_overall}).

\begin{table}[!h]
\centering
\caption{\label{tab:exclusion_reasons_overall}\textbf{Reasons for Exclusion (Overall).}}
\centering
\begin{tabular}[t]{lrr}
\toprule
Reason & N & Percent\\
\midrule
Included & 203 & 64.4\%\\
Not debated & 91 & 28.9\%\\
Technical/assignment & 14 & 4.4\%\\
Failed consistency check & 7 & 2.2\%\\
\bottomrule
\end{tabular}
\end{table}

\begin{figure}[h]
  \centering
  \includegraphics[width=0.8\linewidth]{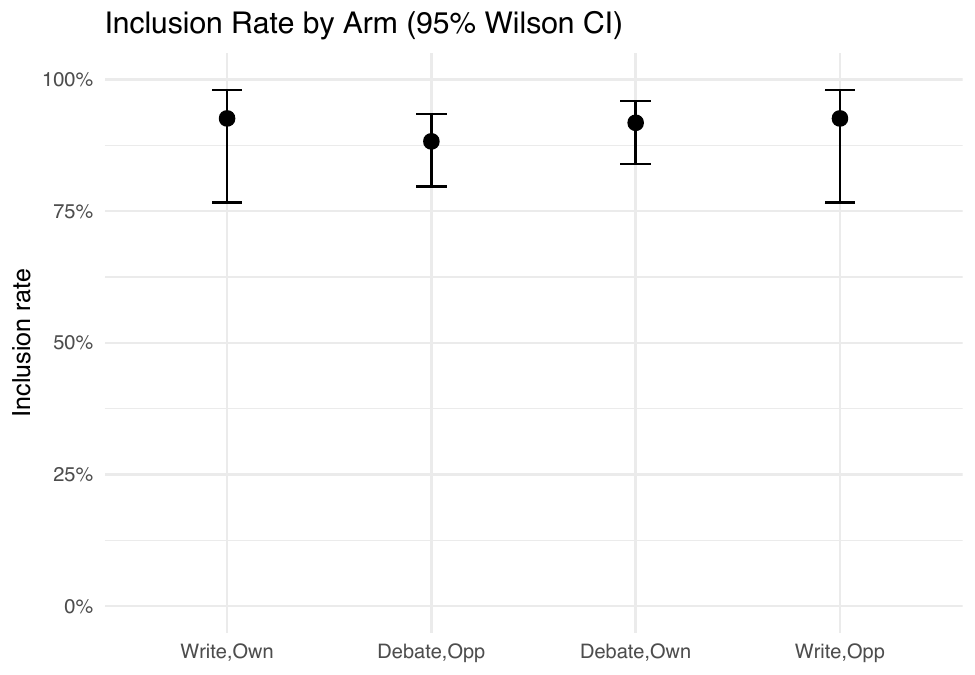}
  \caption{\textbf{Inclusion rate by arm with 95\% Wilson confidence intervals.}}
  \label{fig:inclusionarm}
\end{figure}

Inclusion rates were similar across arms. For example, the Write/Own arm had 92.6\% inclusion (95\% CI: 76.6\%--97.9\%), Debate/Opp 88.2\% (79.7\%--93.5\%), Debate/Own 91.8\% (84.0\%--96.0\%), and Write/Opp 92.6\% (76.6\%--97.9\%). A test of equality of proportions finds no significant differences ($p=0.81$). These rates are shown in \cref{fig:inclusionarm}.

\subsection{Demographics and arms in the analysis sample}
Demographics (gender, ethnicity, political viewpoint) were balanced across arms. Chi-square tests of balance all yield $p > 0.2$ (\cref{tab:demographics_by_arm}). No systematic differences are observed.

\begin{table}[!h]
\centering\centering
\caption{\label{tab:demographics_by_arm}\textbf{Demographics by Study Arm.} [Unit: participant. Sample restricted to debated==TRUE, consistent==TRUE, exclude==FALSE. Categories lumped/suppressed as noted.]}
\centering
\fontsize{7.5}{9.5}\selectfont
\setlength{\tabcolsep}{2pt}
\begin{tabular}[t]{lcccccc}
\toprule
\textbf{Characteristic} & \makecell[c]{\textbf{Overall}\\ \\N = 203} & \makecell[c]{\textbf{Write,Own}\\ \\N = 25} & \makecell[c]{\textbf{Debate,Opp}\\ \\N = 75} & \makecell[c]{\textbf{Debate,Own}\\ \\N = 78} & \makecell[c]{\textbf{Write,Opp}\\ \\N = 25} & \textbf{p-value}\\
\midrule
gender, n (\%) &  &  &  &  &  & 0.22\\
\hspace{1em}Male & 91 (45) & 16 (64) & 27 (36) & 35 (45) & 13 (52) & \\
\hspace{1em}Female & 102 (50) & 9 (36) & 43 (57) & 40 (51) & 10 (40) & \\
\hspace{1em}Other & 10 (4.9) & 0 (0) & 5 (6.7) & 3 (3.8) & 2 (8.0) & \\
\hspace{1em}Missing & 0 (0) & 0 (0) & 0 (0) & 0 (0) & 0 (0) \vphantom{2} & \\
ethnic, n (\%) &  &  &  &  &  & 0.15\\
\hspace{1em}White / Caucasian & 80 (39) & 8 (32) & 25 (33) & 34 (44) & 13 (52) & \\
\hspace{1em}Black / Hispanic & 20 (9.9) & 1 (4.0) & 7 (9.3) & 12 (15) & 0 (0) & \\
\hspace{1em}Asian & 79 (39) & 14 (56) & 33 (44) & 24 (31) & 8 (32) & \\
\hspace{1em}Other & 24 (12) & 2 (8.0) & 10 (13) & 8 (10) & 4 (16) & \\
\hspace{1em}Missing & 0 (0) & 0 (0) & 0 (0) & 0 (0) & 0 (0) \vphantom{1} & \\
Political viewpoint, n (\%) &  &  &  &  &  & 0.38\\
\hspace{1em}Neutral & 47 (23) & 7 (28) & 15 (20) & 21 (27) & 4 (16) & \\
\hspace{1em}Prefer not to say & 13 (6.4) & 2 (8.0) & 3 (4.0) & 5 (6.4) & 3 (12) & \\
\hspace{1em}Conservative & 20 (9.9) & 4 (16) & 9 (12) & 3 (3.8) & 4 (16) & \\
\hspace{1em}Liberal & 123 (61) & 12 (48) & 48 (64) & 49 (63) & 14 (56) & \\
\hspace{1em}Missing & 0 (0) & 0 (0) & 0 (0) & 0 (0) & 0 (0) & \\
\bottomrule
\end{tabular}
\par\smallskip
\parbox{0.95\linewidth}{\footnotesize \textsuperscript{1} Pearson's Chi-squared test}
\end{table}

Attrition at follow-up (completion of the follow-up survey) was also not significantly different by arm ($p=0.76$; \cref{tab:attrition_by_arm}). 

\begin{table}[!h]
\centering\centering
\caption{\label{tab:attrition_by_arm}\textbf{Attrition by Study Arm.} [Counts are per participant; same sample restrictions as above.]}
\centering
\fontsize{8}{10}\selectfont
\setlength{\tabcolsep}{3pt}
\begin{tabular}[t]{lcccccc}
\toprule
\textbf{Characteristic} & \makecell[c]{\textbf{Overall}\ \ \\N = 203} & \makecell[c]{\textbf{Write,Own}\ \ \\N = 25} & \makecell[c]{\textbf{Debate,Opp}\ \ \\N = 75} & \makecell[c]{\textbf{Debate,Own}\ \ \\N = 78} & \makecell[c]{\textbf{Write,Opp}\ \ \\N = 25} & \textbf{p-value}\\
\midrule
followup, n (\%) &  &  &  &  &  & 0.86\\
\hspace{1em}FALSE & 85 (42) & 11 (44) & 29 (39) & 33 (42) & 12 (48) & \\
\hspace{1em}TRUE & 118 (58) & 14 (56) & 46 (61) & 45 (58) & 13 (52) & \\
\bottomrule
\end{tabular}
\par\smallskip
\parbox{0.95\linewidth}{\footnotesize \textsuperscript{1} Pearson's Chi-squared test}
\end{table}

\subsection{Intervention topics}
\begin{figure}[h]
  \centering
  \includegraphics[width=0.9\linewidth]{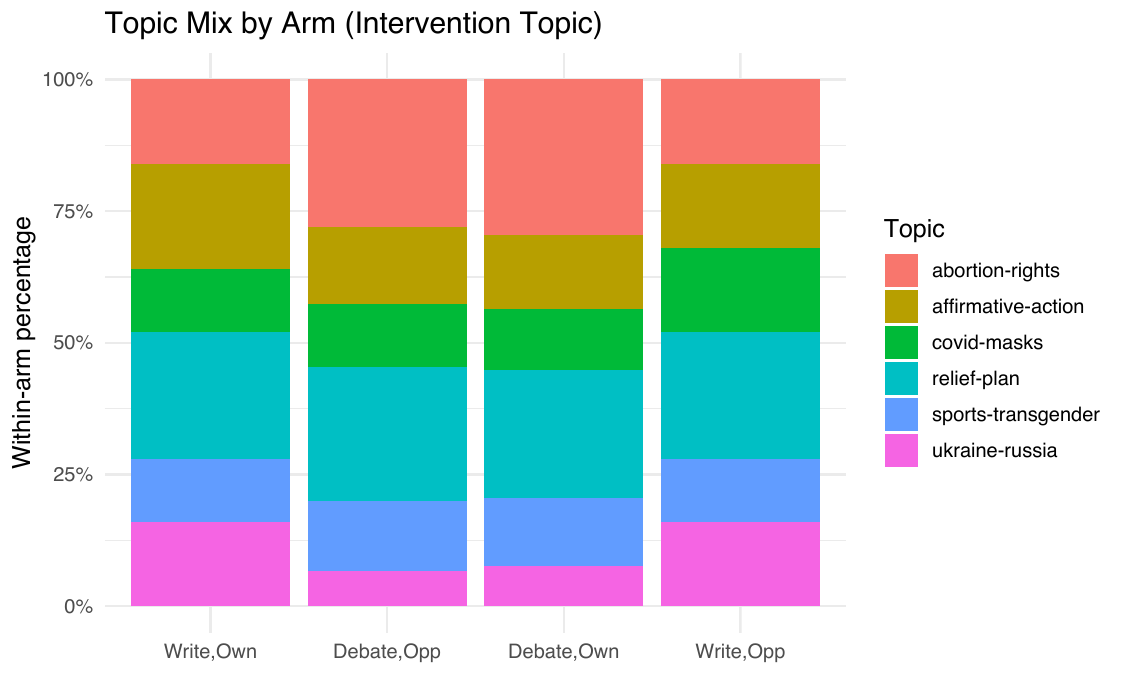}
  \caption{\textbf{Topic mix within each arm (percent within arm).}}
  \label{fig:topicmix}
\end{figure}

The distribution of intervention topics by arm is reported in \cref{tab:topic_by_arm_counts}. The most common topics were abortion rights (52 participants) and relief plan (50 participants). Visual inspection of the topic mix (see \cref{fig:topicmix}) suggests no extreme imbalance across arms.

\begin{table}[!h]
\centering
\caption{\label{tab:topic_by_arm_counts}\textbf{Counts by Topic and Arm (Intervention Topic).} [Unit: participant's intervention topic row; totals include all arms.]}
\centering
\begin{tabular}[t]{lrrrrr}
\toprule
topic & Write,Own & Debate,Opp & Debate,Own & Write,Opp & Total\\
\midrule
abortion-rights & 4 & 21 & 23 & 4 & 52\\
affirmative-action & 5 & 11 & 11 & 4 & 31\\
covid-masks & 3 & 9 & 9 & 4 & 25\\
relief-plan & 6 & 19 & 19 & 6 & 50\\
sports-transgender & 3 & 10 & 10 & 3 & 26\\
\addlinespace
ukraine-russia & 4 & 5 & 6 & 4 & 19\\
Total & 25 & 75 & 78 & 25 & 203\\
\bottomrule
\end{tabular}
\end{table}

\subsection{Follow-up and attrition}
\begin{figure}[h]
  \centering
  \includegraphics[width=0.8\linewidth]{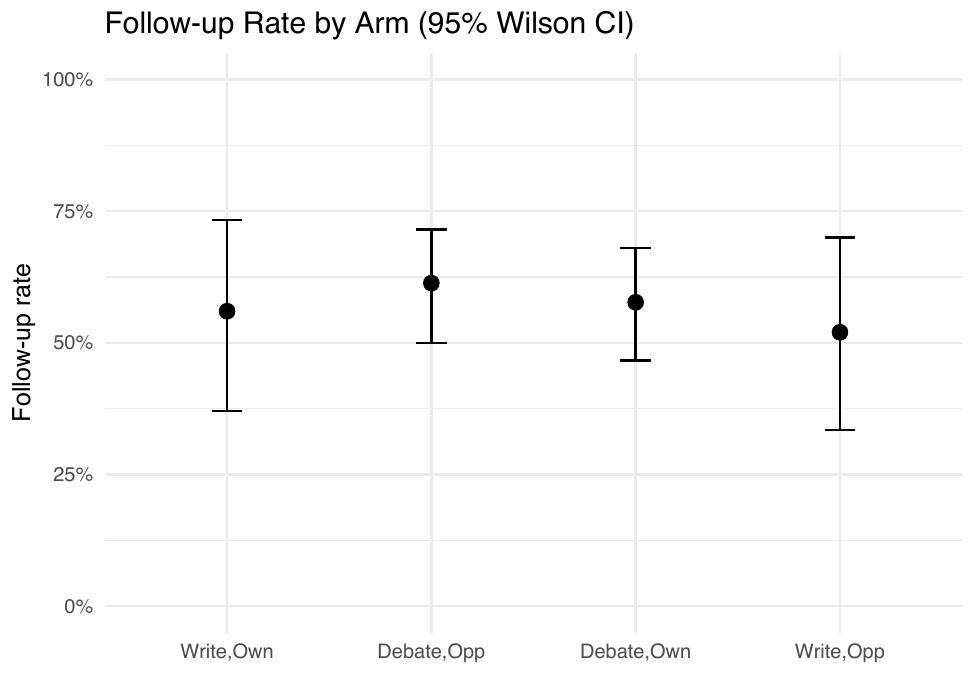}
  \caption{\textbf{Follow-up rates by arm with 95\% Wilson confidence intervals.}}
  \label{fig:followuparm}
\end{figure}

Follow-up rates by arm were as follows: Write/Own 51.9\% (95\% CI: 34.0\%--69.3\%), Debate/Opp 61.3\% (50.0\%--71.5\%), Debate/Own 59.2\% (48.0\%--69.6\%), and Write/Opp 52.0\% (33.5\%--70.0\%) (\cref{tab:followup_by_arm}, \cref{fig:followuparm}). Chi-square test indicates no significant differences ($p=0.76$).

\begin{table}[!h]
\centering
\caption{\label{tab:followup_by_arm}\textbf{Follow-up by Arm (Rate with 95\% CI).}}
\centering
\begin{tabular}[t]{lrrrc}
\toprule
Arm & N & Follow-ups & Rate & 95\% CI\\
\midrule
Write,Own & 25 & 14 & 56.0\% & 37.1\%–73.3\%\\
Debate,Opp & 75 & 46 & 61.3\% & 50.0\%–71.5\%\\
Debate,Own & 78 & 45 & 57.7\% & 46.6\%–68.0\%\\
Write,Opp & 25 & 13 & 52.0\% & 33.5\%–70.0\%\\
\bottomrule
\end{tabular}
\end{table}

We additionally tested whether attrition was associated with baseline covariates (arm, topic, gender, ethnicity, political viewpoint). Logistic regression estimates are reported in \cref{tab:attrition_logit}; no covariate shows significant association with follow-up. Balance tables comparing responders vs. non-responders are shown in \cref{tab:followup_balance}, with no significant differences detected (all $p > 0.1$ except for gender $p=0.09$).

\begin{table}[h]
\caption{\label{tab:attrition_logit}\textbf{Logistic regression of follow-up (ORs, 95\% CI)}}
\fontsize{9.8pt}{11.7pt}\selectfont
\begin{tabular*}{\linewidth}{@{\extracolsep{\fill}}lcc}
\toprule
\textbf{Characteristic} & \textbf{OR} \textbf{(95\% CI)} & \textbf{p-value} \\ 
\midrule\addlinespace[2.5pt]
{\bfseries Arm} &  & 0.81 \\ 
    Write,Own & — &  \\ 
    Debate,Opp & 1.05 (0.39 to 2.77) &  \\ 
    Debate,Own & 1.07 (0.40 to 2.83) &  \\ 
    Write,Opp & 0.67 (0.20 to 2.17) &  \\ 
{\bfseries Topic} &  & 0.57 \\ 
    abortion-rights & — &  \\ 
    affirmative-action & 2.15 (0.79 to 6.21) &  \\ 
    covid-masks & 1.61 (0.57 to 4.71) &  \\ 
    relief-plan & 1.11 (0.49 to 2.54) &  \\ 
    sports-transgender & 1.40 (0.52 to 3.87) &  \\ 
    ukraine-russia & 0.78 (0.25 to 2.42) &  \\ 
{\bfseries Gender} &  & 0.083 \\ 
    Male & — &  \\ 
    Female & 1.27 (0.67 to 2.43) &  \\ 
    Other & 8.12 (1.25 to 164) &  \\ 
{\bfseries Ethnicity} &  & 0.76 \\ 
    White / Caucasian & — &  \\ 
    Black / Hispanic & 0.71 (0.24 to 2.09) &  \\ 
    Asian & 0.88 (0.44 to 1.76) &  \\ 
    Other & 1.40 (0.49 to 4.25) &  \\ 
{\bfseries Political viewpoint} &  & 0.42 \\ 
    Neutral & — &  \\ 
    Prefer not to say & 0.87 (0.19 to 3.97) &  \\ 
    Conservative & 2.57 (0.80 to 8.95) &  \\ 
    Liberal & 1.32 (0.62 to 2.82) &  \\ 
\bottomrule
\end{tabular*}
\begin{minipage}{\linewidth}
Abbreviations: CI = Confidence Interval, OR = Odds Ratio\\
\end{minipage}
\end{table}

\begin{table}[!h]
\centering\centering
\caption{\label{tab:followup_balance}\textbf{Baseline characteristics by follow-up status.}}
\centering
\fontsize{8}{10}\selectfont
\begin{tabular}[t]{lcccc}
\toprule
\textbf{Characteristic} & \makecell[c]{\textbf{Overall}\ \ \\N = 203} & \makecell[c]{\textbf{Follow up}\ \ \\N = 118} & \makecell[c]{\textbf{Did not respond}\ \ \\N = 85} & \textbf{p-value}\\
\midrule
arm, n (\%) &  &  &  & 0.86\\
\hspace{1em}Write,Own & 25 (12) & 14 (12) & 11 (13) & \\
\hspace{1em}Debate,Opp & 75 (37) & 46 (39) & 29 (34) & \\
\hspace{1em}Debate,Own & 78 (38) & 45 (38) & 33 (39) & \\
\hspace{1em}Write,Opp & 25 (12) & 13 (11) & 12 (14) & \\
topic, n (\%) &  &  &  & 0.46\\
\hspace{1em}abortion-rights & 52 (26) & 29 (25) & 23 (27) & \\
\hspace{1em}affirmative-action & 31 (15) & 22 (19) & 9 (11) & \\
\hspace{1em}covid-masks & 25 (12) & 16 (14) & 9 (11) & \\
\hspace{1em}relief-plan & 50 (25) & 28 (24) & 22 (26) & \\
\hspace{1em}sports-transgender & 26 (13) & 15 (13) & 11 (13) & \\
\hspace{1em}ukraine-russia & 19 (9.4) & 8 (6.8) & 11 (13) & \\
gender, n (\%) &  &  &  & 0.10\\
\hspace{1em}Male & 91 (45) & 50 (42) & 41 (48) & \\
\hspace{1em}Female & 102 (50) & 59 (50) & 43 (51) & \\
\hspace{1em}Other & 10 (4.9) & 9 (7.6) & 1 (1.2) & \\
\hspace{1em}Missing & 0 (0) & 0 (0) & 0 (0) \vphantom{2} & \\
ethnic, n (\%) &  &  &  & 0.57\\
\hspace{1em}White / Caucasian & 80 (39) & 49 (42) & 31 (36) & \\
\hspace{1em}Black / Hispanic & 20 (9.9) & 10 (8.5) & 10 (12) & \\
\hspace{1em}Asian & 79 (39) & 43 (36) & 36 (42) & \\
\hspace{1em}Other & 24 (12) & 16 (14) & 8 (9.4) & \\
\hspace{1em}Missing & 0 (0) & 0 (0) & 0 (0) \vphantom{1} & \\
Political viewpoint, n (\%) &  &  &  & 0.53\\
\hspace{1em}Neutral & 47 (23) & 24 (20) & 23 (27) & \\
\hspace{1em}Prefer not to say & 13 (6.4) & 8 (6.8) & 5 (5.9) & \\
\hspace{1em}Conservative & 20 (9.9) & 14 (12) & 6 (7.1) & \\
\hspace{1em}Liberal & 123 (61) & 72 (61) & 51 (60) & \\
\hspace{1em}Missing & 0 (0) & 0 (0) & 0 (0) & \\
\bottomrule
\end{tabular}
\par\smallskip
\parbox{0.95\linewidth}{\footnotesize \textsuperscript{1} Pearson's Chi-squared test}
\end{table}

\section{Chat Platform}
\label{app:platform}

\begin{figure}[h!]
  \centering
  \includegraphics[width=0.95\linewidth]{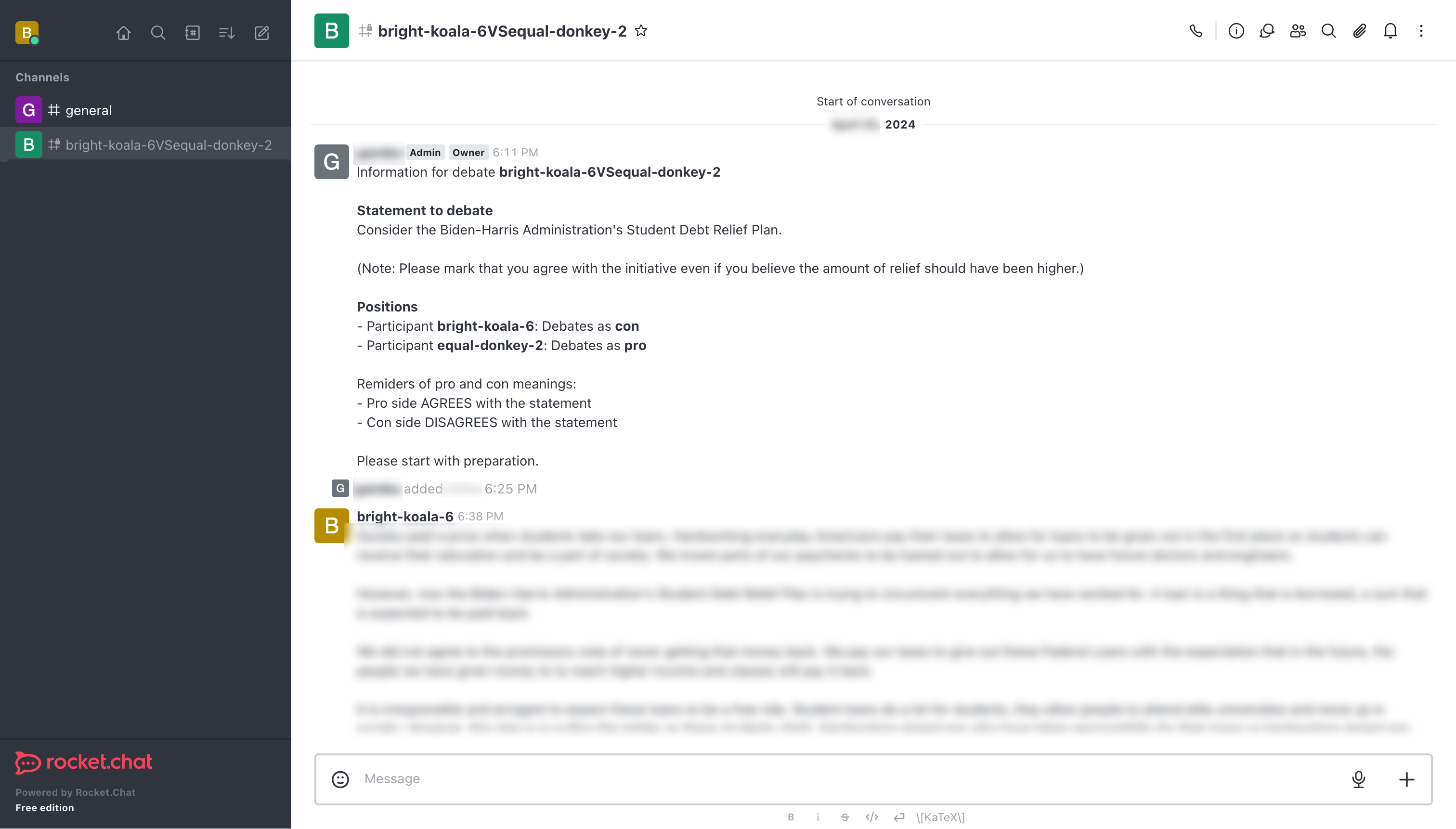}
  \caption{\textbf{Chat platform view in the debate modality.} If participants engaged in the debate modality, they were added to a channel (private chat room) with the other debate party. An automatic message from an admin account wrote the issue and the assigned positions for each username. Sensitive information has been blurred.}
  \label{fig:rocketchat-debating}
\end{figure}

\begin{figure}[h!]
  \centering
  \includegraphics[width=0.95\linewidth]{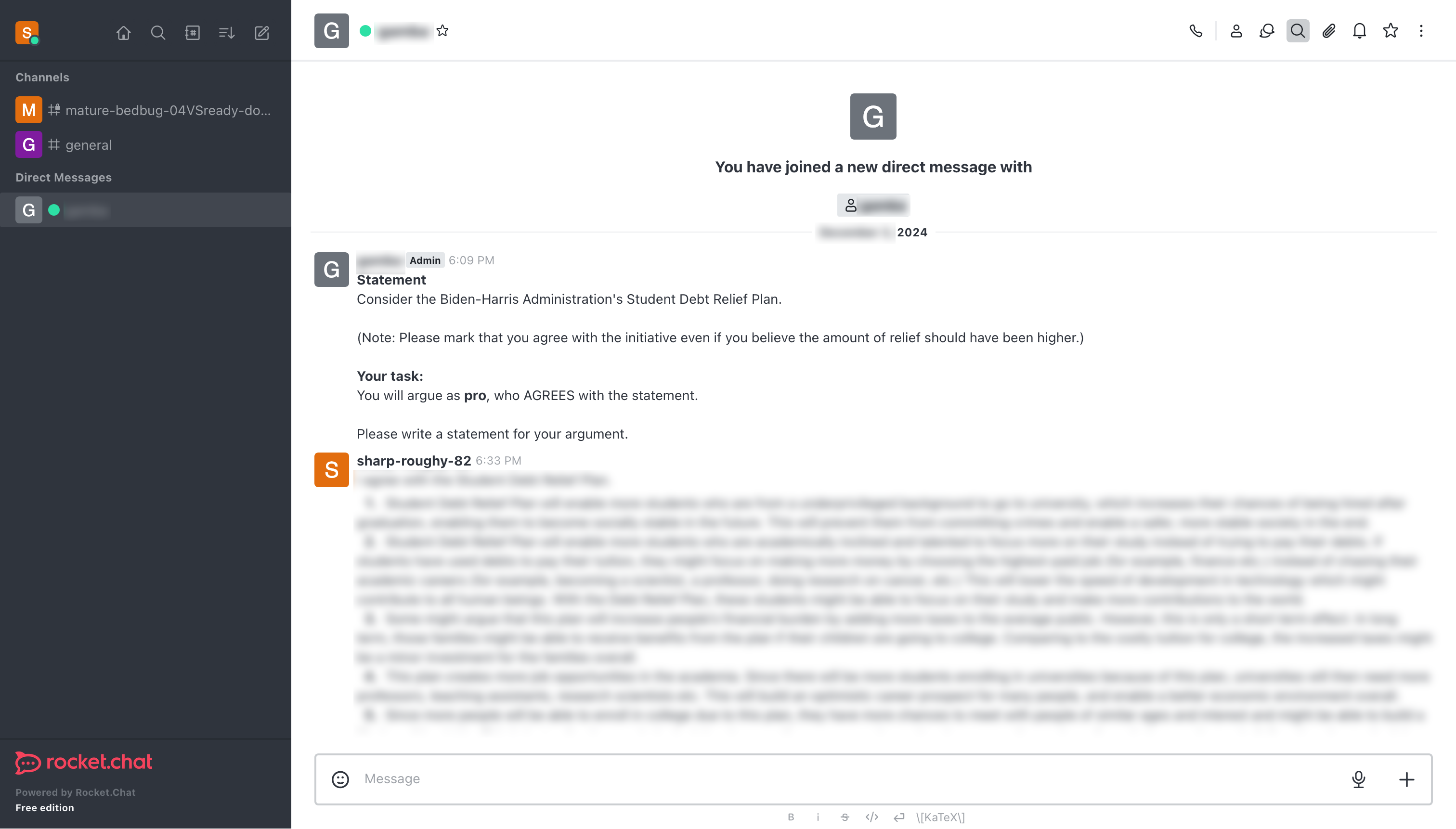}
  \caption{\textbf{Chat platform view in the writing modality.} If participants engaged in the writing modality, they received a private message from the administrator account with the details of the assigned issue and the position to write about. Sensitive information has been blurred.}
  \label{fig:rocketchat-writing}
\end{figure}

We used a custom deployment of the RocketChat messaging server for participant interaction.

Each participant arrived at a computer station that had already logged in with a randomly generated username. We had previously mapped usernames to desks, and we assigned desks randomly at registration. This allowed us to identify the user based on other data (surveys).

If participants engaged in the debate modality, they were added to a channel (private chat room) with the other debate party. An automatic message from an admin account wrote the issue and the assigned positions for each username. An example of this is in figure \cref{fig:rocketchat-debating}. 

For writing modality, the users are not added to a chat room (however, it is created underneath, as judges will see the parties' statements later). Instead, users received a private message from the administrator account with the details of the assigned issue and the position to write about. An illustration of what they would see is in \cref{fig:rocketchat-writing}.

For judging, after the main intervention is complete. Participants, now in the role of judges, are granted viewing permissions to other debate channels. They would read either the debate (if the intervention was a debate) or the statements by the opposing parties (if the intervention was the writing modality).

\section{Methods Details}

\subsection{Measures of Issue-based Animosity and Stance}
\label{app:methods:measures}

We measured our main outcomes---\textbf{issue-based animosity} and \textbf{stance}---at \textit{pre-intervention}, \textit{post-intervention}, and \textit{follow-up} using Qualtrics surveys. Although our scientific target is a change from pre to a later wave, we do not compute individual change scores and then analyze those. Instead, all estimates of change are obtained as model-implied contrasts of estimated marginal means (EMMs) from mixed-effects models fit to the observed outcomes at each wave (see \secref{sec:methods:analysis}). This approach avoids magnifying measurement error and regression-to-the-mean that can arise when subtracting two noisy scores, and it accommodates the correlation structure of repeated measures \citep[e.g.,][]{vickersAnalysingControlledTrials2001,fitzmauriceAppliedLongitudinalAnalysis2012,lenthEmmeansEstimatedMarginal2025}. Reported changes $\Delta_{t}$ therefore denote EMM differences between time $t\in\{\pos,\fol\}$ and \textit{pre} for the relevant arm and outcome, not raw person-level differences.

\paragraph{Issue-based animosity ($Y=\anim$)}
At each wave $s\in\{\pre,\pos,\fol\}$ participants rated their feelings on a thermometer \citep{wilcoxItHotIndividual1989, nelsonFeelingThermometer2008} where participants rated their warmth (0 = ``Very cold/unfavorable'' to 100 = ``Very warm/favorable'') towards two referent groups: people who agree with the policy position and people who disagree. To be conservative and to guard against post-wave ``neutralization'' or shifts in the participant’s reference group, we define the observed affect at wave $s$ as the minimum of the two thermometers,
\[
A_i(s) \;=\; \min\!\big\{T_{i,\text{agree}}(s),\; T_{i,\text{disagree}}(s)\big\},
\qquad A_i(s)\in[0,100].
\]
Higher values indicate warmer feelings toward the least-liked group and, therefore, \textbf{lower issue-based animosity}. The models are fit to $A_i(s)$ on the 0--100 scale; changes are reported as $\Delta_{t}$ EMM contrasts rather than computed for each person. Because our definition yields positive $\Delta$ when animosity decreases, we interpret positive $\Delta$ as improvement.

\paragraph{Stance ($Y=\stance$)}
Agreement with the focal issue statement was collected on an ordered 5-point Likert scale, ranging from 1 (``Strongly Disagree'') to 5 (``Strongly Agree''), and mapped to a numeric score $\tilde S_i(s)\in[-2,2]$, with 0 indicating neutrality. The mapping preserves order with equal spacing across categories. To make increases interpretable as movement toward the assigned opposite stance (and toward moderation when applicable), we reorient each participant’s scale using their baseline sign. Let
\[
s_{i,\pre}=\mathrm{sign}\!\big(\tilde S_i(\pre)\big)\in\{-1,+1\},
\qquad
S_i^{\uparrow}(s)= -\,s_{i,\pre}\,\tilde S_i(s)\in[-2,2].
\]
With this coding, higher values always indicate movement toward the opposite side. Models are fit to $S_i^{\uparrow}(s)$ at pre, post, and follow, and reported changes $\Delta_{t}$ are EMM contrasts from pre-intervention to time $t$ rather than raw individual differences. Positive $\Delta$ indicates movement toward the opposite stance or moderation.

\subsection{Analysis and Models}
\label{app:methods:analysis}

We detail now the statistical approach for how outcomes are constructed and how changes are estimated. Our most important measures are not measures at each timepoint, but measures of the change from pre-intervention to post-intervention or follow-up. We do not analyze raw person-level change scores directly. Instead, we fit models to the observed outcomes at each wave and report model-implied changes as estimated marginal means (EMM) contrasts. This reduces error propagation from subtracting two noisy measurements and respects the correlation structure of repeated measures \citep[e.g.,][]{vickersAnalysingControlledTrials2001, fitzmauriceAppliedLongitudinalAnalysis2012, lenthEmmeansEstimatedMarginal2025}. For any outcome $Y$ and arm $a$, we write
\[
\Delta_{t} \;=\; \EMM\!\big(Y \mid a,t\big)\;-\;\EMM\!\big(Y \mid a,\pre\big),
\qquad t\in\{\pos,\fol\},
\]
on the original response scale of $Y$. Perspective and subgroup comparisons use the notation from \secref{sec:results:measures}.

\subsubsection{Survey response processing}
\label{app:methods:survey-processing}

\paragraph{Issue-based animosity ($Y=\anim$)}
At each wave $s\in\{\pre,\pos,\fol\}$, participants rated their feelings on two thermometers (0 to 100): toward people who agree with the policy statement and toward people who disagree. To be conservative and to guard against post-wave neutralization or shifts in reference group, the observed affect at wave $s$ is the minimum of the two thermometers,
\[
A_i(s) \;=\; \min\!\big\{T_{i,\text{agree}}(s),\; T_{i,\text{disagree}}(s)\big\}\in[0,100].
\]
Higher values indicate warmer feelings toward the least liked group and therefore lower issue-based animosity. Models are fit to $A_i(s)$, and changes are reported as $\Delta_{t}$ EMM contrasts, not raw differences.

\paragraph{Stance ($Y=\stance$)}
Agreement with the focal statement was collected on a 5-point Likert scale and mapped to a numeric score $\tilde S_i(s)\in[-2,2]$ with 0 indicating neutrality. To make increases interpretable as movement toward the assigned opposite stance and toward moderation when applicable, we reorient each participant’s scale using the baseline sign:
\[
s_{i,\pre}=\mathrm{sign}\!\big(\tilde S_i(\pre)\big)\in\{-1,+1\},
\qquad
S_i^{\uparrow}(s)= -\,s_{i,\pre}\,\tilde S_i(s)\in[-2,2].
\]
With this coding, higher values always indicate movement away from the baseline stance and toward the opposite side. Models are fit to $S_i^{\uparrow}(s)$, and reported changes are $\Delta_{t}$ EMM contrasts from pre.

\paragraph{Performance indicators for moderation analyses}
Two binary indicators capture judged performance: Best Argument and Authenticity. Each equals 1 when at least two of three judges selected the participant for that mechanism. We also consider both when both indicators equal 1. Rates by mechanism and arm are reported in the Appendix.

\paragraph{Covariates and coding}
Demographic covariates are effects-coded. Debate topic indicators are included as fixed effects. Random effects structure and additional specifications are presented with each hypothesis below.

\paragraph{Exclusion criteria}
We excluded participants (1.8\%) who failed consistency checks, for example reporting colder feelings toward those who share their own position than toward the opposition at pre. One session ($N{=}12$) was removed due to technical failures that impeded chat functionality and task comprehension.

\subsubsection{Analysis for Hypothesis 1}
\label{app:methods:analysis:h1}

We estimate, for each outcome $Y$ defined in \cref{app:methods:measures}, the change from \textit{pre} to \textit{post} or \textit{follow} within each arm. For arm $a\in\{\arm{\Wrt}{\Own},\arm{\Wrt}{\Opp},\arm{\Dbt}{\Own},\arm{\Dbt}{\Opp}\}$ and time $t\in\{\pos,\fol\}$, we report
\[
\Delta_{t}(a) \;=\; \EMM\!\big(Y \mid a,t\big)\;-\;\EMM\!\big(Y \mid a,\pre\big),
\]
that is, estimated marginal mean contrasts on the response scale. This model-based approach provides covariate adjustment and accounts for repeated measures, and it avoids error amplification from subtracting two noisy scores \citep[e.g.,][]{vickersAnalysingControlledTrials2001,fitzmauriceAppliedLongitudinalAnalysis2012,lenthEmmeansEstimatedMarginal2025}.

\textbf{Model Specification.}
We fit, separately for each outcome $Y$, a mixed model with an arm by time factorial, covariates, and random intercepts:
\[
Y_{i,a}(s) \;=\; \mu \;+\; \alpha_a \;+\; \tau_s \;+\; (\alpha\tau)_{a,s}
\;+\; \mathbf{x}_i^\top\beta \;+\; u_i \;+\; b_{\text{block}(i)} \;+\; \varepsilon_{ias},
\quad s\in\{\pre,\pos,\fol\}.
\]
Covariates $\mathbf{x}_i$ include topic, gender, ethnicity, political viewpoint, and strong\_opinion. Random intercepts are specified for participants ($u_i$) and debate blocks ($b_{\text{block}}$). Outcomes follow the constructions in \secref{sec:methods:measures}: for animosity, $Y=A_i(s)\in[0,100]$ is the within-wave minimum thermometer; for stance, $Y=S_i^{\uparrow}(s)\in[-2,2]$ with
\[
s_{i,\pre}=\mathrm{sign}\!\big(\tilde S_i(\pre)\big)\in\{-1,+1\},
\qquad
S_i^{\uparrow}(s)= -\,s_{i,\pre}\,\tilde S_i(s),
\]
so that higher values indicate movement toward the opposite stance or toward moderation when applicable.

\textbf{Estimation of contrasts.}
From the fitted model we obtain $\EMM(Y\mid a,s)$ and form $\Delta_{\pos}(a)$ and $\Delta_{\fol}(a)$ for each arm. When relevant we also summarize arm-to-arm differences in change,
\[
\diffarms{Y}{t}{a}{a'} \;=\; \Delta_{t}(a) - \Delta_{t}(a'),
\]
with Holm adjustment within each outcome and wave. Confidence intervals and $p$-values are model-implied on the response scale.

\textbf{Sensitivity and robustness.}
Sensitivity to design and missingness is examined by refitting the primary model with inverse-probability weights for modality scheduling ($w_{\text{mod}}$) and for follow-up attrition ($w_{\text{atr}}$), and then recomputing $\Delta_t(a)$.

Robustness analyses appear in the Appendix: (I) Sensitivity analysis of the effect not being driven by attrition or imbalances in modality; (ii) change-score models using $Y(\pos)-Y(\pre)$ and $Y(\fol)-Y(\pre)$ with the same right-hand side; (iii) probability of improvement, where $I_i(t)=\mathbb{I}\{Y_{i,a}(t)>Y_{i,a}(\pre)\}$ is modeled with a logistic mixed model including arm by time; (iv) ordinal models for outcomes treated as ordered categories, recovering marginal effects comparable to $\Delta_t(a)$; (iv) global diagnostics for the arm by time interaction, including an omnibus likelihood-ratio test and the added marginal $R^2$ attributable to the interaction to contextualize variance explained.

\subsubsection{Analysis for Hypothesis 2}
\label{app:methods:analysis:h2}

Our aim is to estimate the perspective advantage within each activity modality, as defined in \secref{sec:results:measures}. For modality $m\in\{\Wrt,\Dbt\}$ and time $t\in\{\pos,\fol\}$, the estimand is the difference in within-arm changes between \Opp\ and \Own\:
\[
\delta_{m,t}
\;=\;
\did{Y}{t}{\text{persp} : \Opp, \Own \mid m}
\;=\;
\Delta_{t}\!\big(\arm{m}{\Opp}\big) - \Delta_{t}\!\big(\arm{m}{\Own}\big),
\]
where each $\Delta_{t}(\cdot)$ is an EMM contrast on the response scale.

\textbf{Model Specification.}
For each outcome $Y\in\{\anim,\stance\}$ we fit a repeated-measures mixed model with a three-way factorial of perspective ($p\in\{\Own,\Opp\}$), modality ($m\in\{\Wrt,\Dbt\}$), and time ($s\in\{\pre,\pos,\fol\}$), along with covariates and random intercepts:
\[
\begin{aligned}
    Y_{i}(s)
    &= \mu + \alpha_{p} + \beta_{m} + \tau_{s}
    + (\alpha\beta)_{p,m} + (\alpha\tau)_{p,s} + (\beta\tau)_{m,s} + (\alpha\beta\tau)_{p,m,s}\\
    &\quad+ \mathbf{x}_i^\top \gamma + u_i + b_{\text{block}(i)} + \varepsilon_{is}.
\end{aligned}
\]
Covariates $\mathbf{x}_i$ include topic, gender, ethnicity, political viewpoint, and strong opinion. Random intercepts are included for participants and debate blocks. Outcomes follow \secref{sec:methods:measures}: $Y=A_i(s)\in[0,100]$ for animosity; $Y=S_i^{\uparrow}(s)\in[-2,2]$ for stance with $S_i^{\uparrow}(s)=-\,s_{i,\pre}\,\tilde S_i(s)$ so that increases mean movement toward the opposite stance.

\textbf{Estimation of contrasts.}
From the fitted model, obtain $\EMM(Y\mid p,m,s)$ for each $(p,m,s)$. Within each $(p,m)$, form the time contrasts to \pre,
\begin{align*}
\Delta_{\pos}(p,m) &= \EMM(Y\mid p,m,\pos)-\EMM(Y\mid p,m,\pre),\\
\Delta_{\fol}(p,m) &= \EMM(Y\mid p,m,\fol)-\EMM(Y\mid p,m,\pre),
\end{align*}
then compute the perspective DiD within modality,
\[
\delta_{m,t}=\Delta_{t}(\Opp,m)-\Delta_{t}(\Own,m), \qquad t\in\{\pos,\fol\}.
\]
We report $\delta_{m,t}$ with model-implied confidence intervals and Holm adjustment within each outcome and time. For interpretation, we also display the marginal changes $\Delta_{t}(\Own,m)$ and $\Delta_{t}(\Opp,m)$ that compose each $\delta_{m,t}$.

\textbf{Sensitivity and robustness.}
To examine sensitivity to design and missingness, we repeat the primary model with inverse-probability weights: $w_{\text{mod}}$ for scheduling by modality and $w_{\text{atr}}$ for follow-up attrition, and then recompute $\delta_{m,t}$. As a simpler robustness alternative, we also fit change-score versions that model $Y(\pos)-Y(\pre)$ and $Y(\fol)-Y(\pre)$ within the same perspective by modality framework; these yield estimates on the same contrast scale and are reported alongside the EMM-based results in the Appendix.

\subsubsection{Analysis for Hypothesis 3}
\label{app:methods:analysis:h3}

The aim is to test whether judged performance moderates change in the outcomes defined in \secref{sec:methods:measures}. We focus on three win types coded at the participant level using judge decisions at the activity: Best Argument, Authenticity, and Both. A participant is a winner for a mechanism when at least two of three judges selected them; `Lose' indicates neither mechanism. Estimation proceeds on the response scale via estimated marginal means (EMMs) from mixed-effects models, following best practice for longitudinal contrasts \citep{fitzmauriceAppliedLongitudinalAnalysis2012} and for contrast recovery with EMMs \citep{lenthEmmeansEstimatedMarginal2025}.

\textbf{Model Specification.}
For each outcome $Y\in\{\anim,\stance\}$ we fit a repeated-measures mixed model to observations at $s\in\{\pre,\pos,\fol\}$ with time, arm, and mechanism indicators:

\[
\begin{aligned}
Y_i(s) &= \mu + \tau_s + \alpha_a + \alpha_{a,s} \\
&\quad + (\beta_{\text{BA}}+\beta_{\text{BA},s})\,\text{BestArg}_i
       + (\beta_{\text{AU}}+\beta_{\text{AU},s})\,\text{Authentic}_i \\
&\quad + \mathbf{x}_i^\top\gamma + u_i + b_{\text{block}(i)} + \varepsilon_{is}.
\end{aligned}
\]

Covariates $\mathbf{x}_i$ include topic, gender, ethnicity, political viewpoint, and whether the participant expresses a strong opinion on the issue (agreement is marked as strongly agree or strongly disagree). Random intercepts are included for participants and debate blocks. The outcome is $Y=A_i(s)\in[0,100]$ for animosity and $Y=S_i^{\uparrow}(s)\in[-2,2]$ for stance, where $S_i^{\uparrow}(s)=-\,s_{i,\pre}\,\tilde S_i(s)$ so that increases always indicate movement toward the opposite stance. Models are fit with \texttt{lme4} \citep{batesFittingLinearMixedeffects2015} and EMMs are computed with \texttt{emmeans} \citep{lenthEmmeansEstimatedMarginal2025}.

\textbf{Estimands and pooling.}
Let $\Delta_{t}(\text{type},a)$ denote the EMM change from \pre\ to time $t\in\{\pos,\fol\}$ for win type $\text{type}\in\{\text{BestArg},\text{Authentic},\text{Both},\text{Lose}\}$ within arm $a$. The mechanism contrast within arm is
\[
\delta_{t}(\text{win type} \text{ vs. Lose} \mid a)
= \Delta_{t}(\text{win type},a) - \Delta_{t}(\text{Lose},a).
\]
Because the main results summarize the moderation \emph{pooled across arms}, we average these arm-specific contrasts using prespecified weights $v_a$,

\[
\bar{\delta}_{t}^{(\text{type})}
=\sum_{a} v_a \,\delta_{t}(\text{type} \text{ vs Lose} \mid a),
\]

and report $\bar{\delta}_{t}^{(\text{type})}$ with model-implied confidence intervals. Familywise control uses Holm’s method within each outcome and time \citep{holmSimpleSequentiallyRejective1979}. For transparency, the arm-specific decompositions $\delta_{t}(\cdot\mid a)$ are presented in the Appendix.

\textbf{Model selection note.}
We evaluated a fully interacted alternative that allowed mechanism effects to vary by arm and time. Likelihood-ratio tests and information criteria did not support adding mechanism by arm terms, while retaining arm by time as adjustment aligned the specification with H1 and H2. This parsimony reduces variance without changing qualitative conclusions; fit diagnostics and comparisons are documented in the Appendix.

\textbf{Sensitivity and robustness.}
(i) Design and missingness: we repeat the primary model with inverse-probability weights for modality scheduling and for follow-up attrition, then recompute $\bar{\delta}_{t}^{(\text{type})}$.  
(ii) Alternative coding of winning: we analyze continuous judge scores for Best Argument and Authenticity on the $[0,1]$ scale, and a stricter unanimity threshold.  
(iii) Simpler change-score analysis: as a robustness check we model $Y(\pos)-Y(\pre)$ and $Y(\fol)-Y(\pre)$ directly with the same fixed and random effects, recognizing that change scores can propagate measurement error compared to EMM contrasts \citep{vickersAnalysingControlledTrials2001,fitzmauriceAppliedLongitudinalAnalysis2012}. Results are reported alongside the primary specification and are qualitatively consistent.

\subsubsection{Analysis for Hypothesis 4}
\label{app:methods:analysis:h4}

\textbf{Estimands.}
We evaluate willingness to repeat the activity without compensation on the probability scale. Let $W_i\in\{0,1\}$ indicate ``Probably/Definitely yes'' at the post survey. From a covariate-adjusted model we recover marginal (response-scale) arm probabilities $\Pr(W{=}1\mid m,p)$ for modality $m\in\{\Wrt,\Dbt\}$ and perspective $p\in\{\Own,\Opp\}$. The primary estimands are pooled \emph{risk differences}:

\begin{align*}
\Delta^{\text{mod}}
&= \mathbb{E}_{p}\!\left[\Pr(W{=}1 \mid \Dbt,p)-\Pr(W{=}1 \mid \Wrt,p)\right],\\
\Delta^{\text{persp}}
&= \mathbb{E}_{m}\!\left[\Pr(W{=}1 \mid m,\Opp)-\Pr(W{=}1 \mid m,\Own)\right].
\end{align*}

where pooling uses proportional weights (cell sizes). Non-inferiority (NI) is assessed against a margin $\delta=0.05$ on the risk-difference scale: $H_0:\Delta\le -\delta$ vs.\ $H_1:\Delta>- \delta$; NI is declared when the lower 95\% Wald bound for $\Delta$ exceeds $-\delta$ \citep{piaggioReportingNoninferiorityEquivalence2012,walkerUnderstandingEquivalenceNoninferiority2011}.

\textbf{Primary model and marginalization.}
We fit a logistic mixed model with modality $\times$ perspective and baseline covariates,

\begin{align*}
\operatorname{logit}\Pr(W_i{=}1)
&= \mu + \beta_m m_i + \beta_p p_i + \beta_{mp}(m_i\!\times\!p_i) \\
&\quad + \mathbf{x}_i^\top\gamma + b_{\text{pair}(i)}, \\
b_{\text{pair}} &\sim \mathcal N(0,\sigma_b^2).
\end{align*}

where $\mathbf{x}_i$ includes topic, demographics, political viewpoint, and opinion strength; $b_{\text{pair}}$ is a random intercept for pairing/debate block. We obtain covariate-adjusted arm probabilities and contrasts using estimated marginal means (EMMs) with marginal standardization on the response scale \citep{lenthEmmeansEstimatedMarginal2025}; this yields $\Pr(W{=}1\mid m,p)$, the stratified risk differences within perspective or modality, and their pooled averages $\Delta^{\text{mod}}$ and $\Delta^{\text{persp}}$.

\textbf{Ordinal robustness.}
Because the willingness item is ordinal, we fit a cumulative logit mixed model (CLMM) on a three-level collapse (No / Indifferent / Yes) with the same fixed and random effects \citep{christensenOrdinalRegressionModels2023}. Category probabilities are recovered via EMMs and collapsed to $P(\text{Yes})$ to recompute the same pooled risk-difference estimands and NI tests on the probability scale.

\textbf{Missing-item sensitivity.}
To address potential bias from item nonresponse, we estimate stabilized inverse-probability weights for answering the willingness item using a logistic model with modality, perspective, topic, demographics, and political viewpoint as predictors; weights are truncated at extreme tails and used in the primary GLMM, after which EMMs and pooled risk differences are recomputed \citep{seamanReviewInverseProbability2013, leeWeightTrimmingPropensity2011}.

\section{Analysis Results}
\label{app:analysis}

\subsection{Balance Analysis}\label{app:analysis:balance}

Three features of the design may bias estimates if unaddressed. First, the follow-up response rate was about $60\%$, raising concerns of \emph{differential attrition}. Second, the two modalities (Debate vs.\ Write) were not run in a fully simultaneous $2{\times}2$ schedule, motivating a modality balance check. Third, the participants were recruited from two different pool sources. For each, we report standardized mean differences (SMDs), overlap diagnostics of the propensity (``distance''), and quality of the weights.

\paragraph{Setup}
For attrition we define $R_i{=}1$ for follow-up responders and $0$ otherwise and compute SMDs for topic, gender, ethnicity, political viewpoint, session, arm, and the scaled baselines of issue-based animosity and stance. We then estimate the ATE IPW via CBPS using the same predictors (capping the top 1\% of weights). For modality we estimate ATE weights for Debate vs.\ Write using topic, strong opinion, gender, ethnicity, and political viewpoint. Throughout, we target $|\text{SMD}|\le .10$--$.20$ after weighting.

\subsubsection*{Attrition (follow-up)}

\begin{figure}[h]
  \centering
  \includegraphics[width=\textwidth]{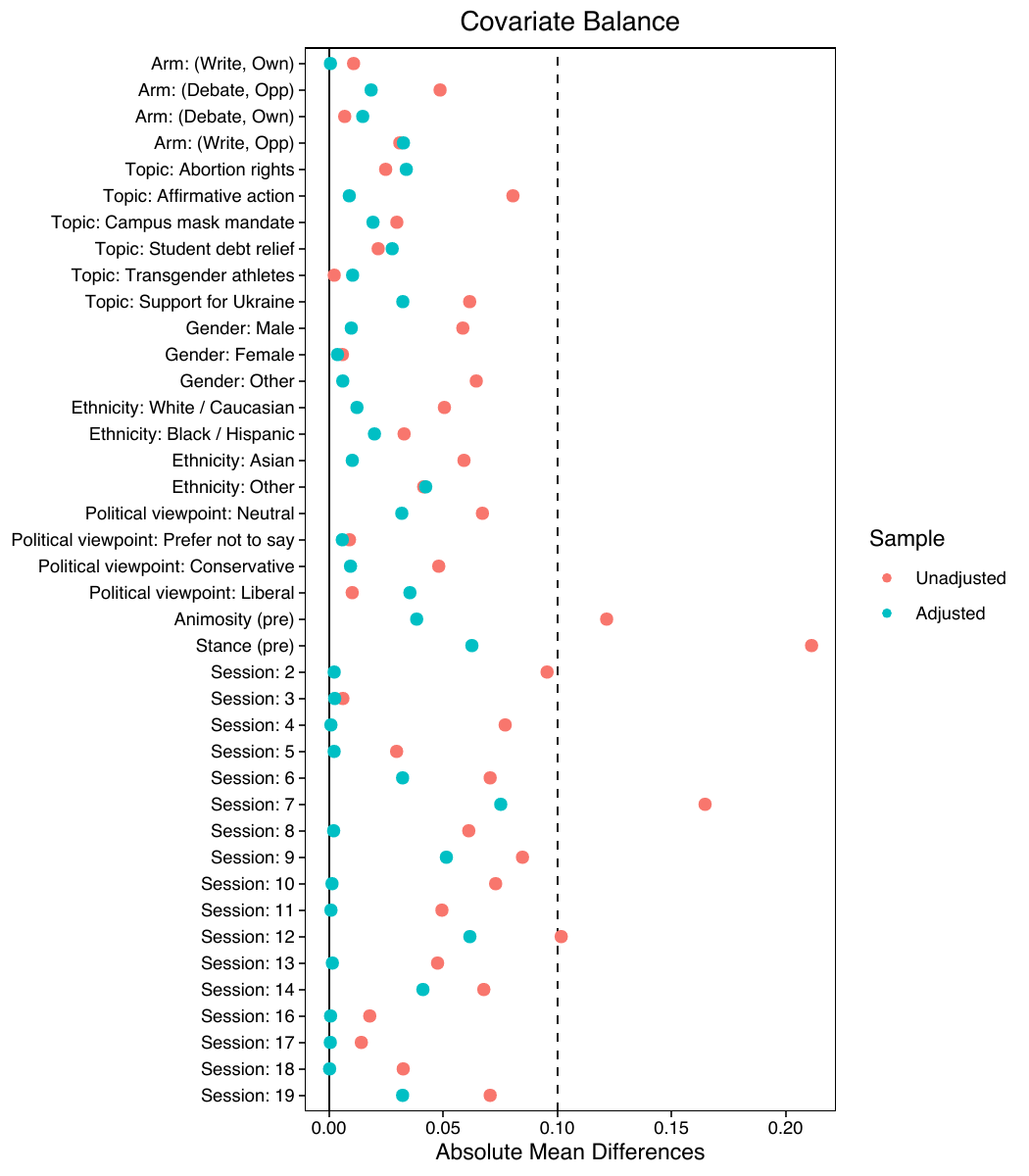}
  \caption{\textbf{Attrition balance: absolute SMDs before/after weighting (love plot; threshold lines at .10).}}
  \label{fig:attrition-love}
\end{figure}

\begin{figure}[h]
  \centering
  \includegraphics[width=.49\textwidth]{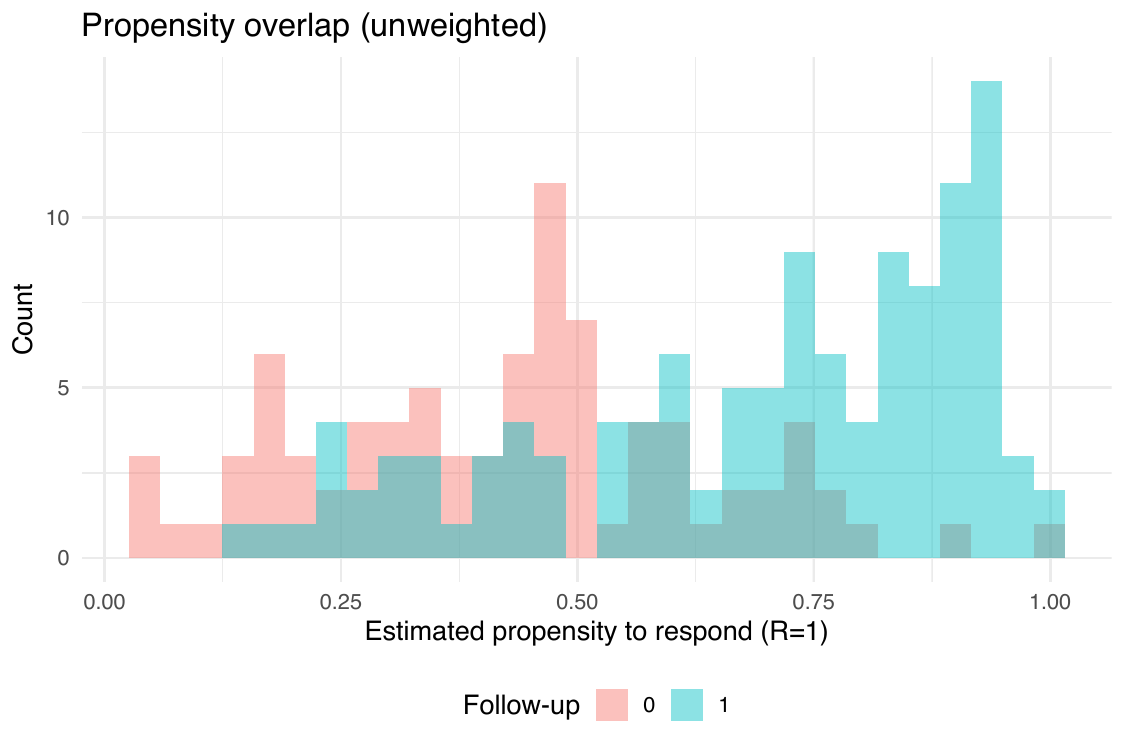}\hfill
  \includegraphics[width=.49\textwidth]{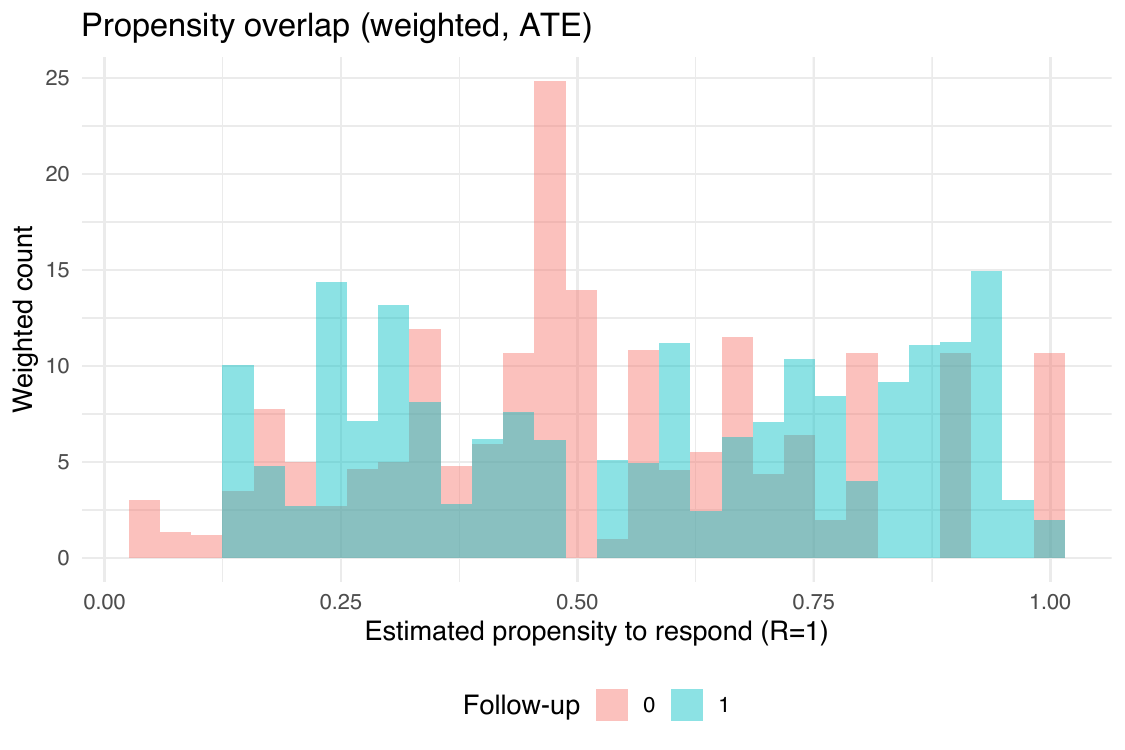}
  \caption{\textbf{Propensity (``distance'') overlap for attrition.} Left: unweighted; Right: ATE-weighted among responders.}
  \label{fig:attrition-ps}
\end{figure}

\begin{figure}[t]
  \centering
  \includegraphics[width=\textwidth]{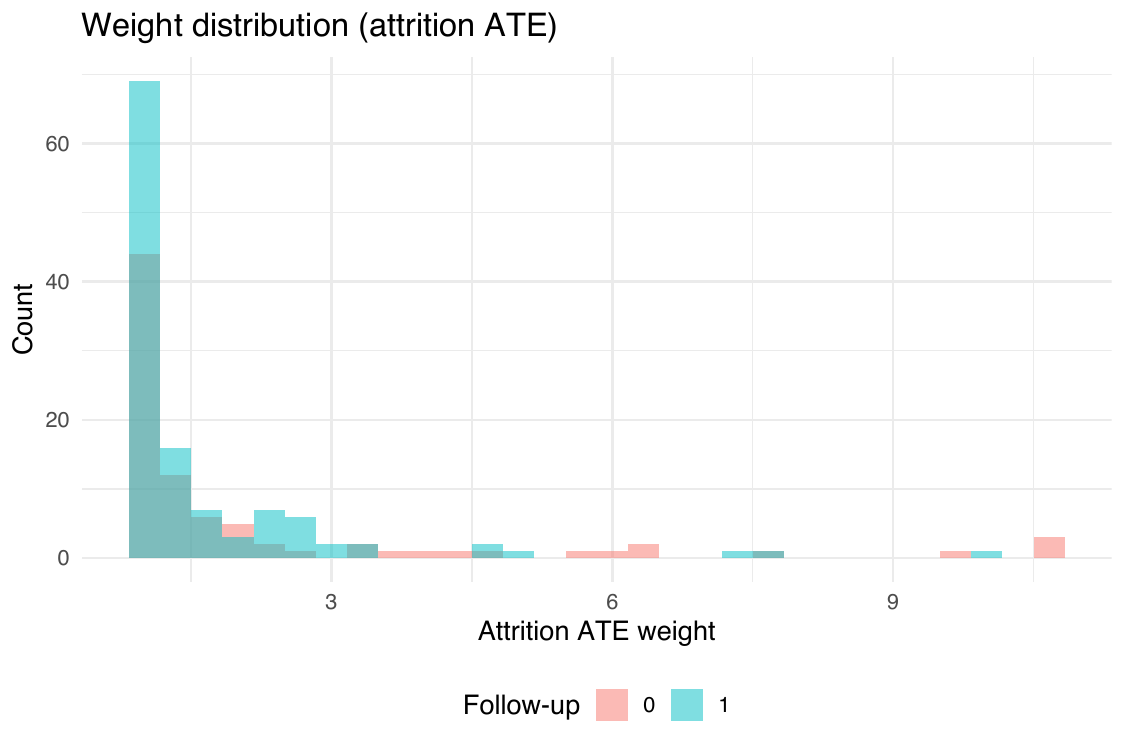}
  \caption{\textbf{ATE weight distribution for attrition, by follow-up group $R\in\{0,1\}$.}}
  \label{fig:attrition-weights}
\end{figure}

Table~\ref{tab:attrition-balance-summary} summarizes balance succinctly (max/median absolute SMD and counts above .10/.20). Figure~\ref{fig:attrition-love} shows the love plot after weighting. To assess positivity/overlap, Figure~\ref{fig:attrition-ps} plots the empirical propensity (``distance'') distributions (unweighted vs.\ weighted). Figure~\ref{fig:attrition-weights} shows the ATE weight distribution by follow-up group.

\begin{table}[h]
  \centering
  \caption{\textbf{Attrition balance summary (responders vs.\ non-responders).}}
  \label{tab:attrition-balance-summary}
  \small

\begin{tabular}[t]{ccccc}
\toprule
\makecell[c]{Max $|$SMD$|$\\(unweighted)} & \makecell[c]{Max $|$SMD$|$\\(weighted)} & \makecell[c]{Median $|$SMD$|$\\(weighted)} & \makecell[c]{Covariates with\\$|$SMD$|$ $>$ 0.10} & \makecell[c]{Covariates with\\$|$SMD$|$ $>$ 0.20}\\
\midrule
0.211 & 0.075 & 0.011 & 0 & 0\\
\bottomrule
\end{tabular}

\end{table}

\noindent
Notes: In our data, the largest unweighted difference occurs on baseline stance (about $0.21$ in absolute SMD); other covariates are smaller. Several topic\,$\times$\,arm cells among responders are sparse (e.g., $n{<}3$), so we present topic counts descriptively and avoid saturated topic fixed effects in small cells.

\subsubsection*{Modality (Debate vs.\ Write)}

\begin{figure}[h]
  \centering
  \includegraphics[width=\textwidth]{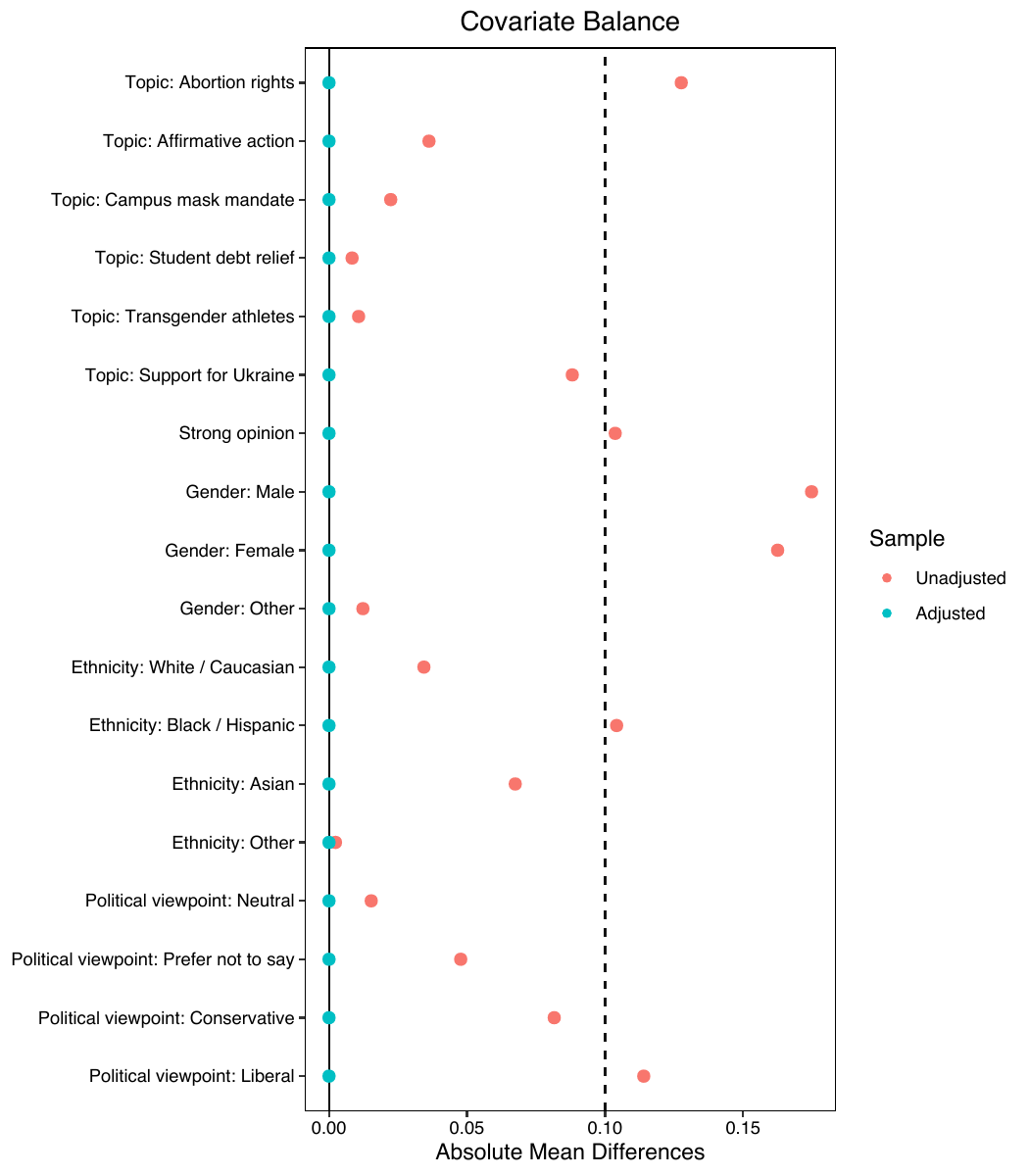}
  \caption{\textbf{Modality balance: absolute SMDs before/after weighting (love plot; threshold at .10).}}
  \label{fig:modality-love}
\end{figure}

\begin{figure}[h]
  \centering
  \includegraphics[width=.49\textwidth]{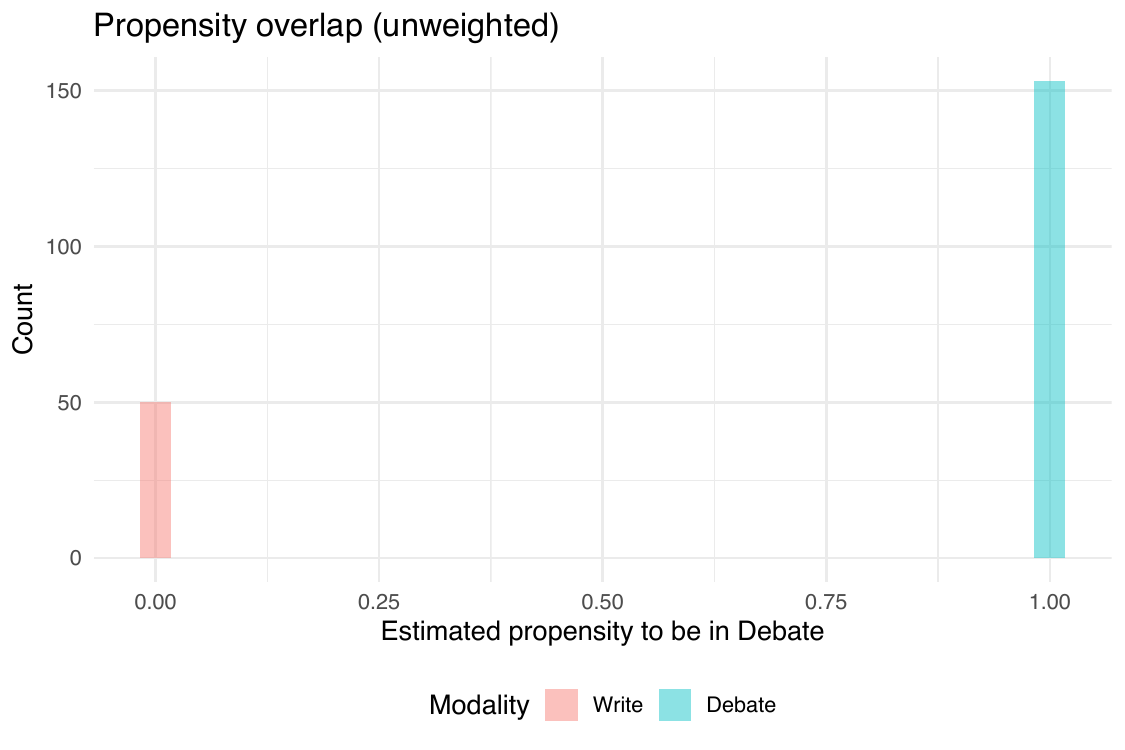}\hfill
  \includegraphics[width=.49\textwidth]{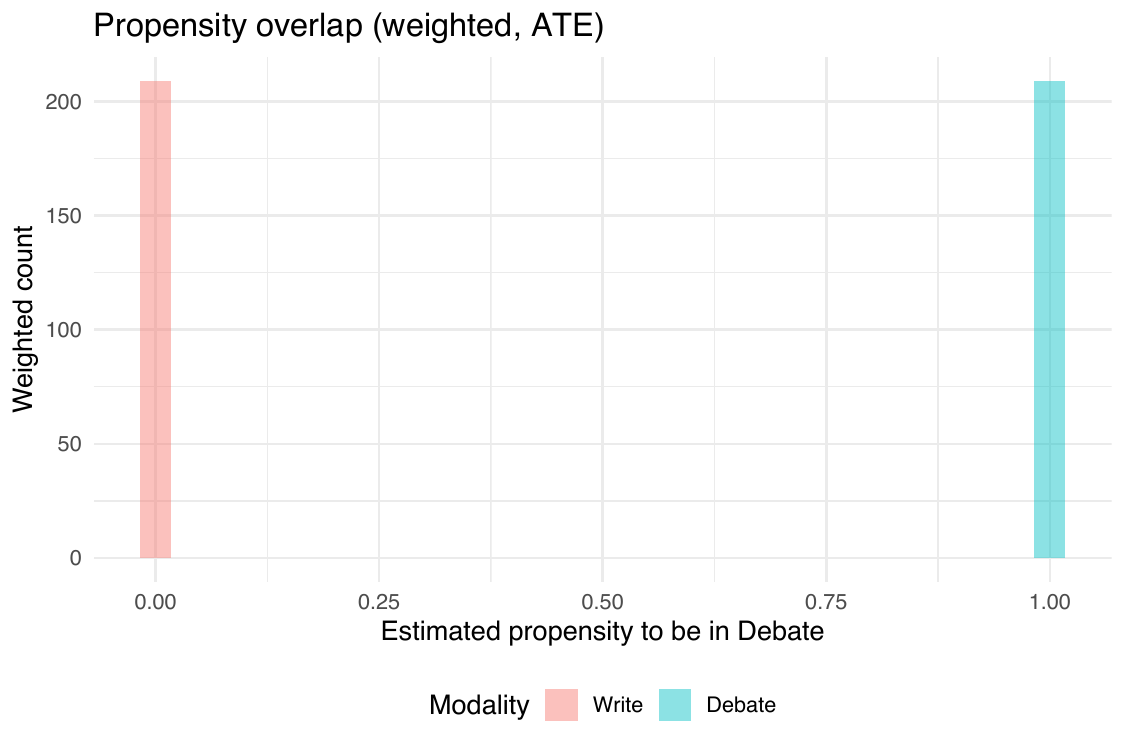}
  \caption{\textbf{Propensity (``distance'') overlap for modality.} Left: unweighted; Right: ATE-weighted.}
  \label{fig:modality-ps}
\end{figure}

\begin{figure}[h]
  \centering
  \includegraphics[width=\textwidth]{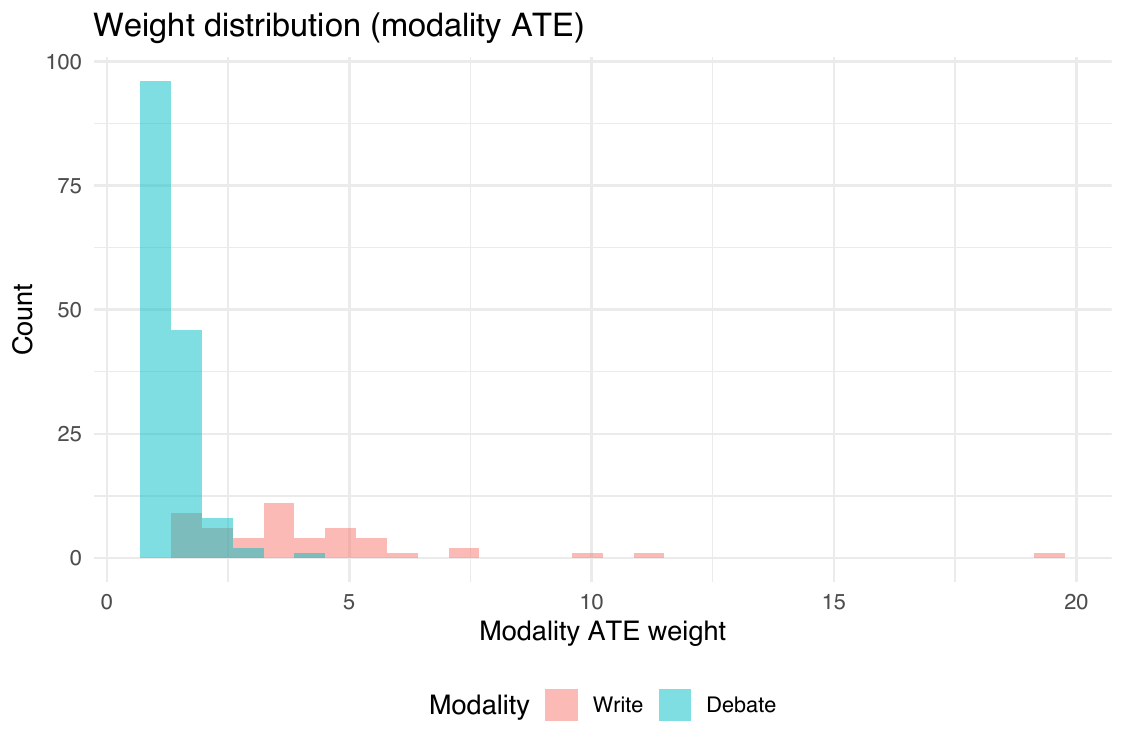}
  \caption{\textbf{ATE weight distribution for modality (Debate vs.\ Write).}}
  \label{fig:modality-weights}
\end{figure}

Table~\ref{tab:modality-balance-summary} provides the one-row summary of SMDs. Figure~\ref{fig:modality-love} shows the love plot. Figure~\ref{fig:modality-ps} reports propensity overlap for Debate vs.\ Write, and Figure~\ref{fig:modality-weights} shows the ATE weight distribution by modality. Residual imbalances are modest (largest on gender and on liberal political viewpoint), and post-weighting SMDs are within the .10--.20 range used as a heuristic threshold.

\begin{table}[h]
  \centering
  \caption{\textbf{Modality balance summary (Debate vs.\ Write).}}
  \label{tab:modality-balance-summary}
  \small

\begin{tabular}[t]{ccccc}
\toprule
\makecell[c]{Max $|$SMD$|$\\(unweighted)} & \makecell[c]{Max $|$SMD$|$\\(weighted)} & \makecell[c]{Median $|$SMD$|$\\(weighted)} & \makecell[c]{Covariates with\\$|$SMD$|$ $>$ 0.10} & \makecell[c]{Covariates with\\$|$SMD$|$ $>$ 0.20}\\
\midrule
0.175 & 0 & 0 & 0 & 0\\
\bottomrule
\end{tabular}

\end{table}

\paragraph{Use in outcome models}
Primary follow-up analyses are estimated on $R{=}1$ with attrition ATE weights; modality ATE weights are included when comparing arms across modalities. Random intercepts for debate account for session-level heterogeneity; topic sparsity is handled by weighting and pooled adjustments rather than saturated topic fixed effects where cells are singular.

\subsubsection{Pool (Class vs. Lab) and Wins(Compensation)}
\label{app:pool-balance}

We conducted comprehensive balance analyses to assess potential differences between participants recruited from class (n=129) and lab (n=74) pools. Table~\ref{tab:pool_demographic_balance} presents demographic characteristics and baseline measures across both pools.

Statistical tests revealed some differences in topic distribution and gender composition between pools. The class pool had higher representation in affirmative action topics, while the lab pool showed greater representation in relief-plan topics ($p < 0.05$). Gender distribution also differed significantly ($p < 0.05$), with the lab pool containing more women and individuals identifying as ``Other'' gender. However, no significant differences were observed in ethnic composition, political viewpoint, or baseline measures of \textbf{issue-based animosity} and \textbf{stance} ($p > 0.05$).

\begin{figure}[h!]
\centering
\includegraphics[width=0.7\textwidth]{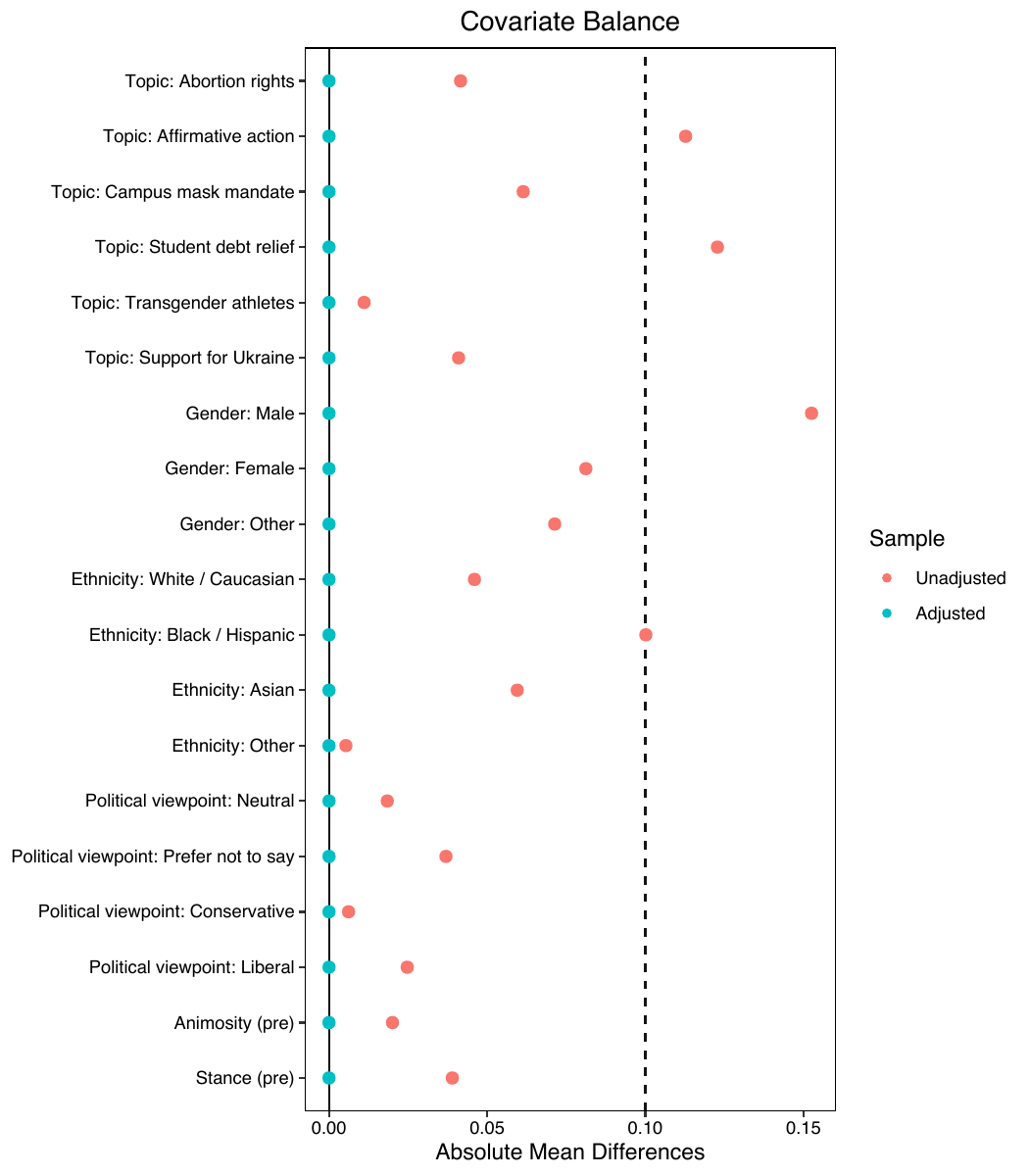}
\caption{\textbf{Standardized Mean Differences Before and After CBPS Weighting.}}
\label{fig:pool-love-plot}
\end{figure}

To address these imbalances, we implemented covariate balancing propensity score (CBPS) weighting targeting the average treatment effect. The weights successfully reduced standardized mean differences, with the maximum absolute difference decreasing from 0.15 (gender) to below 0.10 for all covariates after weighting (Figure~\ref{fig:pool-love-plot}). The weight distribution showed reasonable overlap between pools (range: 1.01-9.67), with effective sample sizes of 111.09 for class and 55.14 for lab participants after weighting.

\begin{figure}[h!]
\centering
\includegraphics[width=0.8\textwidth]{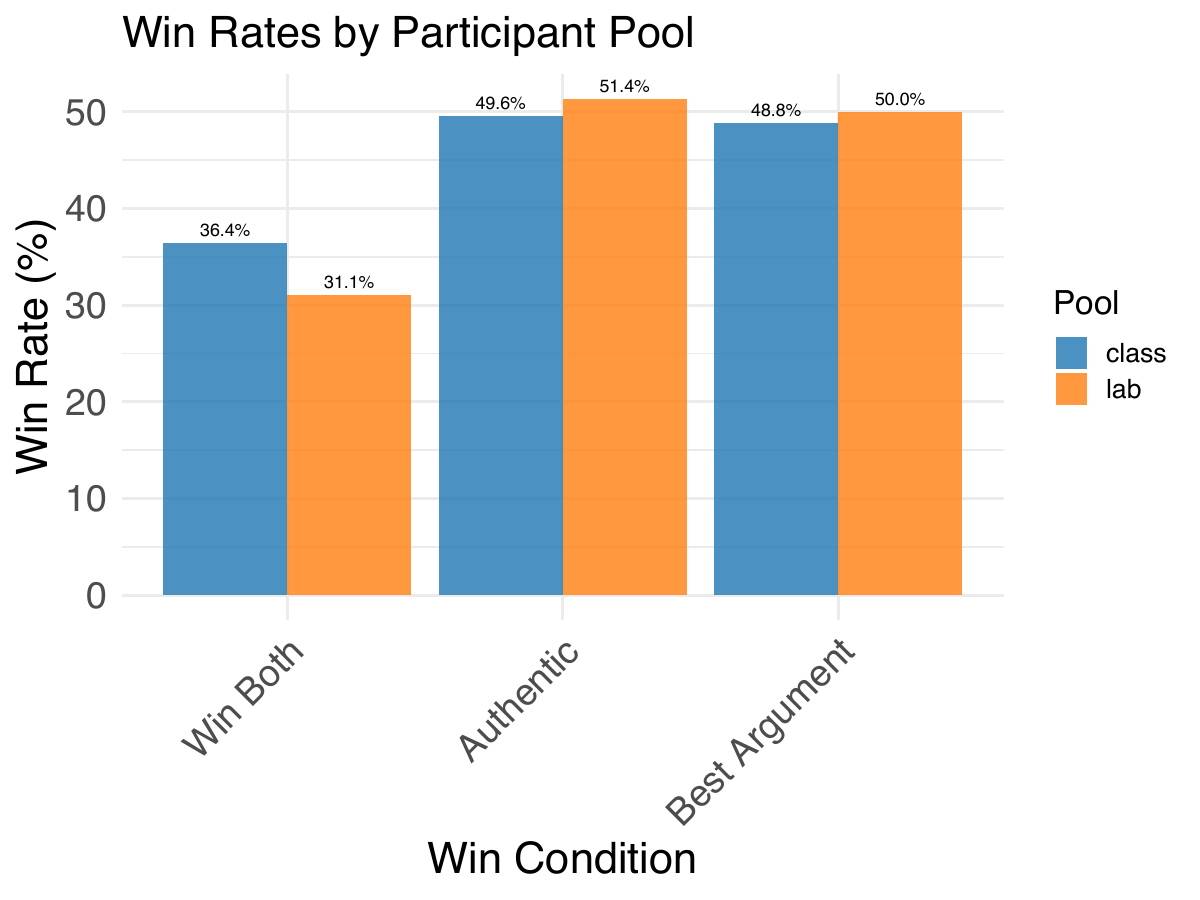}
\caption{\textbf{Win Rates by Participant Pool.}}
\label{fig:pool-win-rates}
\end{figure}

\begin{figure}[h!]
\centering
\includegraphics[width=0.9\textwidth]{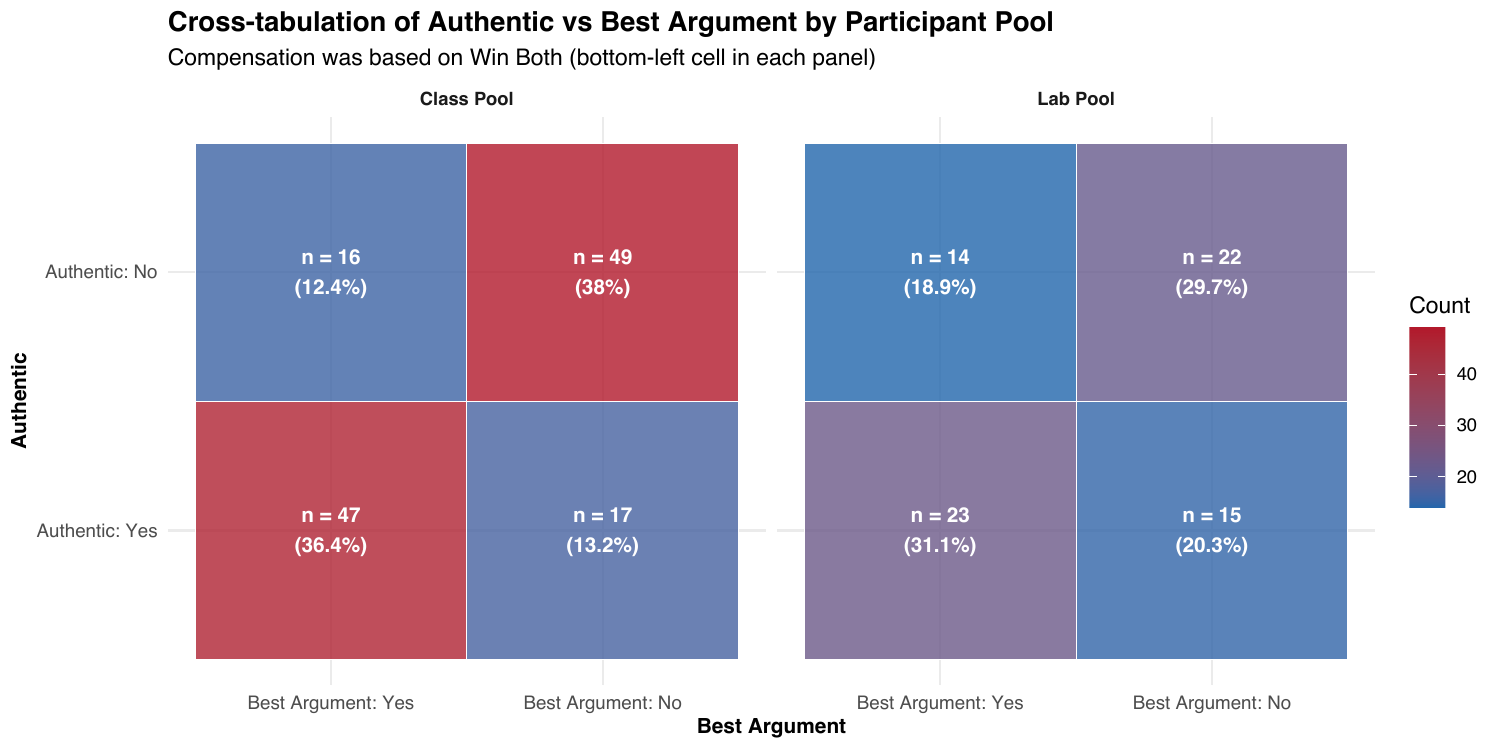}
\caption{\textbf{Cross-tabulation of Authentic vs Best Argument Win Conditions by Pool.}}
\label{fig:pool-crosstab}
\end{figure}

We also examined the win-based compensation mechanism, which was identical across pools and based on achieving both authentic connection and best argument conditions. Table~\ref{tab:pool_win_compensation} shows no significant differences in win rates between pools for any condition ($p > 0.50$). Visual examination of win condition cross-tabulations (Figures~\ref{fig:pool-win-rates} and \ref{fig:pool-crosstab}) confirms similar patterns across pools, indicating equivalent compensation opportunities.

\begin{table}[!h]
\centering\centering
\caption{\label{tab:pool_demographic_balance}\textbf{Demographic and Baseline Balance Between Participant Pools.}}
\centering
\fontsize{8}{10}\selectfont
\begin{tabular}[t]{lccc}
\toprule
\textbf{Variable} & \makecell[c]{\textbf{Class}\ \ \\N = 129} & \makecell[c]{\textbf{Lab}\ \ \\N = 74} & \textbf{p-value}\\
\midrule
topic, n (\%) &  &  & 0.093\\
\hspace{1em}abortion-rights & 35.0 (27.1\%) & 17.0 (23.0\%) & \\
\hspace{1em}affirmative-action & 25.0 (19.4\%) & 6.0 (8.1\%) & \\
\hspace{1em}covid-masks & 13.0 (10.1\%) & 12.0 (16.2\%) & \\
\hspace{1em}relief-plan & 26.0 (20.2\%) & 24.0 (32.4\%) & \\
\hspace{1em}sports-transgender & 16.0 (12.4\%) & 10.0 (13.5\%) & \\
\hspace{1em}ukraine-russia & 14.0 (10.9\%) & 5.0 (6.8\%) & \\
gender, n (\%) &  &  & 0.019\\
\hspace{1em}Male & 65.0 (50.4\%) & 26.0 (35.1\%) & \\
\hspace{1em}Female & 61.0 (47.3\%) & 41.0 (55.4\%) & \\
\hspace{1em}Other & 3.0 (2.3\%) & 7.0 (9.5\%) & \\
ethnic, n (\%) &  &  & 0.14\\
\hspace{1em}White / Caucasian & 53.0 (41.1\%) & 27.0 (36.5\%) & \\
\hspace{1em}Black / Hispanic & 8.0 (6.2\%) & 12.0 (16.2\%) & \\
\hspace{1em}Asian & 53.0 (41.1\%) & 26.0 (35.1\%) & \\
\hspace{1em}Other & 15.0 (11.6\%) & 9.0 (12.2\%) & \\
Political viewpoint, n (\%) &  &  & 0.77\\
\hspace{1em}Neutral & 29.0 (22.5\%) & 18.0 (24.3\%) & \\
\hspace{1em}Prefer not to say & 10.0 (7.8\%) & 3.0 (4.1\%) & \\
\hspace{1em}Conservative & 13.0 (10.1\%) & 7.0 (9.5\%) & \\
\hspace{1em}Liberal & 77.0 (59.7\%) & 46.0 (62.2\%) & \\
Animosity (pre) &  &  & 0.89\\
\hspace{1em}Mean (SD) & 34.88 (18.08) & 34.46 (23.82) & \\
\hspace{1em}Median (Q1, Q3) & 40.00 (20.00, 50.00) & 35.00 (10.00, 50.00) & \\
Stance (pre) &  &  & 0.79\\
\hspace{1em}Mean (SD) & 0.62 (1.36) & 0.68 (1.48) & \\
\hspace{1em}Median (Q1, Q3) & 1.00 (-1.00, 2.00) & 1.00 (-1.00, 2.00) & \\
\bottomrule
\end{tabular}
\par\smallskip
\parbox{0.95\linewidth}{\footnotesize \textsuperscript{1} Pearson's Chi-squared test; Welch Two Sample t-test}
\end{table}

\begin{table}[!h]
\centering\centering
\caption{\label{tab:pool_win_compensation}\textbf{Win-based Compensation Distribution Between Participant Pools.}}
\centering
\fontsize{8}{10}\selectfont
\begin{tabular}[t]{lccc}
\toprule
\textbf{Win Condition} & \makecell[c]{\textbf{Class}\ \ \\N = 129} & \makecell[c]{\textbf{Lab}\ \ \\N = 74} & \textbf{p-value}\\
\midrule
Won both (authentic and best argument), n (\%) &  &  & 0.54\\
\hspace{1em}No & 82.0 (63.6\%) & 51.0 (68.9\%) & \\
\hspace{1em}Yes & 47.0 (36.4\%) & 23.0 (31.1\%) & \\
Judged authentic, n (\%) &  &  & 0.93\\
\hspace{1em}No & 65.0 (50.4\%) & 36.0 (48.6\%) & \\
\hspace{1em}Yes & 64.0 (49.6\%) & 38.0 (51.4\%) & \\
Best argument, n (\%) &  &  & 0.99\\
\hspace{1em}No & 66.0 (51.2\%) & 37.0 (50.0\%) & \\
\hspace{1em}Yes & 63.0 (48.8\%) & 37.0 (50.0\%) & \\
\bottomrule
\end{tabular}
\par\smallskip
\parbox{0.95\linewidth}{\footnotesize \textsuperscript{1} Pearson's Chi-squared test}
\end{table}

\subsection{Additional Details for H1}
\label{app:analysis:h1}

This section presents robustness checks and additional analyses to assess the sensitivity of our main findings to modeling choices and potential sources of bias.

\subsubsection{Main Model Specifications}

Table~\ref{tab:h1_model_table} presents the complete mixed-effects model specification for both \textbf{animosity} and \textbf{stance} outcomes. The models include random intercepts for participants and debates, with fixed effects for treatment arms, time periods, and their interactions, along with control variables for strong opinion, topic, and demographic characteristics.

\begin{table}[!h]
\centering\centering
\caption{\label{tab:h1_model_table}\textbf{Mixed Effects Models: Treatment Effects on Issue-based Animosity and Stance.}}
\centering
\renewcommand{\baselinestretch}{1}\fontsize{8.5}{10}\selectfont\setlength{\tabcolsep}{3pt}
\begin{tabular}[t]{lcc}
\toprule
  & $Y=\anim$ & $Y=\stance$\\
\midrule
Intercept & 0.91 [0.21, 1.60]* & -0.44 [-0.83, -0.05]*\\
Arm: (Debate, Opp) & -0.07 [-0.51, 0.36] & -0.01 [-0.29, 0.27]\\
Arm: (Debate, Own) & -0.27 [-0.70, 0.17] & 0.00 [-0.28, 0.28]\\
Arm: (Write, Opp) & -0.20 [-0.70, 0.29] & 0.00 [-0.33, 0.34]\\
Time: post & -0.11 [-0.47, 0.24] & 0.12 [-0.19, 0.42]\\
Time: follow & -0.01 [-0.45, 0.42] & -0.06 [-0.43, 0.30]\\
Strong opinion & -0.53 [-0.78, -0.28]*** & -0.62 [-0.75, -0.49]***\\
Ethnicity: Black / Hispanic & -0.00 [-0.42, 0.41] & -0.11 [-0.33, 0.12]\\
Ethnicity: Asian & 0.01 [-0.26, 0.29] & -0.01 [-0.16, 0.14]\\
Ethnicity: Other & 0.17 [-0.22, 0.56] & 0.14 [-0.07, 0.35]\\
Gender: Female & -0.27 [-0.52, -0.02]* & -0.01 [-0.14, 0.13]\\
Gender: Other & -0.54 [-1.11, 0.03] & -0.05 [-0.35, 0.24]\\
Political viewpoint: Prefer not to say & -0.35 [-0.91, 0.21] & -0.36 [-0.66, -0.06]*\\
Political viewpoint: Conservative & -0.17 [-0.60, 0.26] & -0.18 [-0.42, 0.05]\\
Political viewpoint: Liberal & -0.28 [-0.57, -0.00]* & -0.24 [-0.39, -0.08]**\\
Topic: Affirmative action & 0.27 [-0.14, 0.68] & 0.03 [-0.18, 0.23]\\
Topic: Campus mask mandate & 0.56 [0.13, 1.00]* & 0.15 [-0.07, 0.37]\\
Topic: Student debt relief & 0.71 [0.35, 1.06]*** & 0.14 [-0.04, 0.32]\\
Topic: Transgender athletes & 0.32 [-0.11, 0.75] & -0.02 [-0.24, 0.20]\\
Topic: Support for Ukraine & 0.44 [-0.05, 0.92] & -0.09 [-0.34, 0.16]\\
(Debate, Opp) × post & 0.37 [-0.04, 0.78] & 0.49 [0.14, 0.84]**\\
(Debate, Own) × post & 0.28 [-0.13, 0.68] & 0.06 [-0.29, 0.41]\\
(Write, Opp) × post & 0.56 [0.06, 1.06]* & 0.79 [0.36, 1.22]***\\
(Debate, Opp) × follow & 0.38 [-0.12, 0.88] & 0.56 [0.14, 0.98]**\\
(Debate, Own) × follow & 0.17 [-0.33, 0.67] & 0.35 [-0.07, 0.77]\\
(Write, Opp) × follow & -0.13 [-0.76, 0.49] & 0.79 [0.27, 1.31]**\\
SD (participant intercept) & 0.61 & 0.23\\
SD (debate intercept) & 0.34 & 0.11\\
SD (residual) & 0.63 & 0.54\\
Observations & 523 & 523\\
ICC & 0.6 & 0.2\\
\bottomrule
\end{tabular}
\par\smallskip
\parbox{0.95\linewidth}{\footnotesize 95\% confidence intervals in brackets. Random effects for participant and debate included.}
\end{table}

\subsubsection{Pairwise Comparisons Between Arms}

While our primary hypothesis focuses on within-arm changes over time, we also examined whether treatment arms differed significantly from each other in their effectiveness. Table~\ref{tab:h1_arm_differences} presents pairwise comparisons between arms for each contrast period.

\begin{table}[!h]
\centering
\caption{\label{tab:h1_arm_differences}\textbf{Pairwise Differences Between Treatment Arms.}}
\small
\begin{tabular}[t]{llcrr}
\toprule
\textbf{Outcome} & \textbf{Arm Comparison} & \textbf{Difference} & \textbf{95\% CI} & \textbf{$p$}\\
\midrule
\addlinespace[0.3em]
\multicolumn{5}{l}{\textbf{Post $-$ Pre}}\\
\hspace{1em}Animosity & (Write, Own) $-$ (Debate, Opp) & -0.374 & [-0.913, 0.164] & 0.278\\
\hspace{1em}Animosity & (Write, Own) $-$ (Debate, Own) & -0.276 & [-0.812, 0.260] & 0.546\\
\hspace{1em}Animosity & (Write, Own) $-$ (Write, Opp) & -0.565 & [-1.222, 0.092] & 0.120\\
\hspace{1em}Animosity & (Debate, Opp) $-$ (Debate, Own) & 0.099 & [-0.274, 0.471] & 0.904\\
\hspace{1em}Animosity & (Debate, Opp) $-$ (Write, Opp) & -0.190 & [-0.723, 0.342] & 0.792\\
\hspace{1em}Animosity & (Debate, Own) $-$ (Write, Opp) & -0.289 & [-0.819, 0.241] & 0.494\\
\hspace{1em}Stance & (Write, Own) $-$ (Debate, Opp) & -0.489* & [-0.952, -0.025] & 0.034\\
\hspace{1em}Stance & (Write, Own) $-$ (Debate, Own) & -0.064 & [-0.525, 0.397] & 0.984\\
\hspace{1em}Stance & (Write, Own) $-$ (Write, Opp) & -0.793** & [-1.358, -0.227] & 0.002\\
\hspace{1em}Stance & (Debate, Opp) $-$ (Debate, Own) & 0.425** & [0.103, 0.746] & 0.004\\
\hspace{1em}Stance & (Debate, Opp) $-$ (Write, Opp) & -0.304 & [-0.763, 0.155] & 0.321\\
\hspace{1em}Stance & (Debate, Own) $-$ (Write, Opp) & -0.729*** & [-1.185, -0.272] & $<$0.001\\
\addlinespace[0.3em]
\multicolumn{5}{l}{\textbf{Follow $-$ Pre}}\\
\hspace{1em}Animosity & (Write, Own) $-$ (Debate, Opp) & -0.378 & [-1.034, 0.277] & 0.444\\
\hspace{1em}Animosity & (Write, Own) $-$ (Debate, Own) & -0.172 & [-0.828, 0.484] & 0.905\\
\hspace{1em}Animosity & (Write, Own) $-$ (Write, Opp) & 0.135 & [-0.685, 0.954] & 0.974\\
\hspace{1em}Animosity & (Debate, Opp) $-$ (Debate, Own) & 0.206 & [-0.244, 0.657] & 0.639\\
\hspace{1em}Animosity & (Debate, Opp) $-$ (Write, Opp) & 0.513 & [-0.156, 1.182] & 0.197\\
\hspace{1em}Animosity & (Debate, Own) $-$ (Write, Opp) & 0.307 & [-0.363, 0.977] & 0.638\\
\hspace{1em}Stance & (Write, Own) $-$ (Debate, Opp) & -0.564* & [-1.117, -0.011] & 0.043\\
\hspace{1em}Stance & (Write, Own) $-$ (Debate, Own) & -0.350 & [-0.903, 0.203] & 0.362\\
\hspace{1em}Stance & (Write, Own) $-$ (Write, Opp) & -0.791* & [-1.480, -0.102] & 0.017\\
\hspace{1em}Stance & (Debate, Opp) $-$ (Debate, Own) & 0.214 & [-0.166, 0.595] & 0.466\\
\hspace{1em}Stance & (Debate, Opp) $-$ (Write, Opp) & -0.227 & [-0.789, 0.335] & 0.725\\
\hspace{1em}Stance & (Debate, Own) $-$ (Write, Opp) & -0.441 & [-1.004, 0.122] & 0.182\\
\bottomrule
\end{tabular}
\par\smallskip
\parbox{0.95\linewidth}{\footnotesize Differences in standardized change scores between treatment arms. Positive values indicate larger increases in the first arm relative to the second. *** $p<0.001$, ** $p<0.01$, * $p<0.05$.}
\end{table}

Most pairwise differences between arms are not statistically significant, likely reflecting limited statistical power given our sample sizes. However, for stance movement, several patterns emerge consistently: Write/Opp shows larger positive effects compared to other arms, particularly relative to Write/Own and Debate/Own. The directionality of estimates aligns with our main results, with opposing perspective conditions generally showing larger effects than own perspective conditions.

\subsubsection{Model Specification: Adjusted vs. Unadjusted}

We assessed whether including control variables meaningfully improved model fit compared to a parsimonious model with only treatment effects. The likelihood ratio test strongly favors the adjusted model ($\chi^2 = 62.17$, $df = 14$, $p < 0.001$), with marginal $R^2$ increasing from 0.010 (unadjusted) to 0.043 (adjusted). 

The control variables explain substantial variance: strong opinion and topic controls contribute $\Delta R^2 = 0.026$, while demographic variables add $\Delta R^2 = 0.008$. The arm×time interaction contributes $\Delta R^2 = 0.008$ beyond main effects and controls. While this interaction effect is small in variance terms (Cohen's $f^2 = 0.01$), it represents the systematic treatment-specific changes that are the focus of our hypothesis.

\subsubsection{Sensitivity to Survey Modality}

\begin{figure}[htbp]
\centering
\includegraphics[width=\textwidth]{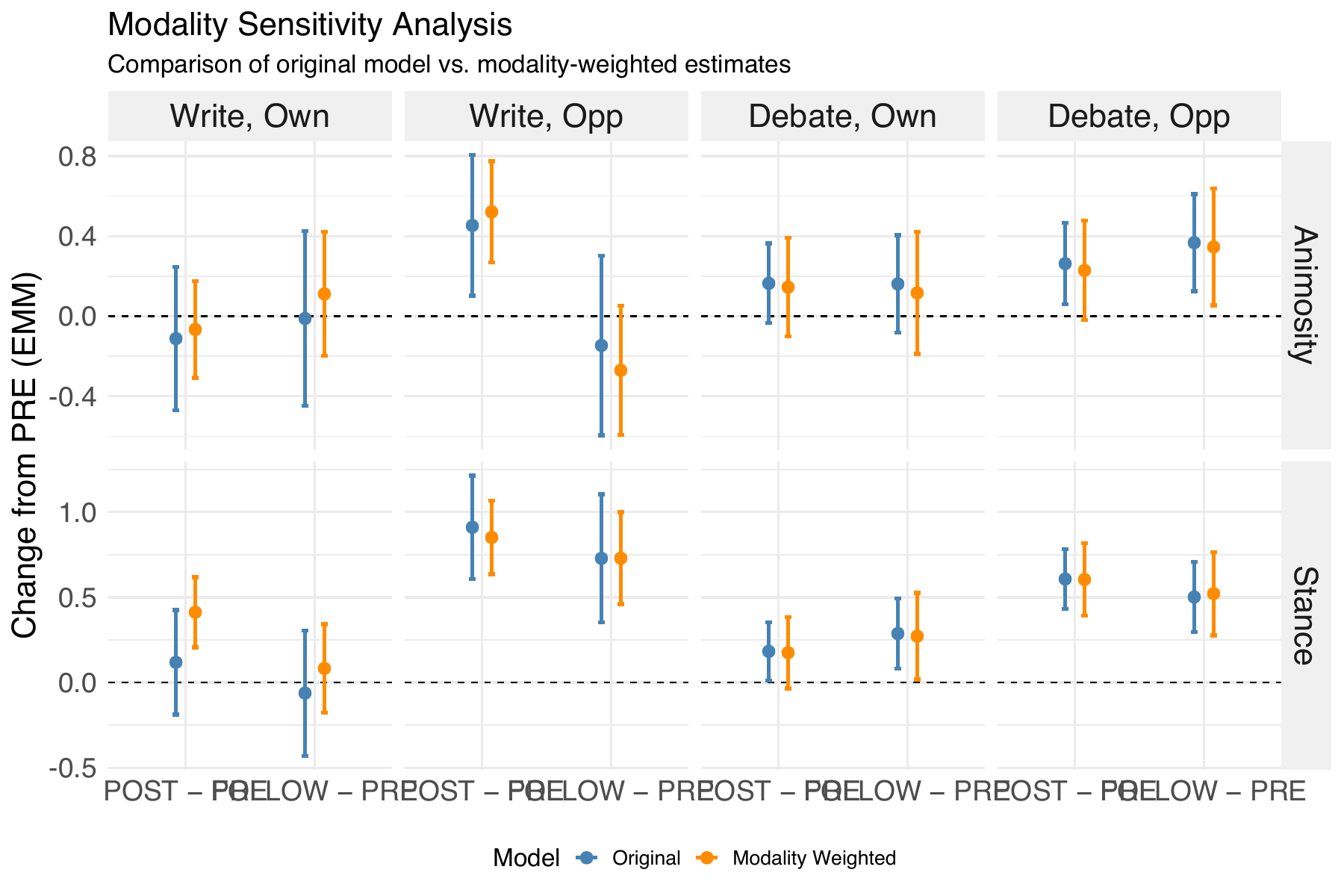}
\caption{\textbf{Modality Sensitivity Analysis.}}
\label{fig:h1_modality_sensitivity}
\end{figure}

To address potential confounding by survey modality (online vs. in-person), we re-estimated our models using inverse probability weights for modality assignment \citep{seamanReviewInverseProbability2013, leeWeightTrimmingPropensity2011}. Figure~\ref{fig:h1_modality_sensitivity} compares the original and modality-weighted estimates.

The modality-weighted results are qualitatively similar to the original estimates, with one notable difference: in the stance outcome, the Write/Own change at post is larger and significant under modality weighting (0.41 [0.21, 0.62], versus 0.12 [$-0.19$, 0.43] in the main model), because the weights give a few of its 25 participants large influence. This suggests our main conclusions are robust to potential modality-related selection effects.

\subsubsection{Sensitivity to Attrition}

\begin{figure}[htbp]
\centering
\includegraphics[width=\textwidth]{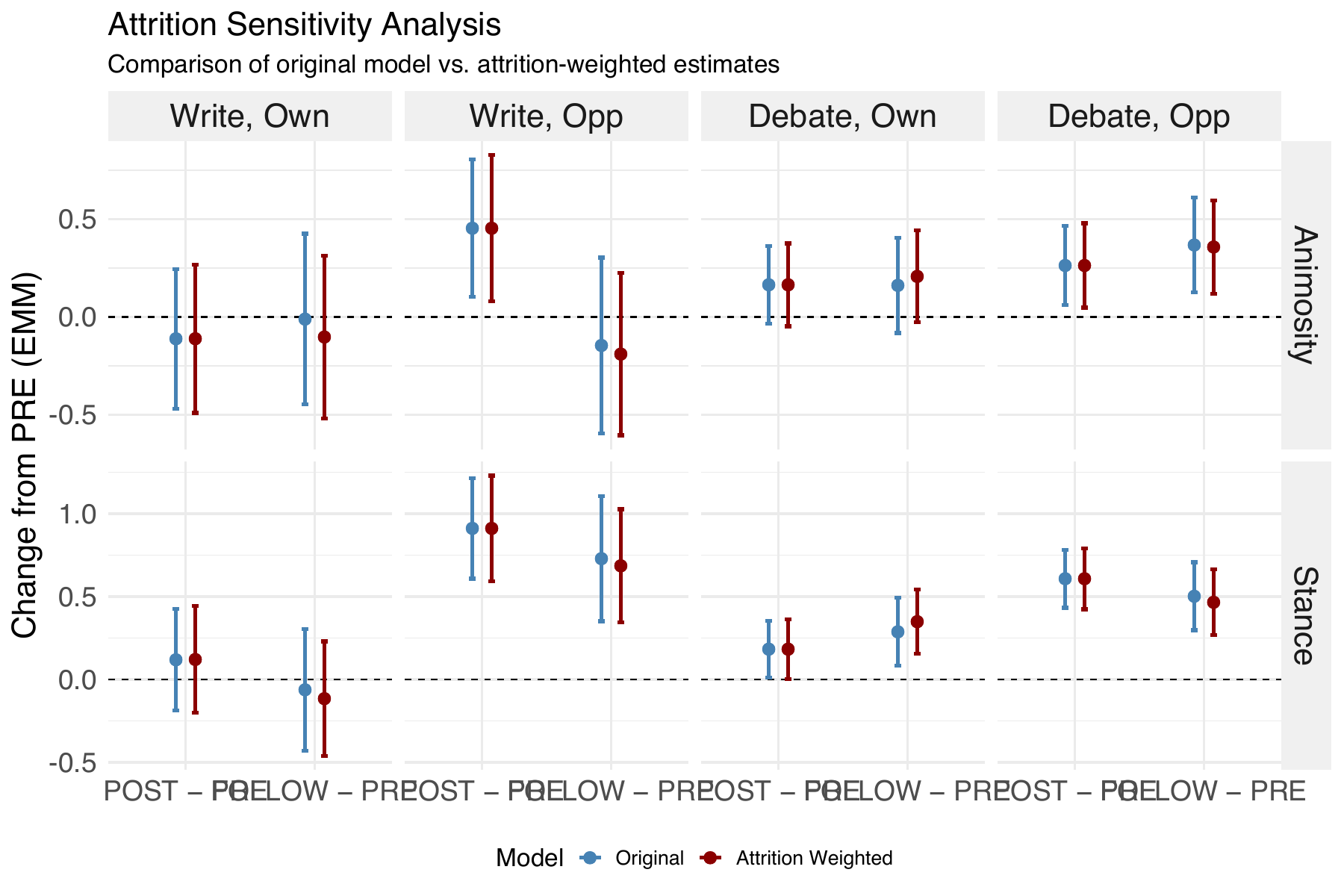}
\caption{\textbf{Attrition Sensitivity Analysis.}}
\label{fig:h1_attrition_sensitivity}
\end{figure}

We assessed sensitivity to differential attrition using inverse probability weights based on baseline characteristics. Figure~\ref{fig:h1_attrition_sensitivity} shows the comparison between original and attrition-weighted estimates.

Results remain substantively unchanged under attrition weighting, indicating that our findings are not driven by systematic differences between participants who completed follow-up surveys and those who did not. The attrition weights had minimal impact because baseline characteristics were well-balanced across arms and attrition rates were relatively modest.

\subsubsection{Sensitivity to Pool Source}

\begin{figure}[htbp]
\centering
\includegraphics[width=\textwidth]{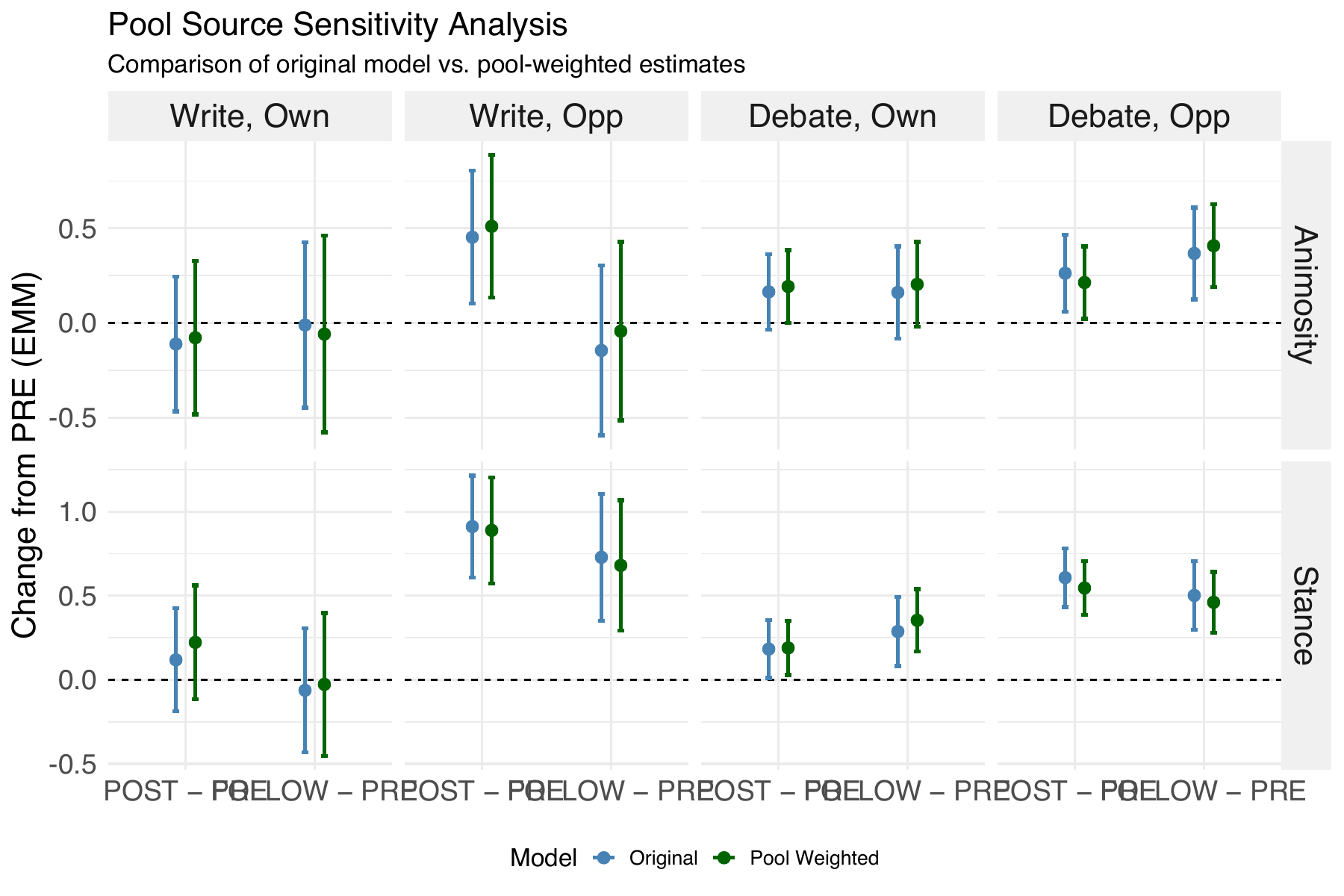}
\caption{\textbf{Pool Sensitivity Analysis.}}
\label{fig:h1_pool_sensitivity}
\end{figure}

As we recruited participants from two different types of pools, compensated with money and compensated with extra credit; similar to the previous analysis, we also assess potential confounding due to the pool type that the participant originated from. Figure~\ref{fig:h1_pool_sensitivity} compares the original and pool-weighted estimates.

We find no substantive differences when weighting by pool source, confirming that the pool composition does not drive our results. This analysis is also complemented by earlier analysis on balance due to the pools.

\subsubsection{Alternative Model: Change Score Approach}

\begin{figure}[htbp]
\centering
\includegraphics[width=\textwidth]{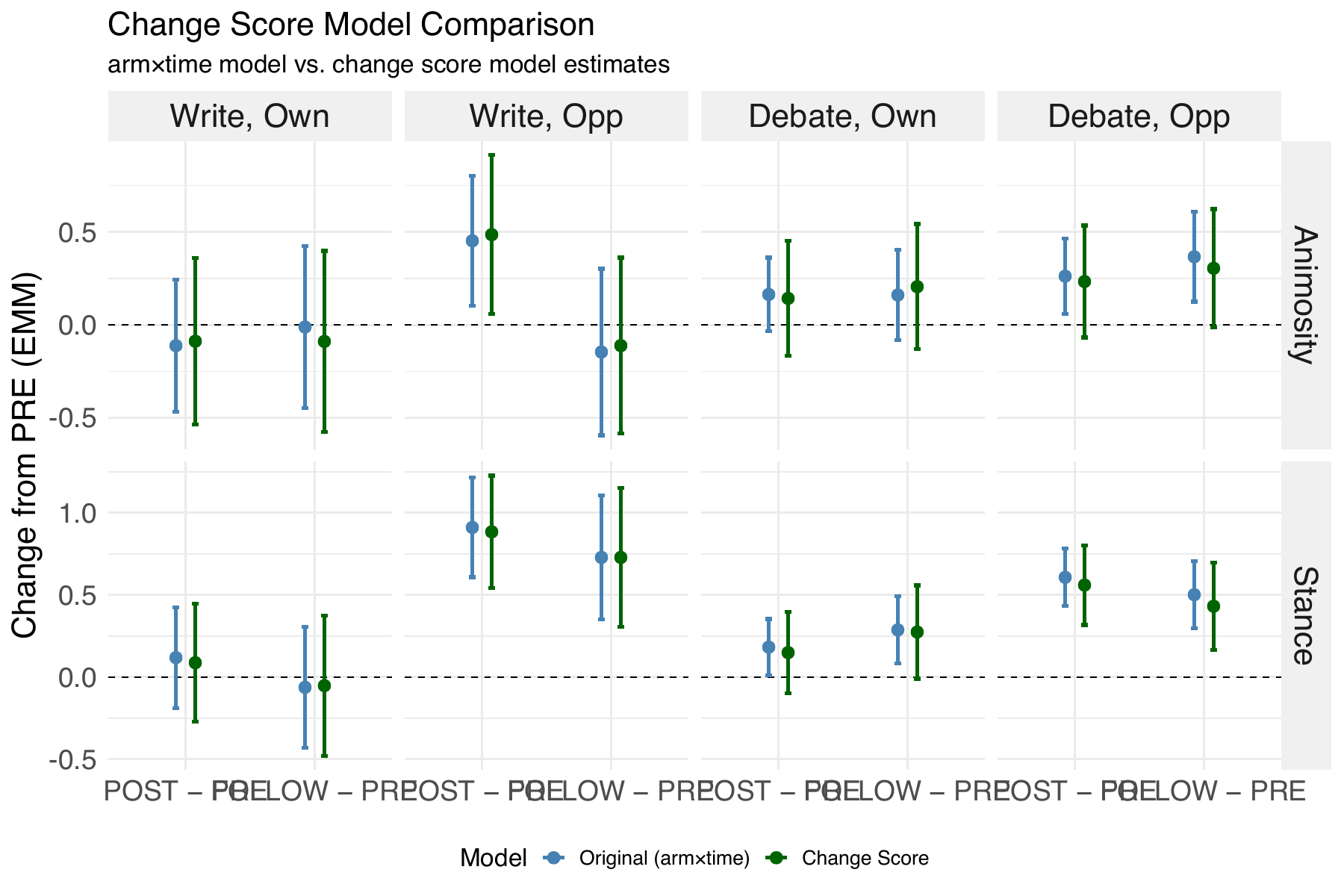}
\caption{\textbf{Change Score Model Comparison.}}
\label{fig:h1_changeScore_comparison}
\end{figure}

As an alternative to the arm×time interaction model, we estimated change score models that directly model post-pre and follow-pre differences. Figure~\ref{fig:h1_changeScore_comparison} compares estimates from both approaches.

The change score approach yields similar patterns but with somewhat smaller effect sizes (approximately 25\% reduction). For animosity, the Debate/Opp effect becomes marginally significant ($p \approx 0.10$) rather than clearly significant. This difference likely reflects the change score model's susceptibility to measurement error propagation and reduced statistical power compared to the repeated measures approach. The arm×time interaction model is preferred for its superior handling of within-person correlation and measurement precision.

\subsubsection{Summary of Robustness for H1}

Table~\ref{tab:h1_robustness_summary} provides a comprehensive comparison across all modeling approaches for both outcomes.

\begin{sidewaystable}[h]
\caption{\label{tab:h1_robustness_summary}
\textbf{{Robustness Check Summary: All Modeling Approaches}.}}
\fontsize{8pt}{10pt}\selectfont\setlength{\tabcolsep}{3pt}
\begin{tabular*}{0.95\linewidth}{@{\extracolsep{\fill}}llllll}
\toprule
Arm \& Contrast & Main Model & Modality Weighted & Pool Weighted & Attrition Weighted & Change Score \\ 
\midrule\addlinespace[2.5pt]
\multicolumn{6}{l}{{\bfseries Animosity}} \\[2.5pt] 
\midrule\addlinespace[2.5pt]
Write, Own \(\Delta_{\text{pos}}\) & -0.11 [-0.47, 0.25] & -0.07 [-0.31, 0.18] & -0.08 [-0.48, 0.33] & -0.11 [-0.49, 0.27] & -0.09 [-0.54, 0.36] \\ 
Write, Own \(\Delta_{\text{fol}}\) & -0.01 [-0.45, 0.43] & 0.11 [-0.20, 0.42] & -0.06 [-0.58, 0.46] & -0.10 [-0.52, 0.31] & -0.09 [-0.58, 0.40] \\ 
Debate, Opp \(\Delta_{\text{pos}}\) & 0.26 [0.06, 0.47]* & 0.23 [-0.02, 0.48] & 0.21 [0.02, 0.40]* & 0.26 [0.05, 0.48]* & 0.23 [-0.07, 0.54] \\ 
Debate, Opp \(\Delta_{\text{fol}}\) & 0.37 [0.12, 0.61]* & 0.35 [0.05, 0.64]* & 0.41 [0.19, 0.63]* & 0.36 [0.12, 0.59]* & 0.31 [-0.01, 0.62] \\ 
Debate, Own \(\Delta_{\text{pos}}\) & 0.16 [-0.03, 0.36] & 0.15 [-0.10, 0.39] & 0.19 [0.00, 0.38]* & 0.16 [-0.05, 0.38] & 0.14 [-0.17, 0.45] \\ 
Debate, Own \(\Delta_{\text{fol}}\) & 0.16 [-0.08, 0.40] & 0.12 [-0.19, 0.42] & 0.20 [-0.02, 0.43] & 0.21 [-0.03, 0.44] & 0.21 [-0.13, 0.54] \\ 
Write, Opp \(\Delta_{\text{pos}}\) & 0.45 [0.10, 0.80]* & 0.52 [0.27, 0.77]* & 0.51 [0.13, 0.89]* & 0.45 [0.08, 0.83]* & 0.49 [0.06, 0.91]* \\ 
Write, Opp \(\Delta_{\text{fol}}\) & -0.15 [-0.59, 0.30] & -0.27 [-0.59, 0.05] & -0.04 [-0.52, 0.43] & -0.19 [-0.60, 0.22] & -0.11 [-0.59, 0.36] \\ 
\midrule\addlinespace[2.5pt]
\multicolumn{6}{l}{{\bfseries Stance}} \\[2.5pt] 
\midrule\addlinespace[2.5pt]
Write, Own \(\Delta_{\text{pos}}\) & 0.12 [-0.19, 0.43] & 0.41 [0.21, 0.62]* & 0.22 [-0.12, 0.56] & 0.12 [-0.20, 0.44] & 0.09 [-0.27, 0.45] \\ 
Write, Own \(\Delta_{\text{fol}}\) & -0.06 [-0.43, 0.31] & 0.08 [-0.18, 0.34] & -0.03 [-0.45, 0.40] & -0.12 [-0.46, 0.23] & -0.05 [-0.48, 0.37] \\ 
Debate, Opp \(\Delta_{\text{pos}}\) & 0.61 [0.43, 0.78]* & 0.60 [0.39, 0.82]* & 0.55 [0.38, 0.71]* & 0.61 [0.42, 0.79]* & 0.56 [0.32, 0.80]* \\ 
Debate, Opp \(\Delta_{\text{fol}}\) & 0.50 [0.30, 0.71]* & 0.52 [0.28, 0.77]* & 0.46 [0.28, 0.64]* & 0.47 [0.27, 0.66]* & 0.43 [0.17, 0.70]* \\ 
Debate, Own \(\Delta_{\text{pos}}\) & 0.18 [0.01, 0.35]* & 0.18 [-0.03, 0.38] & 0.19 [0.03, 0.35]* & 0.18 [0.00, 0.36]* & 0.15 [-0.10, 0.40] \\ 
Debate, Own \(\Delta_{\text{fol}}\) & 0.29 [0.08, 0.49]* & 0.27 [0.02, 0.53]* & 0.35 [0.17, 0.54]* & 0.35 [0.15, 0.54]* & 0.27 [-0.01, 0.56] \\ 
Write, Opp \(\Delta_{\text{pos}}\) & 0.91 [0.61, 1.21]* & 0.85 [0.64, 1.07]* & 0.89 [0.57, 1.20]* & 0.91 [0.59, 1.23]* & 0.88 [0.54, 1.23]* \\ 
Write, Opp \(\Delta_{\text{fol}}\) & 0.73 [0.35, 1.11]* & 0.73 [0.46, 1.00]* & 0.68 [0.29, 1.07]* & 0.69 [0.34, 1.03]* & 0.73 [0.31, 1.15]* \\ 
\bottomrule
\end{tabular*}
\begin{minipage}{0.95\linewidth}
* indicates 95\% CI excludes zero. Estimates show change from PRE with 95\% confidence intervals.\\
\end{minipage}
\end{sidewaystable}

Across all robustness checks, the core pattern remains consistent: Write/Opp and Debate/Opp show the largest and most reliable effects for both outcomes, with effects generally persisting through follow-up. The magnitude of effects varies somewhat across approaches, but the substantive conclusions about which interventions are most effective remain unchanged.

\subsubsection{Probability of Individual Improvement}

While our main analyses focus on average changes, we also examined how many participants improved. We coded each participant as improved at post and at follow-up if the outcome was above its pre-intervention value (no change counts as not improved) and fit Bayesian mixed-effects logistic models with the same covariates as H1 and random intercepts for participant and debate. From these models we computed covariate-standardized shares of participants who improved in each arm, and the posterior probability that each arm's share exceeds that of Write/Own (Table~\ref{tab:h1_improvement_prob}).

\begin{table}[t]
\caption{\label{tab:h1_improvement_prob}
\textbf{{\large Probability of Improvement by Arm and Time}.}}
\fontsize{9pt}{11pt}\selectfont\setlength{\tabcolsep}{4pt}
\begin{tabular*}{\linewidth}{@{\extracolsep{\fill}}llrrlr}
\toprule
Timepoint & Arm & n & \makecell[l]{Observed:\\improved / same / worse} & \makecell[l]{Model: improved\\(95\% CrI)} & \makecell[r]{P($>$\\Write/Own)} \\ 
\midrule\addlinespace[2.5pt]
\multicolumn{6}{l}{Animosity} \\[2.5pt] 
\midrule\addlinespace[2.5pt]
post & Write/Own & 24 & 38\% / 21\% / 42\% & 41\% [26\%, 58\%] & -- \\ 
post & Write/Opp & 25 & 60\% / 24\% / 16\% & 59\% [41\%, 75\%] & 94\% \\ 
post & Debate/Own & 78 & 41\% / 32\% / 27\% & 40\% [31\%, 49\%] & 44\% \\ 
post & Debate/Opp & 75 & 44\% / 33\% / 23\% & 44\% [35\%, 54\%] & 61\% \\ 
follow & Write/Own & 14 & 29\% / 14\% / 57\% & 37\% [19\%, 58\%] & -- \\ 
follow & Write/Opp & 13 & 38\% / 15\% / 46\% & 45\% [23\%, 67\%] & 69\% \\ 
follow & Debate/Own & 45 & 36\% / 29\% / 36\% & 37\% [26\%, 49\%] & 50\% \\ 
follow & Debate/Opp & 46 & 52\% / 30\% / 17\% & 52\% [40\%, 63\%] & 87\% \\ 
\midrule\addlinespace[2.5pt]
\multicolumn{6}{l}{Stance} \\[2.5pt] 
\midrule\addlinespace[2.5pt]
post & Write/Own & 24 & 21\% / 54\% / 25\% & 29\% [16\%, 43\%] & -- \\ 
post & Write/Opp & 25 & 68\% / 32\% / 0\% & 60\% [43\%, 76\%] & 100\% \\ 
post & Debate/Own & 78 & 27\% / 59\% / 14\% & 28\% [20\%, 36\%] & 47\% \\ 
post & Debate/Opp & 75 & 48\% / 47\% / 5\% & 48\% [38\%, 57\%] & 98\% \\ 
follow & Write/Own & 14 & 14\% / 57\% / 29\% & 25\% [11\%, 44\%] & -- \\ 
follow & Write/Opp & 13 & 46\% / 54\% / 0\% & 47\% [26\%, 69\%] & 95\% \\ 
follow & Debate/Own & 45 & 33\% / 49\% / 18\% & 36\% [26\%, 48\%] & 85\% \\ 
follow & Debate/Opp & 46 & 48\% / 48\% / 4\% & 47\% [36\%, 59\%] & 97\% \\ 
\bottomrule
\end{tabular*}
\begin{minipage}{\linewidth}
\footnotesize Improved: outcome above its baseline value (no change counts as not improved). Observed: shares of participants who improved / did not change / worsened. Model: covariate-standardized share who improved, from a Bayesian mixed-effects logistic model with participant and debate random intercepts (posterior median, 95\% credible interval). P($>$ Write/Own): posterior probability that the arm's share exceeds Write/Own's at the same time point.\\
\end{minipage}
\end{table}

\begin{figure}[h]
\centering
\includegraphics[width=\textwidth]{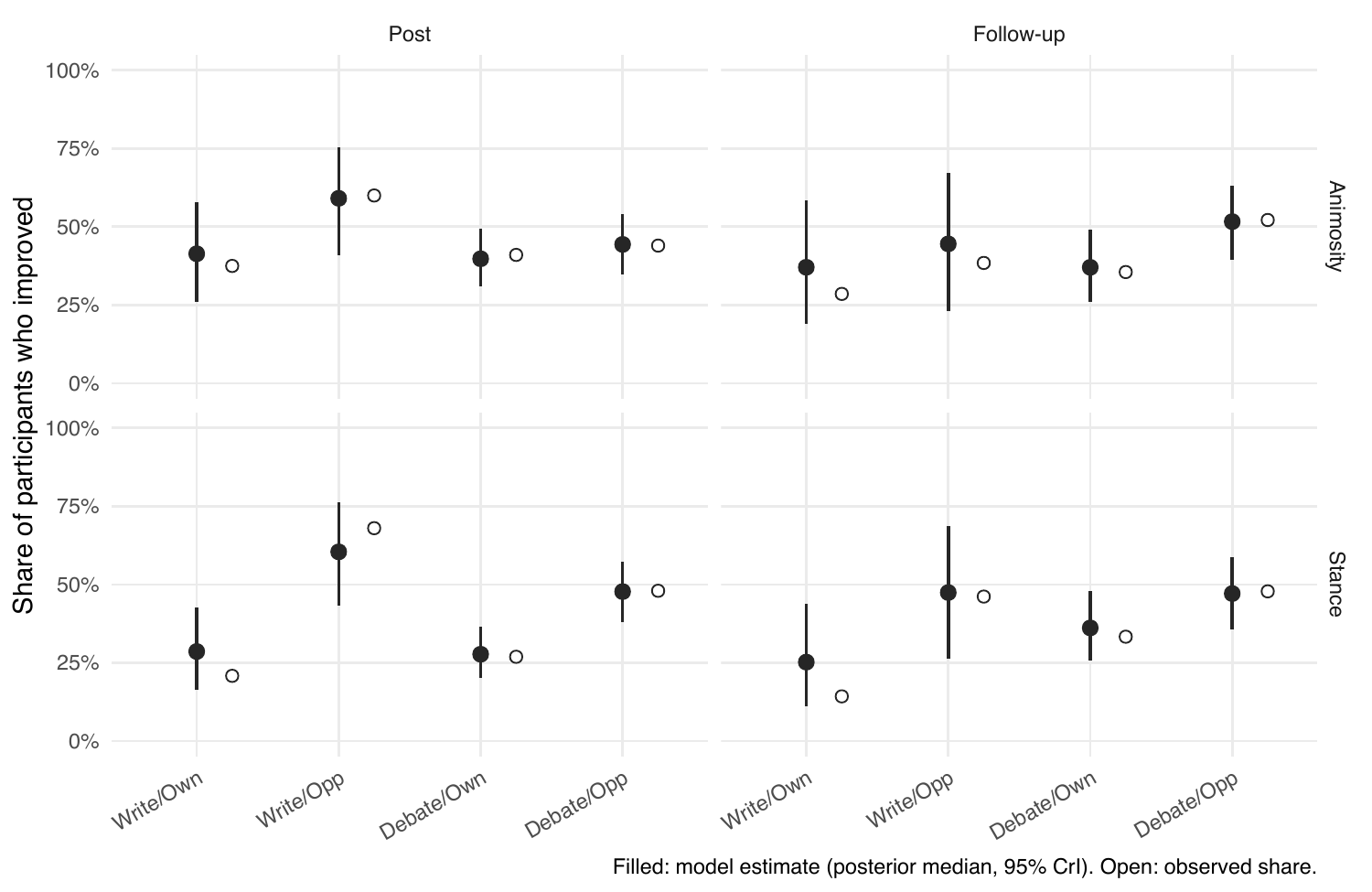}
\caption{\textbf{Share of participants who improved, by arm and time point.} Filled points: covariate-standardized model estimate (posterior median, 95\% credible interval); open points: observed share. Improvement means the outcome was above its pre-intervention value.}
\label{fig:h1_improvement_shares}
\end{figure}

Figure~\ref{fig:h1_improvement_shares} shows the model-based shares with their 95\% credible intervals next to the observed shares.

The shares mirror the average effects. For stance at post, 60\% [43\%, 76\%] of Write/Opp participants and 48\% [38\%, 57\%] of Debate/Opp participants improved, compared with 29\% [16\%, 43\%] in Write/Own (posterior probabilities of exceeding Write/Own: 100\% and 98\%); the two own-perspective arms were similar (28--29\%). Both opposing-perspective arms kept these shares at follow-up (47\% each), while Write/Own stayed lowest (25\% [11\%, 44\%]). Few opposing-perspective participants moved away from the other side's position (at most 5\% worsened on stance). For animosity, Write/Opp had the largest share improving at post (59\% [41\%, 75\%]; 94\% probability of exceeding Write/Own), and at follow-up Debate/Opp had the largest (52\% [40\%, 63\%]; 87\%), consistent with the durability pattern in the main results. Estimates for the writing arms at follow-up rest on 13--14 participants each and are correspondingly imprecise.

\subsection{Additional Details for H2}
\label{app:analysis:h2}

This section presents robustness checks for Hypothesis 2 (H2), which examined perspective taking effects across experimental conditions. The primary difference-in-differences specification was compared against change score models to assess the sensitivity of findings to alternative modeling approaches.

\subsubsection{Main Model Specifications}

Table~\ref{tab:h2_model_table} presents the complete mixed-effects model specification for both \textbf{animosity} and \textbf{stance} outcomes. The models include random intercepts for participants and debates, with fixed effects for treatment arms, time periods, and their interactions, along with control variables for strong opinion, topic, and demographic characteristics.

\begin{table}[h]
\centering\centering
\caption{\label{tab:h2_model_table}\textbf{Mixed Effects Models for H2: Perspective Taking and Modality Effects on Issue-based Animosity and Stance.} Random effects for participant and debate included. Reference categories: Own Perspective, Write Modality, PRE time period.}
\centering
\renewcommand{\baselinestretch}{1}\fontsize{8.5}{10}\selectfont\setlength{\tabcolsep}{3pt}
\begin{tabular}[t]{lcc}
\toprule
  & $Y=\anim$ & $Y=\stance$\\
\midrule
Intercept & 0.91 [0.21, 1.60]* & -0.44 [-0.83, -0.05]*\\
Perspective: Opp & -0.20 [-0.70, 0.29] & 0.00 [-0.33, 0.34]\\
Modality: Debate & -0.27 [-0.70, 0.17] & 0.00 [-0.28, 0.28]\\
Time: post & -0.11 [-0.47, 0.24] & 0.12 [-0.19, 0.42]\\
Time: follow & -0.01 [-0.45, 0.42] & -0.06 [-0.43, 0.30]\\
Strong opinion & -0.53 [-0.78, -0.28]*** & -0.62 [-0.75, -0.49]***\\
Topic: Affirmative action & 0.27 [-0.14, 0.68] & 0.03 [-0.18, 0.23]\\
Topic: Campus mask mandate & 0.56 [0.13, 1.00]* & 0.15 [-0.07, 0.37]\\
Topic: Student debt relief & 0.71 [0.35, 1.06]*** & 0.14 [-0.04, 0.32]\\
Topic: Transgender athletes & 0.32 [-0.11, 0.75] & -0.02 [-0.24, 0.20]\\
Topic: Support for Ukraine & 0.44 [-0.05, 0.92] & -0.09 [-0.34, 0.16]\\
Ethnicity: Black / Hispanic & -0.00 [-0.42, 0.41] & -0.11 [-0.33, 0.12]\\
Ethnicity: Asian & 0.01 [-0.26, 0.29] & -0.01 [-0.16, 0.14]\\
Ethnicity: Other & 0.17 [-0.22, 0.56] & 0.14 [-0.07, 0.35]\\
Gender: Female & -0.27 [-0.52, -0.02]* & -0.01 [-0.14, 0.13]\\
Gender: Other & -0.54 [-1.11, 0.03] & -0.05 [-0.35, 0.24]\\
Political viewpoint: Prefer not to say & -0.35 [-0.91, 0.21] & -0.36 [-0.66, -0.06]*\\
Political viewpoint: Conservative & -0.17 [-0.60, 0.26] & -0.18 [-0.42, 0.05]\\
Political viewpoint: Liberal & -0.28 [-0.57, -0.00]* & -0.24 [-0.39, -0.08]**\\
Debate × Opp & 0.40 [-0.18, 0.97] & -0.01 [-0.40, 0.37]\\
Opp × post & 0.56 [0.06, 1.06]* & 0.79 [0.36, 1.22]***\\
Opp × follow & -0.13 [-0.76, 0.49] & 0.79 [0.27, 1.31]**\\
Debate × post & 0.28 [-0.13, 0.68] & 0.06 [-0.29, 0.41]\\
Debate × follow & 0.17 [-0.33, 0.67] & 0.35 [-0.07, 0.77]\\
Debate × Opp × post & -0.47 [-1.04, 0.11] & -0.37 [-0.86, 0.13]\\
Debate × Opp × follow & 0.34 [-0.37, 1.05] & -0.58 [-1.17, 0.02]\\
SD (participant intercept) & 0.61 & 0.23\\
SD (debate intercept) & 0.34 & 0.11\\
SD (residual) & 0.63 & 0.54\\
Observations & 523 & 523\\
ICC & 0.6 & 0.2\\
\bottomrule
\end{tabular}
\par\smallskip
\parbox{0.95\linewidth}{\footnotesize 95\% confidence intervals in brackets.}
\end{table}

\subsubsection{Robustness Check: Change Score Specification}
\label{supp:h2:changescore}

We implemented change score models that directly estimate changes from baseline rather than leveraging the full longitudinal structure of the difference-in-differences approach. Both specifications used identical standardization procedures and covariate adjustments.

\begin{figure}[h]
\centering
\includegraphics[width=0.9\textwidth]{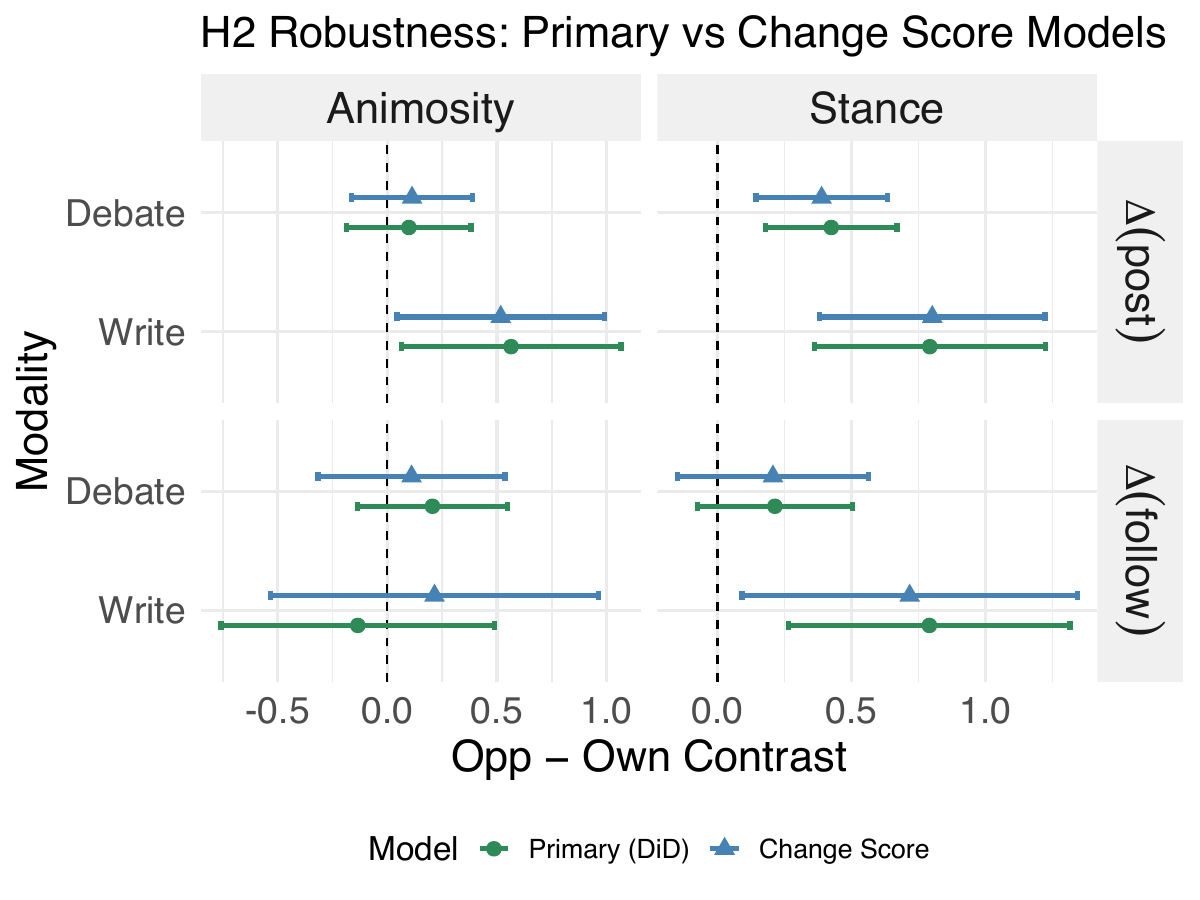}
\caption{\textbf{Comparison of Opp - Own contrast estimates between primary difference-in-differences and change score specifications.} Error bars represent 95\% confidence intervals.}
\label{fig:h2-robustness}
\end{figure}

Comparison between specifications revealed strong agreement (\Cref{fig:h2-robustness}). Across 8 comparisons (2 modalities × 2 time points × 2 outcomes), 87.5\% showed directional consistency and 100\% maintained identical significance patterns after Holm correction. The single directional discrepancy occurred for animosity in written debates at follow-up, where estimates differed in sign but both were non-significant with overlapping confidence intervals.

This consistency between modeling approaches strengthens confidence in the primary H2 findings, indicating they are robust to alternative specifications of longitudinal change.

\subsection{Additional Details for H3}
\label{app:analysis:h3}

This section presents supplementary analyses for Hypothesis 3 (H3), which examined whether self-perceived winning during debates was associated with changes in \textbf{animosity} and \textbf{stance}. The primary analysis used a difference-in-differences approach with linear mixed-effects models to estimate contrasts between different types of winning conditions (best argument, authentic connection, both) relative to losing, while adjusting for experimental arms and covariates.

\subsubsection{Main Model Specifications}

Table~\ref{tab:h3_model_table} presents the complete mixed-effects model specification for both outcomes. The models include random intercepts for participants and debates, with fixed effects for treatment arms, time periods, and their interactions, along with control variables for strong opinion, topic, and demographic characteristics. Note that we report the full model, but the key measures are obtained via estimated marginal means of this model.

\begin{table}[h]
\centering\centering
\caption{\label{tab:h3_model_table}\textbf{Mixed Effects Models for H3: Winning Effects on Issue-based Animosity and Stance.} Random effects for participant and debate are included. Reference categories: No Best Argument Win, No Authentic Win, PRE time period, Write Own arm. Controls: strong opinion, topic, demographics.}
\centering
\renewcommand{\baselinestretch}{1}\fontsize{8.5}{10}\selectfont\setlength{\tabcolsep}{3pt}
\begin{tabular}[t]{lcc}
\toprule
  & $Y=\anim$ & $Y=\stance$\\
\midrule
Intercept & 0.82 [0.11, 1.53]* & -0.46 [-0.86, -0.06]*\\
Best argument & 0.09 [-0.18, 0.37] & 0.04 [-0.14, 0.22]\\
Judged authentic & 0.14 [-0.14, 0.42] & 0.03 [-0.16, 0.21]\\
Time: post & -0.04 [-0.43, 0.34] & 0.26 [-0.07, 0.59]\\
Time: follow & 0.06 [-0.42, 0.54] & -0.15 [-0.55, 0.25]\\
Arm: (Debate, Opp) & -0.05 [-0.49, 0.39] & -0.01 [-0.28, 0.27]\\
Arm: (Debate, Own) & -0.30 [-0.74, 0.14] & -0.01 [-0.29, 0.27]\\
Arm: (Write, Opp) & -0.20 [-0.70, 0.30] & 0.00 [-0.33, 0.33]\\
Strong opinion & -0.52 [-0.78, -0.27]*** & -0.63 [-0.76, -0.49]***\\
Topic: Affirmative action & 0.25 [-0.16, 0.66] & 0.01 [-0.20, 0.22]\\
Topic: Campus mask mandate & 0.57 [0.14, 1.01]** & 0.15 [-0.07, 0.37]\\
Topic: Student debt relief & 0.71 [0.36, 1.06]*** & 0.14 [-0.04, 0.31]\\
Topic: Transgender athletes & 0.32 [-0.11, 0.74] & -0.02 [-0.24, 0.19]\\
Topic: Support for Ukraine & 0.42 [-0.06, 0.91] & -0.09 [-0.33, 0.16]\\
Ethnicity: Black / Hispanic & 0.01 [-0.41, 0.43] & -0.12 [-0.34, 0.11]\\
Ethnicity: Asian & 0.02 [-0.26, 0.30] & -0.01 [-0.16, 0.14]\\
Ethnicity: Other & 0.20 [-0.20, 0.60] & 0.13 [-0.08, 0.35]\\
Gender: Female & -0.27 [-0.52, -0.01]* & -0.01 [-0.15, 0.12]\\
Gender: Other & -0.52 [-1.09, 0.05] & -0.04 [-0.34, 0.25]\\
Political viewpoint: Prefer not to say & -0.44 [-1.01, 0.14] & -0.36 [-0.67, -0.05]*\\
Political viewpoint: Conservative & -0.22 [-0.66, 0.22] & -0.20 [-0.44, 0.04]\\
Political viewpoint: Liberal & -0.33 [-0.62, -0.03]* & -0.24 [-0.40, -0.08]**\\
Best argument × post & -0.01 [-0.28, 0.26] & -0.07 [-0.30, 0.16]\\
Best argument × follow & -0.22 [-0.55, 0.11] & -0.18 [-0.46, 0.09]\\
Judged authentic × post & -0.12 [-0.40, 0.16] & -0.21 [-0.44, 0.03]\\
Judged authentic × follow & 0.07 [-0.28, 0.42] & 0.31 [0.02, 0.60]*\\
(Debate, Opp) × post & 0.35 [-0.06, 0.77] & 0.45 [0.09, 0.80]*\\
(Debate, Opp) × follow & 0.40 [-0.11, 0.90] & 0.64 [0.22, 1.06]**\\
(Debate, Own) × post & 0.29 [-0.12, 0.70] & 0.10 [-0.25, 0.44]\\
(Debate, Own) × follow & 0.18 [-0.32, 0.69] & 0.31 [-0.10, 0.73]\\
(Write, Opp) × post & 0.56 [0.06, 1.06]* & 0.78 [0.36, 1.21]***\\
(Write, Opp) × follow & -0.11 [-0.74, 0.51] & 0.84 [0.32, 1.35]**\\
SD (participant intercept) & 0.62 & 0.24\\
SD (debate intercept) & 0.33 & 0.10\\
SD (residual) & 0.63 & 0.54\\
Observations & 523 & 523\\
ICC & 0.6 & 0.2\\
\bottomrule
\end{tabular}
\par\smallskip
\parbox{0.95\linewidth}{\footnotesize 95\% confidence intervals in brackets.}
\end{table}

\subsubsection{Simplified Anywin Model}
\label{app:models:h3_app:anywin}

\begin{figure}[h]
\centering
\includegraphics[width=0.5\textwidth]{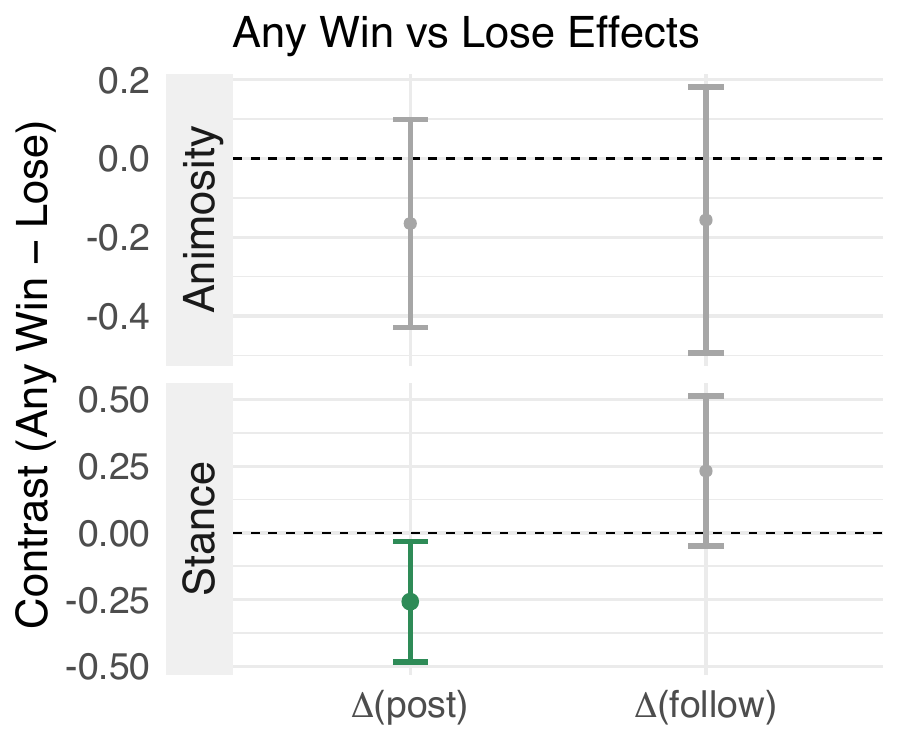}
\caption{\textbf{Simplified model checking the OR win condition.} Most contrasts don't have evidence of significance, except for a slightly significant negative effect for winning regarding stance on the post condition. The same effect cannot be detected at follow-up.}
\label{fig:h3_anywin_plot}
\end{figure}

To test whether simpler operationalizations of winning yielded similar results, we estimated models using a binary \textit{anywin} variable (winning on either best argument OR authentic connection dimensions). Mixed-effects models with the same structure as the primary analysis revealed no statistically significant effects of winning versus losing on either animosity or stance at post-debate or follow-up (Figure~\ref{fig:h3_anywin_plot}). The only marginal effect was observed for stance at post-debate ($\beta = -0.26$, $p = 0.092$), though this did not persist at follow-up and did not survive multiple comparison correction. These results suggest that the distinctions between types of winning captured in our primary analysis provide more meaningful insights than a simple win-lose dichotomy.

\subsubsection{Change Score Specification}
\label{app:models:h3_app:changescore}

We further validated our primary difference-in-differences approach by comparing it with change score models, which directly model post-pre and follow-up-pre differences rather than leveraging the full longitudinal structure. The change score models included the same win mechanism contrasts, arm adjustments, and covariates as the primary analysis.

\begin{figure}[h!]
\centering
\includegraphics[width=\textwidth]{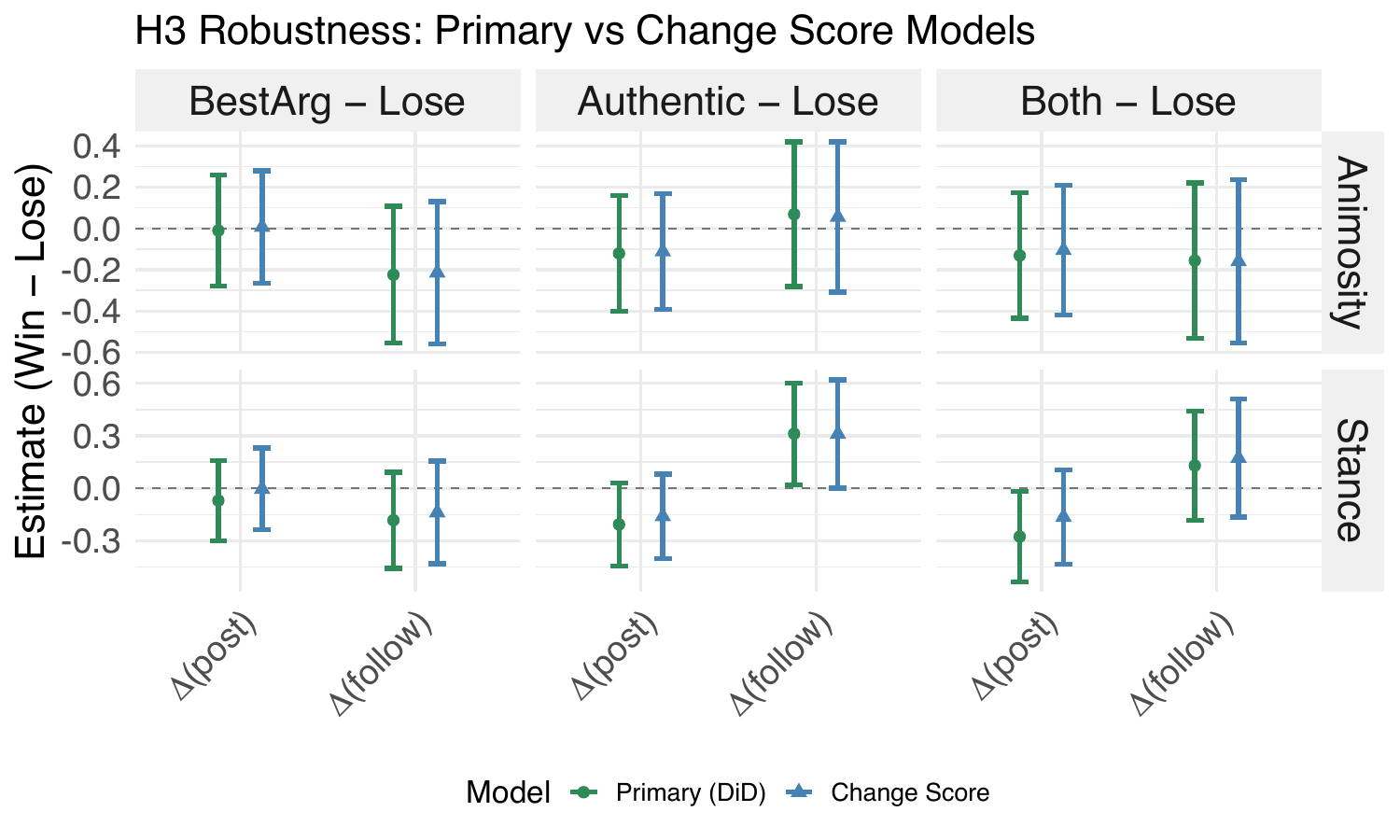}
\caption{\textbf{Comparison of effect estimates between primary difference-in-differences and change score specifications for H3 win mechanism analyses.} Error bars represent 95\% confidence intervals.}
\label{fig:h3-model-comparison}
\end{figure}

Comparison between the primary and change score specifications revealed substantial agreement in effect directions and magnitudes (\Cref{fig:h3-model-comparison}). Of the 12 mechanism-by-time contrasts examined (3 mechanisms × 2 time points × 2 outcomes), 10 (83\%) showed directional agreement between models, with only 2 contrasts (17\%) showing disagreement in sign. No contrasts reached statistical significance in either specification after Holm correction for multiple comparisons, consistent with the primary analysis findings. The similarity in results across difference-in-differences and change score specifications strengthens confidence in our primary conclusions regarding the limited and nuanced associations between self-perceived winning and outcome changes.

\subsubsection{Model Specification Sensitivity Analysis}
\label{app:models:h3_app:specification}

We conducted formal model comparisons to determine the optimal specification for win mechanism effects. The goal was to simplify the model as the specification started to become quite complicated for the time variant per arm model. Comparing three nested models:

\begin{itemize}
\item Model A: Full interaction between win mechanisms, arm, and time
\item Model B: Win mechanisms × time + arm × time (primary specification)
\item Model C: Win mechanisms × time only
\end{itemize}

Likelihood ratio tests revealed no evidence for mechanism-by-arm heterogeneity ($\chi^2(18) = 14.92$, $p = 0.668$), supporting the pooled-over-arms specification. The improvement from adding arm-by-time adjustment was marginal ($\chi^2(9) = 14.97$, $p = 0.092$) and information criteria favored the more parsimonious Model C. However, we retained arm-by-time adjustment in our primary specification to maintain consistency with H1/H2 analyses and ensure complete accounting for experimental design factors.

\subsection{Additional Details for H4}
\label{app:analysis:h4}

This section presents supplementary analyses for Hypothesis 4 (H4), which examined participants' willingness to repeat the experience without compensation. Across multiple robustness checks---including inverse probability weighting for non-response and ordinal modeling of the full response scale---we found consistent evidence supporting the primary conclusion of no meaningful differences in willingness to repeat across experimental conditions. Non-inferiority tests remained inconclusive across all specifications.

\subsubsection{Main Model Specification}

Table~\ref{tab:h4_model_table_or} presents the logistic mixed-effects model for willingness to repeat without compensation, with coefficients expressed as odds ratios for interpretability. Values greater than 1 indicate increased odds of willingness to repeat. The model includes a random intercept for debates and fixed effects for treatment conditions and control variables. Key probability differences and non-inferiority tests were derived from estimated marginal means of this model specification.

\begin{table}[h]
\centering\centering
\caption{\label{tab:h4_model_table_or}\textbf{Logistic Mixed Model for H4: Willingness to Repeat (Odds Ratios).} Values >1 indicate increased odds of willingness to repeat. Random intercept for debate included. Reference categories: Write modality, Own perspective, No strong opinion, Topic: Abortion rights, Ethnicity: White, Gender: Male, Political viewpoint: Neutral.}
\centering
\begin{tabular}[t]{lc}
\toprule
  & Willingness to Repeat\\
\midrule
Intercept & 0.06 [0.01, 0.45]**\\
Modality: Debate & 2.10 [0.66, 6.72]\\
Perspective: Opp & 2.55 [0.65, 10.04]\\
Strong opinion & 2.70 [1.29, 5.66]**\\
Topic: Affirmative action & 0.51 [0.16, 1.63]\\
Topic: Campus mask mandate & 1.53 [0.53, 4.44]\\
Topic: Student debt relief & 0.89 [0.36, 2.15]\\
Topic: Transgender athletes & 0.24 [0.06, 0.99]*\\
Topic: Support for Ukraine & 0.87 [0.25, 3.01]\\
Ethnicity: Black / Hispanic & 1.56 [0.49, 4.95]\\
Ethnicity: Asian & 1.76 [0.79, 3.92]\\
Ethnicity: Other & 0.55 [0.16, 1.92]\\
Gender: Female & 0.69 [0.33, 1.45]\\
Gender: Other & 0.72 [0.15, 3.50]\\
Political viewpoint: Prefer not to say & 1.45 [0.29, 7.21]\\
Political viewpoint: Conservative & 0.85 [0.23, 3.12]\\
Political viewpoint: Liberal & 1.10 [0.47, 2.58]\\
Debate × Opp & 0.28 [0.06, 1.33]\\
SD (debate intercept) & 0.00\\
Observations & 203\\
AIC & 263.4\\
BIC & 326.4\\
\bottomrule
\end{tabular}
\par\smallskip
\parbox{0.95\linewidth}{\footnotesize Odds ratios with 95\% confidence intervals in brackets.}
\end{table}

\subsubsection{Response Pattern Analysis}

Table~\ref{tab:h4_response_rates} shows response rates for the willingness-to-repeat question across experimental conditions. Response rates were balanced, ranging around 95\% across arms, with no systematic pattern by modality or perspective. 

\begin{table}[h]
\centering
\caption{\label{tab:h4_response_rates}\textbf{Response Rates by Experimental Condition.}}
\centering
\begin{tabular}[t]{lccc}
\toprule
Condition & Total N & Responded & Response Rate\\
\midrule
Write/Own & 25 & 24 & 96.0\%\\
Write/Opp & 25 & 25 & 100.0\%\\
Debate/Own & 78 & 75 & 96.2\%\\
Debate/Opp & 75 & 72 & 96.0\%\\
\bottomrule
\end{tabular}
\end{table}

\begin{figure}[h!]
\centering
\includegraphics[width=0.8\textwidth]{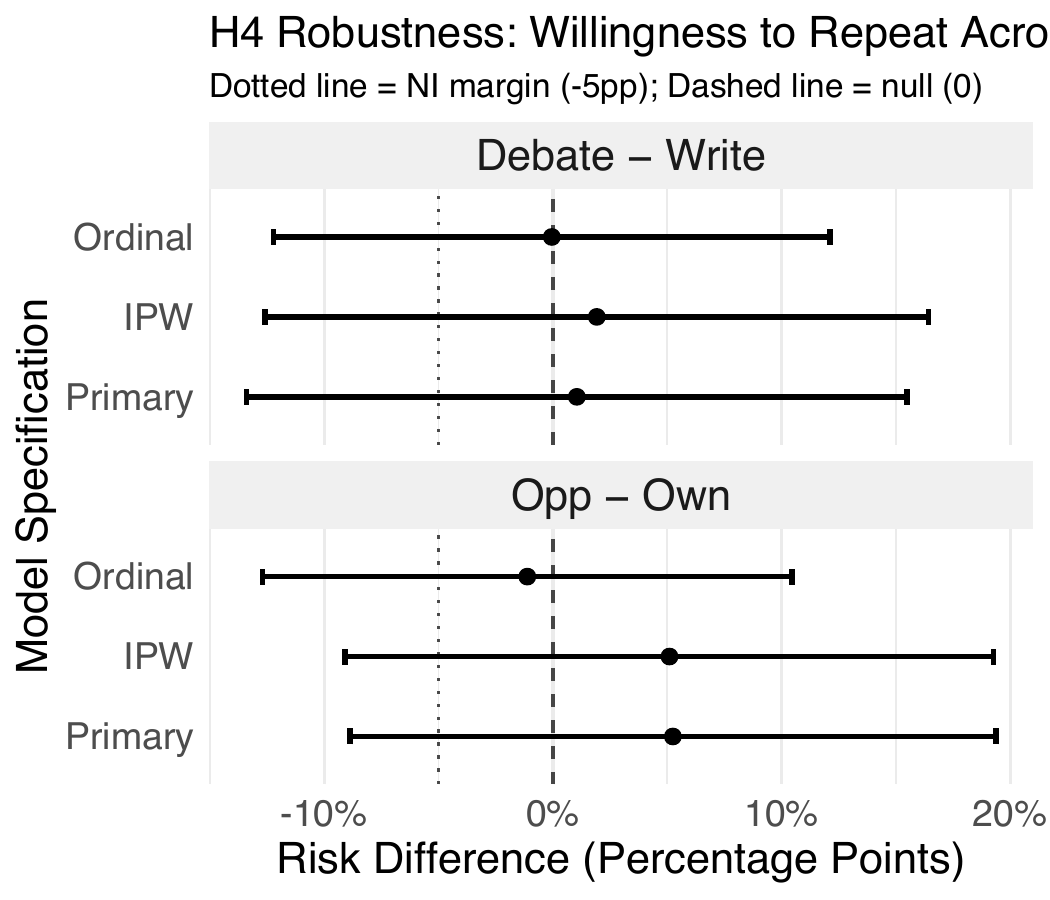}
\caption{\textbf{Comparison of risk difference estimates across primary, IPW, and ordinal specifications for H4.} Error bars represent 95\% confidence intervals. The dotted line indicates the -5 percentage point non-inferiority margin.}
\label{fig:h4-robustness}
\end{figure}

\subsubsection{Non-Response Analysis with Inverse Probability Weighting}
\label{app:analysis:h4:ipw}

Some participants did not respond to the willingness-to-repeat question, raising potential concerns about selection bias. To address this, we implemented inverse probability weighting (IPW) using a propensity model that predicted response based on experimental conditions and baseline covariates. Weights were stabilized and truncated at the 1st and 99th percentiles to limit extreme values.

The IPW sensitivity analysis yielded results nearly identical to the primary specification (\cref{fig:h4-robustness}). For the Debate vs Write contrast, the IPW estimate was 0.6 percentage points (95\% CI: -14.0, 15.2) compared to the primary estimate of 1.0 percentage points (95\% CI: -14.1, 16.1). Similarly, for the Opp vs Own contrast, the IPW estimate was 4.5 percentage points (95\% CI: -9.9, 19.0), compared with the primary estimate of 3.8 percentage points (95\% CI: -11.3, 18.9). The consistency across specifications suggests minimal bias from non-response.

\subsubsection{IPW Implementation Details}

Table~\ref{tab:h4_ipw_diagnostics} provides diagnostics for the inverse probability weighting analysis. Weights were well-behaved with minimal variation (mean = 1.00, SD = 0.15), and truncation affected only extreme values. The balanced weights across conditions support the robustness of IPW adjustment.

\begin{table}[h]
\centering
\caption{\label{tab:h4_ipw_diagnostics}\textbf{IPW Diagnostics for Non-Response Adjustment.}}
\centering
\begin{tabular}[t]{ll}
\toprule
Diagnostic Metric & Value\\
\midrule
Response Rate & 96.6\%\\
Mean Weight & 1.000\\
SD Weight & 0.065\\
Weight Range & {}[0.970, 1.296]\\
Truncation Points & {}[0.970, 1.296]\\
\bottomrule
\end{tabular}
\end{table}

\subsubsection{Ordinal Model}
\label{app:analysis:h4:ordinal}

To recover information lost by dichotomizing the original 5-point Likert scale response, we fitted a cumulative logit mixed model (CLMM) with three ordered categories (No, Indifferent, Yes). This approach leverages the full ordinal structure of the response variable and typically provides more precise estimates.

The ordinal model produced slightly tighter confidence intervals while maintaining similar point estimates (\cref{fig:h4-robustness}). For the Debate vs Write contrast, the ordinal estimate was 0.3 percentage points (95\% CI: -11.9, 12.5) and for the Opp vs Own contrast, 1.2 percentage points (95\% CI: -12.0, 14.4). Despite the improved precision, the substantive conclusions remained unchanged: no evidence of meaningful differences across experimental conditions.

\subsubsection{Ordinal Model Specification}

Table~\ref{tab:h4_ordinal_coefficients} presents the full ordinal model coefficients. The cumulative logit model estimates thresholds between response categories and provides coefficients on the log-odds scale. While the model leverages the full ordinal information, the substantive conclusions align with the primary binary specification.

\begin{table}[!h]
\centering
\caption{\label{tab:h4_ordinal_coefficients}\textbf{Ordinal Model Coefficients for Willingness to Repeat.}}
\centering
\small
\begin{tabular}[t]{lrrr}
\toprule
Predictor & Log-odds & SE & 95\% CI\\
\midrule
Threshold: No | Indifferent & 1.904 & 0.905 & {}[0.131, 3.677]\\
Threshold: Indifferent | Yes & 3.002 & 0.942 & {}[1.155, 4.849]\\
Modality: Debate & 0.227 & 0.476 & {}[-0.706, 1.161]\\
Perspective: Opp & 0.168 & 0.581 & {}[-0.971, 1.306]\\
Strong opinion & 1.203 & 0.361 & {}[0.495, 1.911]\\
\addlinespace
Topic: Affirmative action & -0.043 & 0.514 & {}[-1.051, 0.965]\\
Topic: Campus mask mandate & 0.742 & 0.546 & {}[-0.328, 1.812]\\
Topic: Student debt relief & 0.008 & 0.433 & {}[-0.841, 0.857]\\
Topic: Transgender athletes & -1.101 & 0.565 & {}[-2.209, 0.007]\\
Topic: Support for Ukraine & 0.097 & 0.591 & {}[-1.062, 1.255]\\
\addlinespace
Ethnicity: Black / Hispanic & 0.434 & 0.560 & {}[-0.663, 1.531]\\
Ethnicity: Asian & 0.775 & 0.386 & {}[0.019, 1.531]\\
Ethnicity: Other & -0.221 & 0.529 & {}[-1.258, 0.815]\\
Gender: Female & -0.266 & 0.338 & {}[-0.928, 0.397]\\
Gender: Other & -0.513 & 0.744 & {}[-1.972, 0.945]\\
\addlinespace
Political viewpoint: Prefer not to say & 0.312 & 0.782 & {}[-1.219, 1.844]\\
Political viewpoint: Conservative & -0.338 & 0.573 & {}[-1.462, 0.786]\\
Political viewpoint: Liberal & 0.241 & 0.372 & {}[-0.488, 0.971]\\
Debate × Opp & -0.466 & 0.672 & {}[-1.783, 0.852]\\
\bottomrule
\end{tabular}
\end{table}

\subsubsection{Comprehensive Non-Inferiority Testing}

Table~\ref{tab:h4_ni_comprehensive} extends the non-inferiority analysis to multiple margins across all model specifications. The consistent failure to demonstrate non-inferiority across specifications and margins underscores the inconclusive nature of these tests, regardless of the analytical approach.

\begin{table}[!h]
\centering
\caption{\label{tab:h4_ni_comprehensive}\textbf{Non-Inferiority Test Results Across Specifications.}}
\centering
\begin{tabular}[t]{llcccl}
\toprule
Model & Contrast & NI Margin & Estimate & Lower CI & NI Result\\
\midrule
Primary & Debate - Write & 3 pp & 0.010 & -0.134 & Fail\\
Primary & Debate - Write & 5 pp & 0.010 & -0.134 & Fail\\
Primary & Debate - Write & 7 pp & 0.010 & -0.134 & Fail\\
Primary & Opp - Own & 3 pp & 0.052 & -0.089 & Fail\\
Primary & Opp - Own & 5 pp & 0.052 & -0.089 & Fail\\
\addlinespace
Primary & Opp - Own & 7 pp & 0.052 & -0.089 & Fail\\
IPW & Debate - Write & 3 pp & 0.019 & -0.126 & Fail\\
IPW & Debate - Write & 5 pp & 0.019 & -0.126 & Fail\\
IPW & Debate - Write & 7 pp & 0.019 & -0.126 & Fail\\
IPW & Opp - Own & 3 pp & 0.051 & -0.091 & Fail\\
\addlinespace
IPW & Opp - Own & 5 pp & 0.051 & -0.091 & Fail\\
IPW & Opp - Own & 7 pp & 0.051 & -0.091 & Fail\\
Ordinal & Debate - Write & 3 pp & -0.001 & -0.122 & Fail\\
Ordinal & Debate - Write & 5 pp & -0.001 & -0.122 & Fail\\
Ordinal & Debate - Write & 7 pp & -0.001 & -0.122 & Fail\\
\addlinespace
Ordinal & Opp - Own & 3 pp & -0.011 & -0.127 & Fail\\
Ordinal & Opp - Own & 5 pp & -0.011 & -0.127 & Fail\\
Ordinal & Opp - Own & 7 pp & -0.011 & -0.127 & Fail\\
\bottomrule
\end{tabular}
\par\smallskip
\parbox{0.95\linewidth}{\footnotesize Pass = lower confidence interval above NI margin}
\end{table}

\subsection{Partner Effects in Debates}
\label{app:analysis:partner}

Each debate paired one \Own{} and one \Opp{} participant on the same topic, so a participant's change could depend on their partner. If partners influenced each other, people who debated together should change more alike (convergence) or less alike (divergence) than people who did not. We tested this in the debate arms ($n=153$), using changes from pre to post and from pre to follow-up in baseline standard-deviation units. Both partners must be observed, which leaves 75 dyads at post and 35 at follow-up.

\textbf{Partner correlation.} We regressed each change on perspective, topic, gender, ethnicity, political viewpoint, and opinion strength, and let the two partners' residuals correlate (compound-symmetry generalized least squares, REML). The partner correlation $\rho$ can range from $-1$ to $1$. We report profile-likelihood 95\% confidence intervals and likelihood-ratio tests of $\rho=0$. Because $\rho$ also absorbs anything else partners shared, such as the session and the conversation, it is an upper bound on convergent influence.

\textbf{Permutation test.} Keeping each \Own{} participant fixed, we randomly reassigned \Opp{} participants among debates on the same topic 5,000 times and recomputed the correlation of \Own{} and \Opp{} residuals. The two-sided $p$-value is the share of shuffles with an absolute correlation at least as large as the observed one. Shuffling within session and topic gave the same conclusions (smallest $p=0.17$).

\textbf{Partner characteristics.} In separate models we added the partner's baseline animosity and extremity (absolute baseline stance) as predictors of a participant's own change. We applied Holm corrections within each family of tests.

\textbf{Results.} No partner correlation differed from zero (Table~\ref{tab:partner_rho}). The largest was for animosity at post ($\rho=0.12$, 95\% CI $[-0.12, 0.35]$), and every observed correlation fell inside the range produced by random re-pairing. Partners' baseline animosity and extremity did not predict own change (all $p\ge0.11$, Holm-adjusted $p\ge0.89$). An exploratory model with the partner's judged performance, which is measured after the debate, was also null. This does not show that partner influence is zero. At post, the data exclude correlations above 0.35 for animosity and 0.24 for stance; at follow-up, with 35 dyads, they exclude only $|\rho|$ above about 0.5.

\begin{table}[ht]
\centering
\caption{\textbf{Correlation between debate partners' changes.} $\rho$: compound-symmetry GLS partner correlation, with profile-likelihood 95\% CI and likelihood-ratio $p$. Perm.\ $r$: \Own{}--\Opp{} residual correlation; shuffled range: central 95\% of 5,000 within-topic re-pairings. All Holm-adjusted $p$-values equal 1.00.}
\label{tab:partner_rho}
\small\setlength{\tabcolsep}{4pt}
\begin{tabular}{llrrcrrcr}
\hline
Outcome & Time & Dyads & $\rho$ & 95\% CI & $p$ & Perm.\ $r$ & Shuffled range & $p$ perm \\
\hline
Animosity & Post & 75 & 0.12 & $[-0.12, 0.35]$ & 0.32 & 0.08 & $[-0.24, 0.22]$ & 0.49 \\
Animosity & Follow-up & 35 & 0.05 & $[-0.36, 0.45]$ & 0.81 & $-0.04$ & $[-0.42, 0.24]$ & 0.82 \\
Stance & Post & 75 & 0.00 & $[-0.24, 0.24]$ & 0.99 & $-0.04$ & $[-0.27, 0.21]$ & 0.75 \\
Stance & Follow-up & 35 & $-0.13$ & $[-0.49, 0.30]$ & 0.56 & $-0.18$ & $[-0.39, 0.33]$ & 0.38 \\
\hline
\end{tabular}
\end{table}

\section{Participant Recruitment, Experimental Protocols, and Logistics}
\label{app:protocols}

This section gives an overview of the experimental procedures, including recruitment, session logistics, randomization protocols, and quality control measures, to ensure full transparency and reproducibility.

\subsubsection{Recruitment Implementation Details}
\begin{itemize}
    \item \textbf{Pool 1 (Course-based):} Students enrolled in information science courses from Fall 2023 to Fall 2024. Instructors announced participation early to the middle of the semester. Sessions were run by the latter half of the fall semester.
    \item \textbf{Pool 2 (Volunteer pool):} Departmental research participation system (ORSEE). A subset of 200 eligible participants was sampled and notified.
    \item \textbf{Pool separation rationale:} Different compensation structures (course credit vs. monetary payment) required separate administrative handling and logistics.
    \item \textbf{Session composition:} No mixing of pools within sessions to maintain compensation consistency.
    \item \textbf{Recruitment timeline:} from Fall 2023 to Fall 2024.
\end{itemize}

\subsubsection{Pre-Session Survey Coordination}
\begin{itemize}
    \item \textbf{Time window:} The pre-intervention survey was completed 1-2 weeks (and up to 1 day) before the session.
    \item \textbf{Reminder procedures:} Participants were reminded via email 3 days and 1 day before their session if the survey was incomplete.
    \item \textbf{Handling non-completion:} Participants who attended without completing the pre-survey could not be matched for debates. They were offered the show-up compensation and could optionally serve as judges (their data was not used in the primary analysis).
\end{itemize}

\subsubsection{Pool Equivalence Verification}

We verified the equivalence of class and lab recruitment pools to justify pooling their data in analyses. Although some demographic differences were observed---specifically in topic distribution and gender composition---several factors support pooling. First, the win-based compensation mechanism (requiring both authentic connection and best argument conditions) showed no significant differences between pools ($p > 0.50$), ensuring equivalent incentive structures. Second, baseline measures of \textbf{issue-based animosity} and \textbf{stance} were balanced across pools. Third, we implemented CBPS weighting that successfully addressed observed imbalances, reducing all standardized mean differences below 0.10. Fourth, the experimental design maintained pool purity across sessions, preventing cross-contamination. Sensitivity analyses confirmed that results remained substantively unchanged when adjusting for pool source using these weights (see Section~\ref{app:analysis:balance} for detailed balance assessment).

\subsection{Registration, Setup \& Matching}  
\label{app:protocols:intro-matching}  

Participants arrived at the lab and were assigned anonymized credentials to access workstations with the RocketChat messaging interface. A 10-minute standardized presentation introduced the activity, the compensation structure (emphasizing rewards for argument quality and authenticity), and the rules, including not revealing identifiable details in their statements or conversations. Crucially, participants were not informed whether they would debate their own or opposing viewpoints.  

During the presentation, a stochastic matching algorithm ran in the background to pair participants:  
\begin{itemize}
    \item 50\% of participants were randomly assigned to defend their \textit{opposite} pre-survey position (pretenders).  
    \item The algorithm matched pretenders with non-pretenders (truthful debaters) who shared the \textit{same original stance} on a topic, ensuring debates always paired participants with identical baseline views but divergent assigned roles.  
    \item Topics were selected from participants' pre-survey eligible issues (non-neutral positions).  
\end{itemize}  

For example, in a session with 15 participants, the algorithm generated 7 debate pairs (14 participants) after 3-5 iterations, leaving 1 unmatched individual. Unmatched participants received partial compensation and were given the option to assist in post-debate judging. In such cases, they would be compensated for their time spent waiting for the debates to conclude or for writing.  

\subsection{Debate Protocol}  
\label{app:protocols:debate}  

\paragraph{Preparation Phase (25 minutes)}  
After matching, participants were notified via chat of their assigned topic and position (either their own or the opposite). They prepared opening statements using any resources, including LLMs, though participants were likewise reminded that they would be judged by others on \emph{Authenticity}: how convincingly they embody their assigned position, which would significantly impact their rewards. Instructions for the preparation also included a soft limit of 250 words and suggested focusing on the main points of argumentation for their assigned position.

\paragraph{Conversation/Debate Phase (25 minutes)}  
Participants posted opening statements to a private chat channel with their matched pair. They then engaged in free-form dialogue, guided only by:  

\begin{itemize}  
    \item Periodic announcements of remaining time for the debate 
    \item Periodic reminders to ``maintain a conversational tone''
    \item A ``wrap-up'' warning at 5 minutes remaining
\end{itemize}  

Post-debate surveys assessed self-perceived performance and changes in attitude. The interface preserved anonymity throughout, with no personal identifiers or post-session interaction.

\subsection{Writing Intervention}  
\label{app:protocols:writing}  

The writing intervention paralleled the debate condition in recruitment, matching, and preparation phases, diverging only after argument preparation. Participants were assigned to defend their own or opposing perspectives on a topic, with identical preparation materials and time constraints (25 minutes). Critically, recruitment materials, instructions, and session framing avoided all references to debate, especially adversarial interaction. Instead, participants were told their essays would be evaluated by judges who would compare arguments across sessions, thereby incentivizing a persuasive defense of their assigned position without implying direct competition. This was done to avoid triggering a defensive stance or an overly adversarial attitude from the writing modality. 

Following preparation, participants composed essays defending their assigned stance, guided by the prompt:  
\begin{quote}  
    ``Craft a compelling, self-contained defense of your assigned position. Judges will reward clarity, coherence, and authenticity in representing this viewpoint.''  
\end{quote}  

A 300-word guideline (``about 1.5 pages'') was suggested to standardize depth, although no technical enforcement was implemented. Participants could use LLMs but were reminded that authenticity, judged via stylistic consistency and argument plausibility, constituted 50\% of performance rewards.  

While sharing matching and incentive structures with debates, the writing condition eliminated three debate-specific elements: (1) real-time interaction, replaced by isolated composition; (2) conversational framing, with tasks presented as standalone persuasive exercises; and (3) synchronous performance pressure, allowing iterative drafting. This design isolated the cognitive demands of perspective-taking from the social dynamics of debate.  

\subsection{Judging}  
\label{app:protocols:judging}  

All participants transitioned to judging immediately after completing their intervention task (debate or writing) and the post-activity survey. This concurrent design ensured every debate and written argument received evaluations while minimizing delays in compensation calculation. Judges reviewed anonymized outputs from the \textit{same session}( debate chat logs or essays) through a secure platform, with each output evaluated by at least three peers to establish majority consensus. Judges accessed these materials via individualized survey links, submitting one evaluation for each assigned debate or writing. Additionally, it is worth noting that judges never evaluated the person they debated, but rather assessed the interactions between other pairs.

The judging task required participants to: (1) identify which debater (if any) was inauthentically defending an opposing position, and (2) select the most persuasive argument. Compensation for modality performance (debate or writing) depended on these peer judgments: debaters and writers received bonuses if a majority of judges deemed their arguments both persuasive and authentic.

On the other hand, participants were also compensated for their judging efforts. For each answer that they answered correctly, either by detecting the inauthentic or by detecting the best argument. For authenticity, we had ground truth, and for argument quality, we used a criterion where the judge was compensated if they agreed with the majority. The reason for this is the simplicity of implementation, and it was also simple to understand for participants who are already paying attention to other aspects of the compensation structure.

This approach strikes a balance between ecological validity, mirroring real-world contexts where authenticity and evaluation coexist, and experimental control. 

\subsection{Session Logistics and In-Lab Procedures}
\label{app:protocols:logistics}

\subsubsection{Lab Setup and Computer Assignment}
\begin{itemize}
    \item \textbf{Physical setup:} The lab contained 20 computer stations arranged with physical separators to minimize participant interaction and ensure privacy.
    \item \textbf{Assignment method:} Participants were assigned to a numbered computer station via a random card draw upon arrival.
    \item \textbf{Username generation:} Memorable, randomized usernames (e.g., \texttt{happy-badger-01}) were generated using the python-petname package \citep{kirklandDustinkirklandPythonpetname2025} to facilitate easy identification during judging.
    \item \textbf{Login security:} Participants accessed pre-logged systems; no credentials were shared or stored on browsers.
\end{itemize}

\subsubsection{Standardized Session Introduction}
\begin{itemize}
    \item \textbf{Duration:} Approximately 10 minutes.
    \item \textbf{Content:} A standardized presentation provided an overview of activities without revealing treatment assignments, technical instructions for the platform, time expectations, and the question protocol.
    \item \textbf{Standardization:} The same lead experimenter delivered the presentation in all sessions to ensure consistency.
\end{itemize}

\subsubsection{Real-Time Monitoring and Support}
Experimenters were present to monitor progress, answer procedural questions, and handle technical issues (e.g., participants accidentally logging out). Time reminders were given verbally and via the presentation slides. Experimenters were trained to assist without influencing participants' behaviour.

\subsection{Treatment Implementation and Randomization}
\label{app:protocols:treatments}

\subsubsection{Condition Protocols}
\begin{itemize}
    \item \textbf{Debating Condition:}
    \begin{itemize}
        \item After the introduction, participants were matched into pairs and assigned a topic and side (Pro/Con) via the matching algorithm.
        \item The activity had a 20-minute preparation phase (to draft an opening statement) followed by a 25-minute live, text-based debate conducted via RocketChat.
        \item Debate rules (e.g., no personal attacks, no personal information) were provided and enforced.
    \end{itemize}
    \item \textbf{Writing Condition:}
    \begin{itemize}
        \item Participants were individually assigned a topic and a side (Pro/Con).
        \item They were given 20 minutes to write a persuasive text arguing for that side, which was submitted via a private RocketChat message.
        \item There was no interactive component.
    \end{itemize}
    \item \textbf{Judging Phase (All Participants):}
    \begin{itemize}
        \item After the intervention, all participants were assigned to judge 1-2 debates/written arguments from other participants in the same session (never their own).
        \item Judges were assigned using a balanced algorithm (see Section~\ref{app:algorithms}).
    \end{itemize}
\end{itemize}

\subsubsection{Randomization Execution}
\label{app:protocols:randomization}
\begin{itemize}
    \item \textbf{Point of randomization:} Occurred in real-time after the introduction via a custom matching algorithm.
    \item \textbf{Allocation concealment:} Staff were unaware of final assignments until the algorithm executed. The algorithm's output determined debates.
    \item \textbf{Stratification variables:} None used. Matching was based solely on topic positions and the random assignment of the ``pretender'' role.
\end{itemize}

\subsection{Algorithmic Details}
\label{app:algorithms}

\subsubsection{Debate Matching Algorithm}
\label{app:algorithms:debate-algo}
\textbf{Purpose:} To pair participants into pro-con debates on a specific topic, while randomly assigning half to argue for their own view and half to \textit{pretend} to hold the opposite view. The goal is to maximize the number of valid debates.

\textbf{High-Level Description:}
The algorithm takes the list of participants and their previously reported positions on various topics. It first randomly assigns each participant to be either a \textit{sincere} or \textit{pretender} arguer. A \textit{valid debate} is one where two participants are matched on the same topic, and their \textit{apparent} positions (their real position if sincere, the opposite if pretending) are pro vs. con. The problem is framed as a graph matching problem: participants are nodes, and an edge connects two participants if they form a valid debate. The algorithm runs this process multiple times to find the random seed that creates the graph allowing for the maximum number of matches (i.e., the largest maximal matching).

\begin{algorithm}
\caption{\textbf{Debate Matching Algorithm.}}
\label{alg:debate-matching}
\begin{algorithmic}[1]
\Require List of Participants $P$, List of their Positions on Topics
\Ensure List of Debates $D$, List of Unmatched Participants $U$
\State $bestRun \gets \emptyset$
\State $maxMatches \gets 0$
\For{$i \gets 1$ to $N_{\text{runs}}$}
    \State $debatants \gets \emptyset$
    \For{each participant $p \in P$}
        \State $p.isPretender \gets \text{Bernoulli}(0.5)$ \Comment{Randomly assign pretender role}
        \State $debatants \gets debatants \cup p$
    \EndFor
    \State $G \gets \text{BuildGraph}(debatants)$ \Comment{Node per participant, edge for valid debate pairs}
    \State $D_i, U_i \gets \text{MaximalMatching}(G)$ \Comment{Find largest set of valid edges (debates)}
    \If{$|D_i| > maxMatches$}
        \State $maxMatches \gets |D_i|$
        \State $bestRun \gets i$
    \EndIf
\EndFor
\State \Return $D_{bestRun}, U_{bestRun}$
\end{algorithmic}
\end{algorithm}

\subsubsection{Balanced Judge Assignment Algorithm}
\label{app:algorithms:judge-algo}
\textbf{Purpose:} To assign three judges to each debate, ensuring no judge is assigned to their own debate and that the workload is balanced so all participants judge a roughly equal number of debates before any participant judges a second one.

\textbf{High-Level Description:}
The algorithm iterates through each debate. For each debate, it considers all participants who are not in that debate as potential judges. It calculates a probability weight for each judge: judges who have been assigned fewer times are given a higher weight, ensuring they are chosen first. This creates a balanced distribution of the judging task across all participants.

\begin{algorithm}
\caption{\textbf{Balanced Judge Assignment Algorithm.}}
\label{alg:judge-assignment}
\begin{algorithmic}[1]
\Require List of Judges $J$, List of Debates $D$, $judgesPerDebate = 3$
\Ensure Assignment of judges to debates
\State Initialize $assignmentCount[j] \gets 0$ for all $j \in J$
\For{each debate $d \in D$}
    \State $validJudges \gets \{ j \in J \mid j \notin d \}$ \Comment{Exclude debate participants}
    \State $weights \gets \emptyset$
    \State $minCount \gets \min(assignmentCount[validJudges])$
    \For{each judge $j \in validJudges$}
        \State $weight \gets \max(0, 1 - (assignmentCount[j] - minCount)) + \epsilon$
        \State $weights[j] \gets weight$ \Comment{Prefer judges used least}
    \EndFor
    \State $normalizedWeights \gets \text{Normalize}(weights)$
    \State $selectedJudges \gets \text{SampleWithoutReplacement}(validJudges,$
    \Statex \hspace{3em}$judgesPerDebate, normalizedWeights)$
    \For{each judge $j \in selectedJudges$}
        \State $assignmentCount[j] \gets assignmentCount[j] + 1$
    \EndFor
    \State Assign $selectedJudges$ to debate $d$
\EndFor
\end{algorithmic}
\end{algorithm}

\subsection{Post-Session and Follow-Up Procedures}

\subsubsection{Immediate Post-Session Protocol}
\begin{itemize}
    \item \textbf{Debriefing:} Conducted after judging, involving a summary of activities and reminder of future surveys without revealing the full study hypotheses.
    \item \textbf{Compensation:} Participants were told compensation (monetary for Pool 2, credit for Pool 1) and anonymized scores would be delivered via email 1-2 days post-session.
    \item \textbf{Question handling:} Experimenters were available for questions after the session, carefully avoiding discussion of expected effects.
\end{itemize}

\subsubsection{Follow-Up Survey Implementation}
\begin{itemize}
    \item \textbf{Timing:} The follow-up survey was sent via Qualtrics approximately 14 days ($\pm$ 3 days) after the participant's lab session; responses came in a median of 15 days after the session, typically within 1--4 weeks.
    \item \textbf{Reminders:} Two reminder emails were sent at weekly intervals to non-respondents.
\end{itemize}

\subsection{Data Management and Security}
\label{app:data_security}
\subsubsection{Data Protection and Linking}
\begin{itemize}
    \item \textbf{Data Linking:} Participant email from Qualtrics was linked to their in-lab random username via a master mapping file.
    \item \textbf{De-identification:} After data collection, the master mapping file was destroyed, leaving only the anonymized username for all analysis datasets.
    \item \textbf{Storage:} All data was stored on encrypted, secure university servers.
    \item The RocketChat database contained only anonymized usernames and chat logs.
\end{itemize}

\subsubsection{Quality Assurance}
\begin{itemize}
    \item \textbf{Real-time monitoring:} Experimenters monitored sessions for technical issues and protocol adherence.
    \item \textbf{Data validation:} Scripts verified that all post-session and judging surveys were completed before participants left the lab.
\end{itemize}

\subsection{Ethical Considerations}

\subsubsection{IRB Approval}
\begin{itemize}
    \item \textbf{Approval:} This study was approved by the University of Michigan Institutional Review Board (HUM00232834) on May 10, 2023.
    \item \textbf{Consent:} Informed consent was obtained electronically within the pre-intervention survey.
    \item \textbf{Withdrawal:} Participants were informed they could withdraw at any time without penalty. One participant withdrew; their data were not included in analyses.
\end{itemize}

\subsubsection{Risk Management}
\begin{itemize}
    \item \textbf{Potential risks:} Included potential distress from debate and risk of de-anonymization.
    \item \textbf{Mitigation:} Participants were instructed not to share personal information. All data was anonymized and secured. Experimenters were trained to handle distressed participants, though no such cases occurred beyond the single withdrawal.
\end{itemize}

\subsection{Technical Specifications}

\subsubsection{Platform Details}
\begin{itemize}
    \item \textbf{Debate/Judging Platform:} A customized instance of RocketChat \citep{RocketChatRocketChat2025}, hosted on secure university infrastructure.
    \item \textbf{Survey Platform:} Qualtrics was used for all surveys.
    \item \textbf{Session Management:} Custom Python scripts handled matching, room creation, and communication with the RocketChat API.
\end{itemize}

\subsubsection{Hardware and Lab Setup}
\begin{itemize}
    \item The lab consisted of 20 standardized computer stations running Windows and Firefox, which were managed by the school's IT department.
    \item The network and virtual machine hosting the RocketChat instance complied with university security protocols.
\end{itemize}

\section{Survey instruments}
\label{app:surveys}
\subsection*{Overview}
Participants completed three Qualtrics surveys---Pre-intervention (PRE), Post-intervention (POST), and Follow-up (FOL)---plus a separate judging instrument (JDG) administered after POST. Exact instruments, flow, and coding are documented here; the modelling of outcomes and change is described in \secref{sec:methods:measures} of the main text.

\subsection{Issues}
\label{app:surveys:issues}
The issues to debate were the following:
{\raggedright
\texttt{ReliefPlan}, \texttt{SportsTransgender}, \texttt{AbortionRights}, \texttt{AffirmativeAction}, \texttt{CovidMasks}, \texttt{UkraineRussia}.
Machine-readable slugs: \texttt{relief-plan}, \texttt{sports-transgender}, \texttt{abortion-rights}, \texttt{affirmative-action}, \texttt{covid-masks}, \texttt{ukraine-russia}. A summary of each is in \cref{tab:issue_list}.
Verbatim prompts are in Supplementary Data~4 (\texttt{issues\_text.csv}).\par}

\begin{table}[h]
\centering
\caption{\textbf{Issue list with machine-readable slugs and lead statements.}}
\label{tab:issue_list}
\begin{tabularx}{\textwidth}{@{} l l X @{}}
\toprule
\textbf{Issue} & \textbf{Slug} & \textbf{Lead statement (abridged)} \\
\midrule
ReliefPlan        & \texttt{relief-plan}        & Consider the Biden--Harris Administration's Student Debt Relief Plan. \\
SportsTransgender & \texttt{sports-transgender} & Allow high school students to join teams matching their gender identity. \\
AbortionRights    & \texttt{abortion-rights}    & Proposal to protect abortion rights in Michigan, modeled on New Jersey’s law. \\
AffirmativeAction & \texttt{affirmative-action} & Proposal to repeal Michigan’s constitutional ban on affirmative action. \\
CovidMasks        & \texttt{covid-masks}        & UM considering update requiring masks in indoor and crowded spaces. \\
UkraineRussia     & \texttt{ukraine-russia}     & The US should maintain or increase support for Ukraine’s war effort against Russia. \\
\bottomrule
\end{tabularx}

\begin{flushleft}\footnotesize
\textbf{Notes:} All six issues were shown to every participant at each wave; issue order was randomized without replacement. Verbatim prompts are archived in Supplementary Data~4 (\texttt{issues\_text.csv}).
\end{flushleft}
\end{table}

\subsection{Pre-intervention (PRE)}
\paragraph{Identification \& consent} Minimal identifiers were collected for linkage; anonymized IDs were used in analysis (no PII released).
\paragraph{Demographics} Baseline demographics collected at PRE (see items in Supplementary Data~1, \texttt{module=demog}).
\paragraph{Issue block} All six issues were shown to every participant; see Section~\ref{app:surveys:shared} for shared item bank and randomization.
\paragraph{Flow} See Supplementary Data~2 (\texttt{instruments/instrument\_flow.csv}, PRE rows).

\subsection{Post-intervention (POST)}
\paragraph{Identification} As above.
\paragraph{Issue block} As in \S E.
\paragraph{Self-assessment module} Prediction of judges’ evaluations and willingness to repeat the activity; wording matched the participant’s modality (writing/debating). Exact items in Supplementary Data~1 (\texttt{module=self}); scoring in Section~\ref{app:surveys:scoring}.
\paragraph{Flow} See Supplementary Data~2 (POST rows).

\subsection{Follow-up (FOL)}
\paragraph{Identification} As above.
\paragraph{Issue block} As in \S E.
\paragraph{Self-assessment module} As in POST (if administered).
\paragraph{Flow} See Supplementary Data~2 (FOL rows).

\subsection{Judging surveys (JDG)}
\paragraph{Assignment} After POST, judges evaluated 1--2 conversations from their session.
\paragraph{Rubric} Items elicited: most persuasive side; persuasiveness ratings for Pro and Con; who was more likely pretending; confidence. Exact wording/options in Supplementary Data~1 (\texttt{survey=JDG, module=judging}); aggregation rules in Section~\ref{app:surveys:scoring}.
\paragraph{Flow} See Supplementary Data~2 (JDG rows).

\subsection{Shared modules \& randomization (PRE/POST/FOL)}\label{app:surveys:shared}

\paragraph{Randomization}
All six issues were shown; issue order was randomized without replacement per participant. Within-issue item order was fixed unless noted.

\paragraph{Common item bank (per issue)}
Each issue included: (i) stance (5-point Likert) and discussion excitement (5-point Likert); (ii) two feeling-thermometer ratings toward people who agree vs.\ disagree with the statement (0--100 slider in steps of 10); and (iii) self-perceived understanding batteries for agree and disagree targets (4 items each; 5-point Likert). See Supplementary Data~1 (\texttt{instruments/instrument\_items.csv}, \texttt{module=issue}).

\subsubsection*{E.1 Instrument revision: affective framing}
\paragraph{Original (self-referential) wording}
Warmth toward people who agree \emph{with you} and who disagree \emph{with you}.
\paragraph{Revised (statement-referential) wording}
Warmth toward people who agree \emph{with the statement} and who disagree \emph{with the statement}, irrespective of the respondent’s stance.
\paragraph{Rationale}
Some respondents moved toward neutrality or switched sides across waves, making the self-referential frame ambiguous. The statement-referential wording fixes the reference point across time.
\paragraph{Flagging}
A \texttt{framing\_version} column in the codebook records items as \texttt{v2\_statement}; early sessions that used \texttt{v1\_self} are documented in the dataset.
\paragraph{Harmonization rule (primary)}
For \textbf{animosity} we compute a conservative outcome as the minimum of the two thermometers:
$A_i(s)=\min\{T_{i,\text{agree}}(s), T_{i,\text{disagree}}(s)\}$ (0--100). See Section~\ref{app:surveys:scoring} and \secref{sec:methods:measures}.

\subsection{Coding, scoring, and derived variables}\label{app:surveys:scoring}
\paragraph{Issue-based animosity}
Primary: $A_{\min}=\min(\text{agree},\text{disagree})$ per issue (0--100). Sensitivity: disagree-only; mean of the two; excluding early sessions.
\paragraph{Stance}
5-point Likert mapped to $[-2,2]$ and reoriented by baseline sign so higher values indicate movement toward the opposite side; see main Methods.
\paragraph{Self-perceived understanding}
Optional indices: mean of agree-target items; mean of disagree-target items (1--5).
\paragraph{Willingness to repeat}
Single item (1--5; modality-specific wording in UI).
\paragraph{Judging aggregation}
Binary side choices (Pro=1/Con=0); persuasiveness ratings (1--4); ``pretending'' side (Pro=1/Con=0); confidence (1--4). Majority aggregation across judges described in the Analysis section if used.
\paragraph{Machine-readable specs}
{\raggedright All formulas, valid ranges, and missingness rules are summarized in Supplementary Data~3 (\texttt{scale\_scoring.csv}).\par}

\subsection{No translations}
All instruments were administered in English; no translations were used.

\end{document}